\documentclass[12pt]{iopart}
\usepackage{iopams} 
\input epsf
\expandafter\let\csname equation*\endcsname\relax
 \expandafter\let\csname endequation*\endcsname\relax
 \usepackage[fleqn]{amsmath}
\usepackage{amsopn}
\usepackage{epsfig}
\usepackage{subfig}
\usepackage{graphics,psfrag,rotating}
\usepackage{graphicx}
\usepackage{dcolumn}
\usepackage{bm}
\usepackage{epstopdf}
\usepackage{color}
\usepackage[usenames,dvipsnames,svgnames]{xcolor}
\usepackage[colorlinks=true,
      linkcolor=red,
      urlcolor=gray,
      citecolor=blue]{hyperref}
 \usepackage[]{cite}

\def\c {\mbox{curl}\,}

\def\3nab{\tilde{\nabla}}

\def \t {\tilde}

\def\c{\mbox{curl}}

\def\hsp5{\hspace{5mm}}

\def\case#1/#2{\textstyle\frac{#1}{#2}}

\def\ber {\begin{eqnarray}}
\def\eer {\end{eqnarray}}
\def\bea {\begin{eqnarray}}
\def\eea {\end{eqnarray}}

\def\bc {\begin{center}}
\def\ec {\end{center}}
\def\case#1/#2{\frac{#1}{#2}}

\newcommand{\be}{\begin{equation}}
\newcommand{\bse}{\begin{subequation}}
\newcommand{\ese}{\end{subequation}}
\newcommand{\ee}{\end{equation}}
\newcommand{\eei}{\end{eqnarray}\indent\indent}
\newcommand{\ba}{\begin{array}}
\newcommand{\ea}{\end{array}}
\newcommand{\bal}{\begin{eqnarray}}
\newcommand{\eal}{\end{eqnarray}} 
\newcommand{\n}{{}^{(3)}\nabla}

\def\case#1/#2{\textstyle\frac{#1}{#2} }

\newcommand{\bver}{\begin{verbatim}}
\newcommand{\ever}{\end{verbatim}}

\begin{document}
\title{Confronting the Chaplygin gas with data: background and perturbed cosmic dynamics}
\author{Shambel Sahlu $^{1,2,3}$\footnote{33659273@mynwu.ac.za}, Heba Sami $^{4}$\footnote{
		Corresponding author: hebasami.abdulrahman@gmail.com},  Renier Hough$^{1}$, Maye Elmardi $^{1}$, Anna-Mia Swart $^{4}$ and Amare Abebe$^{1,5}$\footnote{Amare.Abebe@nithecs.ac.za}} 
\address{$^{1}$Centre for Space Research, North-West University, Potchefstroom, South Africa.}
\address{$^{2}$ Entoto Observatory and Research Center, Space Science and Geospatial Institute, Ethiopia.}
\address{$^{3}$ Department of Physics, Wolkite University, Ethiopia.}
\address{$^{4}$School of Chemistry and Physics, University of KwaZulu-Natal, South Africa.}
\address{$^{5}$ National Institute for Theoretical and Computational Sciences (NITheCS), South Africa.}
\date{\today}

\begin{abstract}
	In this paper, we undertake a unified study of background dynamics and cosmological perturbations in the presence of the Chaplygin gas. This is done by first constraining the background cosmological parameters of different Chaplygin gas models with SNIa and $H(z)$ data for detailed statistical analysis of the CG models. Based on the statistical criteria we followed, none of the models has substantial observational support, but we show that the so-called `original' and `extended/generalised' Chaplygin gas models have {\it some observational support} and {\it less observational support}, respectively, whereas the `modified and `modified generalised Chaplygin gas models miss out on the category {\it less observational support}, but cannot be ruled out. The so-called `generalised cosmic Chaplygin gas model, on the other hand, falls under the {\it no observational support} category of the statistical criterion and can be ruled out. The models which are statistically accepted are considered for perturbation level in both theoretical and observational aspects. We also apply the $1+3$ covariant formalism of perturbation theory and derive the evolution equations of the fluctuations in the matter density contrast of the matter-Chaplygin gas system for the models with some or less statistical support. The solutions to these coupled systems of equations are then computed in both short-wavelength and long-wavelength modes. Then feed these observationally restricted parameters into the analysis of cosmological perturbations {to address the growth of density contrast through redshift}. Using the most recent linear growth of the data $f_{\sigma 8}$, CG models are considered to study the linear growth of the structure.  
\end{abstract}
\vspace{.3cm}
\textbf{PACS} {04.50.Kd, 98.80.Jk, 98.80.-k, 95.36.+x, 98.80.Cq}


\section{Introduction}

The major component of the matter-energy density (about 85\%) is made from dark matter, and currently, our Universe is experiencing an accelerated expansion. The discovery of an accelerating expansion of the universe and the existence of dark matter are the primary indicators of the limitation of our knowledge of the laws of physics, and those two uncharted territories of the universe have opened a new era for the development of modern cosmology. In recent times, one of the most active areas of research in cosmology is trying to understand the nature of dark matter and dark energy.
Different suggestions have been made to understand dark energy/vacuum energy. The first suggestion is that the cosmological constant is responsible for the cosmic acceleration \cite{ade2016planck, peebles1993principles, krauss1995cosmological, ostriker1995observational, allemandi2005dark, riess1998observational, frieman2008dark}. In addition to the cosmological constant many alternative dark energy models such as the quintessence energy model \cite{ratra1988cosmological}, the phantom energy model \cite{farnes2018unifying}, and quantum field theory in curved spacetime \cite{sola2022cosmological,moreno2022equation, moreno2022renormalizing},  just as a few have been suggested to demonstrate dark energy, which is the cause of the accelerating expansion universe. The second approach is the modification of the general theory of relativity (GR) \cite{abdel1997modified, kunz2007dark, clifton2012modified, faraoni2005stability, borowiec2007dark}. Meanwhile, scholars have been proposing the Chaplygin Gas (CG) model to mimic the effects of dark energy at late times and the dark matter before that and can be a possible substitution to our standard model of cosmology \cite{kamenshchik2001alternative, bilic2002unification, saadat2013viscous, dev2003cosmological, akbarieh2020evolution}.   This model comes in several versions \cite{Elmardi:2018ryn} the most common of which are the original Chaplygin gas (OCG), generalised Chaplygin gas (GCG), modified generalised Chaplygin gas (MGCG), Extended Chaplygin gas (ECG) and generalised cosmic Chaplygin gas (GCCG) and has been used mostly to study the Friedmann-Lema\^{\i}tre-Robertson-Walker (FLRW) background expansion history of FLRW without much emphasis on observational viability and implications for large-scale structure formation.

In this manuscript, the study of linear cosmological perturbations is another main point of the work using the covariant formalism $1+3$ to investigate the linear growth of the structure in CG models using growth structure data $f_{\sigma 8}$. Indeed, there are two main approaches to studying linear cosmological perturbations, namely the metric formalism \cite{bardeen1980gauge, kodama1984cosmological} and the $1+3$ covariant formalism \cite{ellis1989covarianta, dunsby1991gauge, abebe2012covariant}. Some works as presented in \cite{fabris2002density, gorini2008gauge} have studied the linear cosmological perturbations in the CG model using the metric formalism \cite{gorini2008gauge, fabris2002density}. But the second approach is local and both covariant and gauge-invariant, and it uses the kinematic quantities, the energy-momentum tensor of the fluid, and the graviton-electromagnetic parts of the Weyl tensor instead of the metric formalism. In this approach, the perturbations defined describe the true physical degree of freedom, and no physical gauge modes exist. For these reasons, we are motivated to consider the $1+3$ covariant formalism to observe the contribution of CG models to the growth structure. To our knowledge, there is no work in the literature that considered the cosmological perturbations of this fluid model in the $1+3$ covariant formalism of perturbation theory \cite{ellis1989covarianta, dunsby1991gauge, abebe2012covariant}. This work attempts to fill this important gap of using background-constrained CG models in the study of large-scale structure formation through the covariant perturbations approach.

In the background cosmology, some observational constraints such as the deceleration parameter, the luminosity distance, and the observational Hubble parameter have been studied in the aforementioned CG models using two different datasets to find the best fits for each of the different models, to be able to make conclusions about the viability of each, as well as to get a better feel for how these models compare to the concordance model of cosmology. For our first dataset, we will use Supernovae Type 1A data. This method is called supernovae cosmology, and with cosmic microwave background (CMB) radiation cosmology, it is the two most popular ways to determine, with observable data, the values of the cosmological parameters. The class of supernovae we will be using is when a binary white dwarf (WD) star has accreted enough hydrogen from the low-mass companion star, to form an outer layer, which when compressed enough results in the WD exploding \cite{Tayler1994}. These types of explosions are regarded as standard candles, since the process is always the same, resulting in their luminosities being relatively similar. Therefore, when measuring the incoming flux, it can be used to approximate the distance to the particular supernova, since distance is the only variable when calculating the flux of a standard candle. We can use the redshift $(z)$ obtained for each supernova to represent the approximated distance. The reason for doing this comes down to the fact that we can use the distance modulus function to measure the expansion of the Universe. {In addition to SNIa data, we also consider observational data of the Hubble parameter $H(z)$ to constrain the best-fit values of different observational parameters in the CG models. We also calculated the statistical analysis. We used 29 $H(z)$ data sets from cosmic chronometers or differential ages of galaxies, 14 $H(z)$ data radial Baryon acoustic oscillations (BAO) from the observable effect of BAO in a total of 43 data points presented \cite{sahni2014model}} from the Sloan Digital Sky Survey (SDSS), DR9, and DR11. 

{As previously mentioned}, the other aspect of this work focusses on linear cosmological perturbations in the $1+3$ covariant formalism in multi-fluid systems \cite{abebe2012covariant, sahlu2020scalar, ntahompagaze2020multifluid} adapted to include CG fluid models. After we derive the density perturbation equations and present the solutions in a step-by-step manner, both analytically and numerically, we will discuss the cosmological implications as far as large-scale structure formation is concerned on both sub- and super-horizon scales and see if there is a breaking of background degeneracy at the level of the perturbations \cite{abebe2012covariant, abebe2015breaking}. {We also use the growth structure data $f_{\sigma 8}$ \cite{barros2019coupled, sagredo2018internal, song2009reconstructing}, where the linear growth rate $f$ and $\sigma_8$ is the root-mean-square mass fluctuations with radius $8 h^{-1}Mpc$, which currently provided from the galaxy surveys, and extended the work to study the growth of matter perturbations in various Chaplygin gas models. For analysis purposes we have used 30 growth structure ($f_{\sigma 8}$) data points as presented in \cite{perenon2019optimising}
	
	The Chaplygin gas model acts as dark matter and dark energy in the early and late universe epochs, respectively. In this manuscript, we are motivated to investigate the behaviour of this exotic gas in the background cosmology to understand how to respond to the late-time accelerating universe presented in Secs. \ref{Sec: Background}, \ref{numerical}. We also extended this motive to study the linear cosmological perturbation to clearly understand the contribution of this exotic gas for the formation of large-scale in Sec. \ref{perturbation}.
	
	The layout of the manuscript is as follows. In the following section, we present the background cosmology in CG models. After reviewing the basic Friedmann equations by unifying the exotic fluid with other components like baryonic matter fluids, we present the numerical analysis of the deceleration parameter to examine the current features of the Universe in OCG and GCG, MGCG, ECG, and GCCG in Sub-Sec. \ref{Nobel} - \ref{OGCG2} respectively. The theoretical distance modulus is investigated in CG models and is computed and compared with results of existing SNIa data in Sec. \ref{numerical}. In Section \ref{perturbation}, we also review the basic spatial gradient variables and derive the linear evolution equations, applying the scalar and harmonic decomposition techniques and obtaining the wave-number-dependent energy density fluctuations for both CG and matter fluids. In this section, we also consider $f_{\sigma 8}$ to study the linear growth of structure.  In Sec. \ref{final} - \ref{final1}, we present and discuss the analytical and numerical solutions of density perturbations by considering the baryonic-CG dominant Universe for both wave-length ranges and provide some concluding remarks.
	
	
	\section{Background cosmology in the CG model}\label{Sec: Background}
	
	The negative pressure associated with the CG model is related to a positive energy density by a characteristic equation of state given as \cite{kamenshchik2001alternative, kahya2014observational}:
	\begin{eqnarray}
		p = - \frac{A}{\rho^{\alpha}}\;,
		\label{CGpressure}
	\end{eqnarray}
	where $p$ is the pressure and $\rho$ is the energy density, both in a co-moving reference frame with $\rho >0$. $A$ and $\alpha$ are positive constants\footnote{The values for $\alpha$ are given by CG model for $0<\alpha \leq 1$, and the original OCG at $\alpha = 1$ \cite{ccgmg}.}.
	
	We assume the energy density of the multifluid universe to consist of CG and baryons. The conservation equations for the $i^{\mathrm{th}}$ fluid is given as
	\begin{eqnarray}
		\dot{\rho}_i + 3\frac{\dot{a}}{a}(\rho_i+p_i) = 0\;, \quad \mbox{$i=\{b, ch\}$}.
	\end{eqnarray}
	
	From Eq. \ref{CGpressure} the energy density of the exotic fluid reads
	\begin{eqnarray}
		\rho_{ch} (a) = \left[ A + \frac{C_2}{a^{3(1+\alpha)}}\right]^{\frac{1}{1+\alpha}}\;,
		\label{chgenergydensity}
	\end{eqnarray}
	where $C_2=e^{C_1(1+\alpha)}$ with $C_1$ a constant of integration, which means that $C_2$ is a positive constant. 
	
	We can now look at equation \ref{chgenergydensity} and determine how the energy density of the CG evolves during different epochs.
	
	In the limiting case where $\frac{C_2}{A}\gg a^{3(1+\alpha)}$ for the early universe, the energy density depends on the scale factor as
	\begin{eqnarray}
		\rho_{ch} (a) = \frac{\sqrt{C_2}}{a^3}\;.
		\label{mattercase}
	\end{eqnarray}
	
	From this we can conclude that the CG corresponds to baryonic-like matter which can be interpreted as the dark matter contribution to dust: $\rho_{ch} \sim a^{-3}$. On the other hand, $\frac{C_2}{A}\ll a^{3(1+\alpha)}$ at late times, the energy density \cite{avelino2003lambdacdm} becomes
	\begin{eqnarray}
		\rho_{ch} (a) \approx \pm \sqrt{A}\;,
		\label{constant}
	\end{eqnarray}
	and this equation clearly shows that $\rho_{ch} \sim$ constant, and therefore the CG corresponds to a cosmological constant \cite{quintessence}. On the basis of Eqs. \ref{mattercase} and \ref{constant}, the exotic fluid CG act as dark matter and dark energy in the early and late times \footnote{We assume a flat Universe ($\Omega_k=0$, where $\Omega_{k}$ is the density due to the curvature of space-time)}.
	
	We can thus write the total energy density as the sum of the different energy density components:
	\begin{eqnarray}
		&&\rho_{tot}(a) = \left[ A + \frac{C_2}{a^{3(1+\alpha)}}\right]^{\frac{1}{1+\alpha}} +\rho_{b0} a^{-3}\;,
		\label{rhotot}
	\end{eqnarray}
	with $\rho_{b0}$ corresponding to the present-day values of the baryonic energy densities, respectively.
	
	The total pressure can also be given similarly:
	\begin{eqnarray}
		p_{tot}(a) =  - \frac{A}{\left[\rho_{ch}(a)\right]^{\alpha}}\;.
		\label{ptot}
	\end{eqnarray}
	
	The fluid equation for the total energy density then becomes
	\begin{eqnarray}
		\dot{\rho}_{tot} + 3\frac{\dot{a}}{a}(\rho_{tot}+p_{tot}) = 0\;.
	\end{eqnarray}
	We can also make a change of variables and use the redshift, as the former is a physically measurable quantity. We do this by substituting $1+z=\frac{a_0}{a}$, where $a_0$ is the scale factor at present which we can normalise to 1. The Friedmann equation consequently reads as:
	\begin{eqnarray}
		3H^2(z) &=& \left[ A + C_2(1+z)^{3(1+\alpha)}\right]^{\frac{1}{1+\alpha}} + \rho_{b0}(1+z)^{3}\;, 
		\label{100}
	\end{eqnarray}
	where $H = \dot{a}/a$ is the Hubble parameter. Dividing throughout by $3H^2_0$ ($H_0$ being the value of the Hubble parameter today) in equation \ref{100}, we obtain
	\begin{eqnarray}
		h^2(z) &=&  \left[D_1 + D_2(1+z)^{3(1+\alpha)}\right]^{\frac{1}{1+\alpha}} +\Omega_{b0}(1+z)^{3}\;,
		\label{eq:chaplygin friedmann model}
	\end{eqnarray}
	where $h\equiv \frac{H}{H_0}$ is the normalised Hubble parameter, and $D_1=\frac{A}{(3H_0^2)^{(1+\alpha)}}$, $D_2=\frac{C_2}{(3H_0^2)^{(1+\alpha)}}$, $\Omega_{r0}\equiv \frac{\rho_{r0}}{3H_0^2} $ and $\Omega_{b0}=\frac{\rho_{b0}}{3H_0^2} $, are fractional (dimensionless) density parameters. The total fractional density parameter for a flat universe at any given redshift can be written as:
	\begin{eqnarray}
		\Omega_b   = 1-\Omega_{ch}\;,
	\end{eqnarray}
	where,
	\begin{eqnarray}
		\Omega_b\equiv \frac{\rho_b}{3H^2}=\frac{\Omega_{b0}(1+z)^3}{\left[D_1 + D_2(1+z)^{3(1+\alpha)}\right]^{\frac{1}{1+\alpha}} + \Omega_{b0}(1+z)^{3}}\;,\\
		\Omega_{ch}\equiv \frac{\rho_{ch}}{3H^2}=\frac{\left[D_1 + D_2(1+z)^{3(1+\alpha)}\right]^{\frac{1}{1+\alpha}}}{\left[D_1 + D_2(1+z)^{3(1+\alpha)}\right]^{\frac{1}{1+\alpha}} + \Omega_{b0}(1+z)^{3}}\;.
	\end{eqnarray}
	
	In the following subsections, we further analyse some cosmological constraints, namely the deceleration parameter and luminosity distance to explain the cosmic history of the late time due to the presence of the CG fluid. We then constrain these parameters by comparing the CG models with $\Lambda$CDM to see which of the CG models best mimics the $\Lambda$CDM model.
	
	\textbf{N.B} We will use the values of all parameters from our MCMC simulations of the values of the best-fit parameters for each model tested including the $\Lambda$ CDM as presented in Table \ref{tab: best-fitting parameter values}. 
	
	
	\subsection{Deceleration parameter in CG models}\label{Nobel}
	
	From the Friedmann equation, we derived the deceleration parameter for OCG and GCG, and it yields as:
	\begin{eqnarray}\label{deceleration1}
		&&q(z) =  \frac{ \left[ A + C_2(1+z)^{3(1+\alpha)}\right]^{\frac{1}{1+\alpha}}}{2\left[A + C_2(1+z)^{3(1+\alpha)}\right]^{\frac{1}{1+\alpha} }+ 2\rho_{b0}(1+z)^{3}}\nonumber\\ 
		&& + \frac{ \rho_{b0}(1+z)^{3} -3A\left[A+C_2(1+z)^{3(1+\alpha)}\right]^{-\frac{\alpha}{1+\alpha}} }{2\left[A + C_2(1+z)^{3(1+\alpha)}\right]^{\frac{1}{1+\alpha} }+ 2\rho_{b0}(1+z)^{3}}\;.
	\end{eqnarray}
	
	In the following, we present the numerical results of the above deceleration parameters for different values of $\alpha$. For the case of $\alpha = 1$ and $\alpha\leq 1$, the results of OCG and GCG models respectively, as presented in Figs. \ref{qvsz} - \ref{fig:GCG3}. 
	
	\begin{figure}[ht!]
		\begin{minipage}{0.45\linewidth}
			\includegraphics[width=1\textwidth]{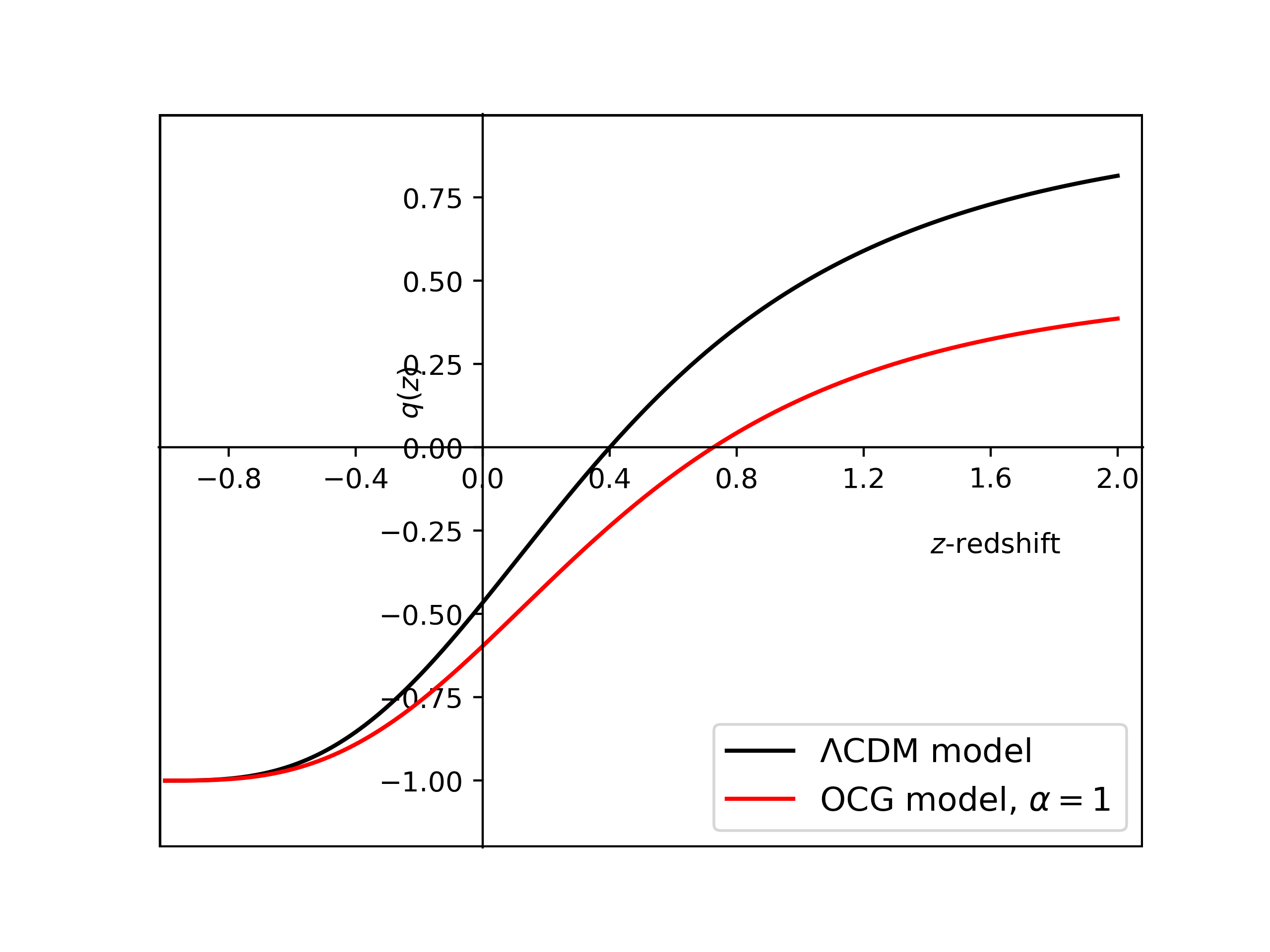}
			\caption{A plot of the deceleration parameter, $q(z)$, versus the redshift, $z$ for Eq. \ref{deceleration1} in the OCG model, $\alpha = 1$.} 
			\label{qvsz}
		\end{minipage} 
		\qquad
		\begin{minipage}{0.45\linewidth}
			\includegraphics[width=1\textwidth]{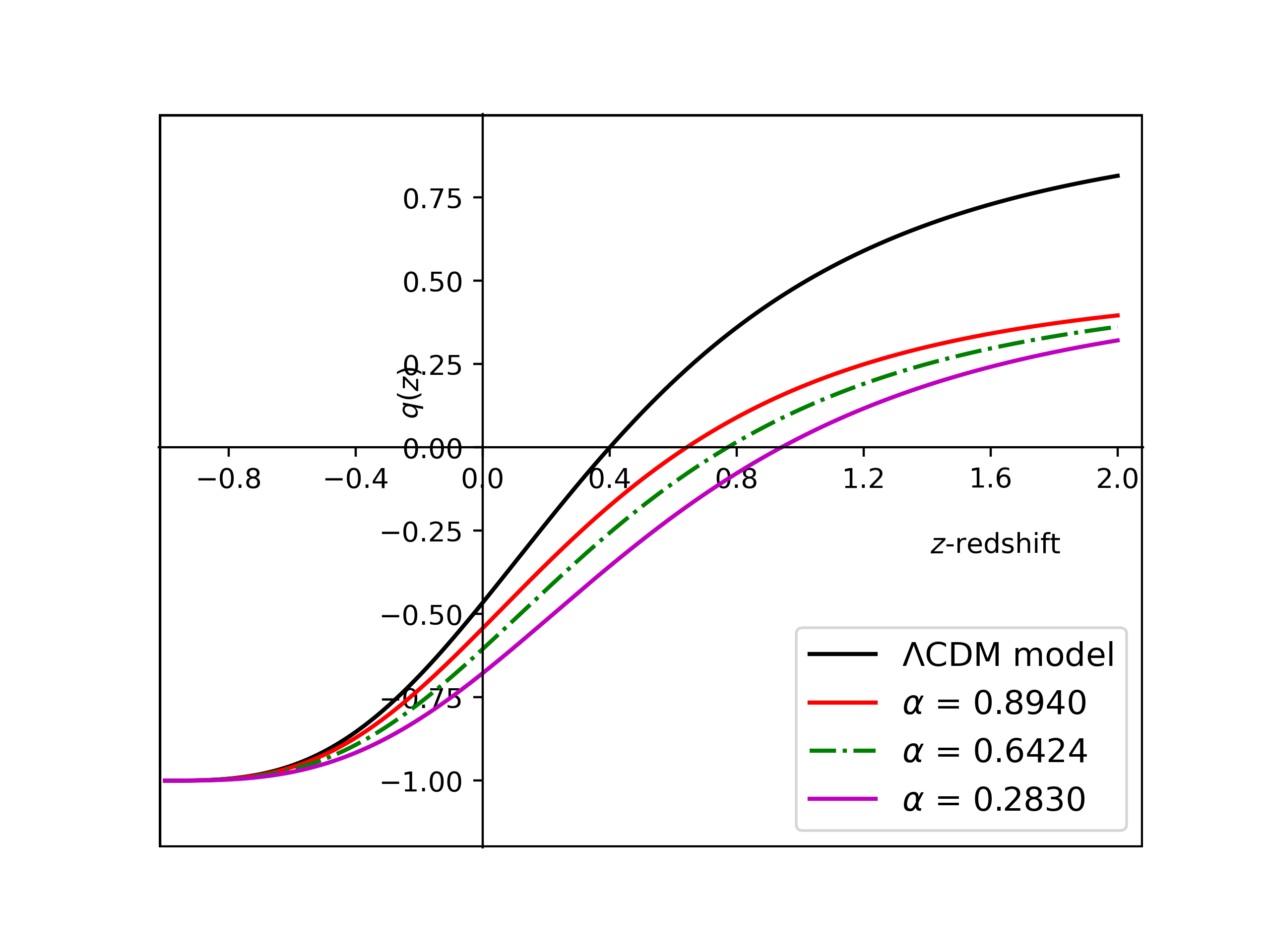}
			\caption{The deceleration parameter $q(z)$ versus $z$ for Eq. \ref{deceleration1} in the GCG model for different values of $\alpha <1$.}
			\label{fig:GCG3}
		\end{minipage} 
	\end{figure}
	
	
	\subsection{The deceleration parameter in MCG cosmology}\label{OGCG1}
	
	Here, we consider the MCG that has an equation of state of the form \cite{Elmardi:2018ryn, kahya2014observational}
	\begin{eqnarray}
		p_{ch}=B\rho_{ch}-\frac{A}{\rho_{ch}^\alpha},
		\label{eqstat}
	\end{eqnarray}
	where $A,~B$ and $\alpha$ are constant parameters. For the case of $A=0$, the well-known equation of state for a perfect fluid is recovered, that is, $p=B\rho$. In addition, for the case of $B = 0$, the equation of state is reduced to GCG and then to OCG at $\alpha = 1$.
	
	The energy density for this model becomes as:
	\begin{eqnarray}
		\rho_{ch}(a)=\left[A \Big( \frac{1}{(B+1)}+ \mathcal{K} a^{-3(\alpha+1)(B +1)} \Big) \right]^{\frac{1}{1+\alpha}},
	\end{eqnarray}
	where $ \mathcal{K}= \frac{C_3}{A}$, and $C_3$ is a constant of integration. Therefore, we now have an equation for the energy density in terms of the scale factor. The total energy density and total pressure are the quantities that we are concerned with, so we would have to add the baryonic matter component by hand (as was done in equations \ref{rhotot} and \ref{ptot} to obtain the normalized Hubble parameter $h(z)$:
	\begin{eqnarray}
		h^2(z)&=& \left[E_0\Big( \frac{1}{(B+1)}+ \mathcal{K}(1+z)^{3(\alpha+1)(B+1)} \Big)\right]^{\frac{1}{1+\alpha}}+\Omega_{b0}(1+z)^{3}\;,
		\label{eq: MCG model}
	\end{eqnarray}
	where $E_0=\frac{A}{(3H_0^2)^{1+\alpha}}$, and the deceleration parameter can be written in terms of redshift dependent as
	
	\begin{eqnarray}\label{decelerationMCG}
		&& q(z) =\frac{ (3B +1)\left[A\Big(\frac{1}{(B+1)}+ \mathcal{K}(1+z)^{3(\alpha+1)(B +1)}\Big)\right]^{\frac{1}{1+\alpha}}}{2\left[A\Big( \frac{1}{(B+1)}+ \mathcal{K}(1+z)^{3(\alpha+1)(B+1)}\Big)\right]^{\frac{1}{1+\alpha}} +2\rho_{b0}(1+z)^{3}} \nonumber \\ 
		&& \quad \quad- \frac{3A\left[ A\Big(\frac{1}{(B+1)}+ \mathcal{K}(1+z)^{3(\alpha+1)(B +1)}\Big)\right]^{-\frac{\alpha}{1+\alpha}}}{2\left[A\Big( \frac{1}{(B+1)}+ \mathcal{K}(1+z)^{3(\alpha+1)(B+1)}\Big)\right]^{\frac{1}{1+\alpha}} +2\rho_{b0}(1+z)^{3}}\nonumber \\
		&& \quad \quad + \frac{\rho_{b0}(1+z)^{3}}{2\left[A\Big( \frac{1}{(B+1)}+ \mathcal{K}(1+z)^{3(\alpha+1)(B+1)}\Big)\right]^{\frac{1}{1+\alpha}}+2\rho_{b0}(1+z)^{3}}\;.
	\end{eqnarray}
	
	In the following, we present the numerical results of the deceleration parameter Eq. \ref{decelerationMCG} for different values of $\alpha$ and $B$ as presented in Figs. \ref{fig:MCG000} - \ref{fig:MCG4}.
	We consider the max and min values of $B$ from Table \ref{tab: best-fitting parameter values} for illustrative purposes.
	\begin{figure}[ht!]
		\begin{minipage}{0.45\linewidth}
			\includegraphics[width=1\textwidth]{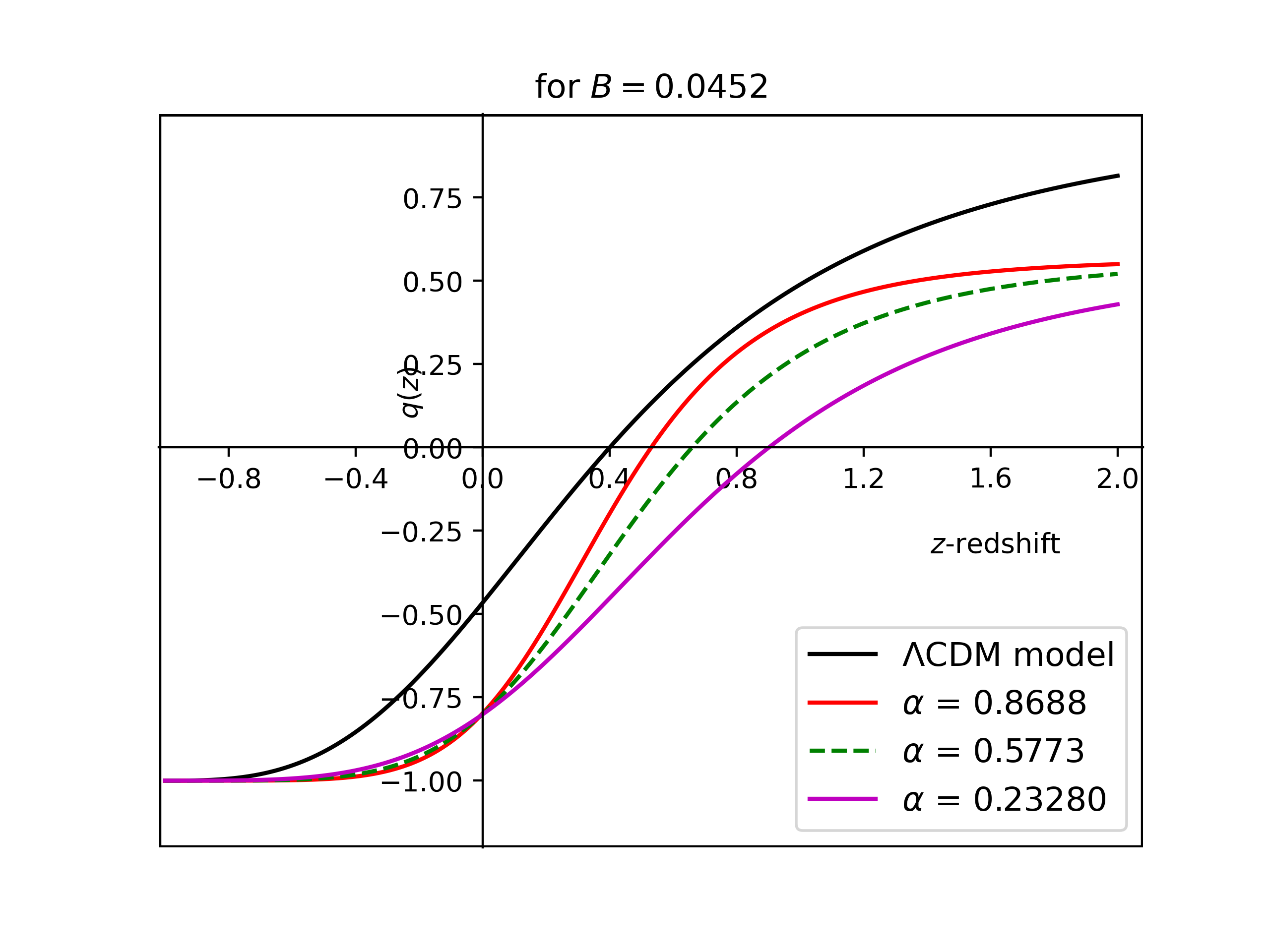}
			\caption{The deceleration parameter $q(z)$ versus $z$ for MCG mode Eq. \ref{decelerationMCG} for different values of $\alpha$ and B=0.0452.}
			\label{fig:MCG000}
		\end{minipage}
		\qquad
		\begin{minipage}{0.45\linewidth}
			\includegraphics[width=1\textwidth]{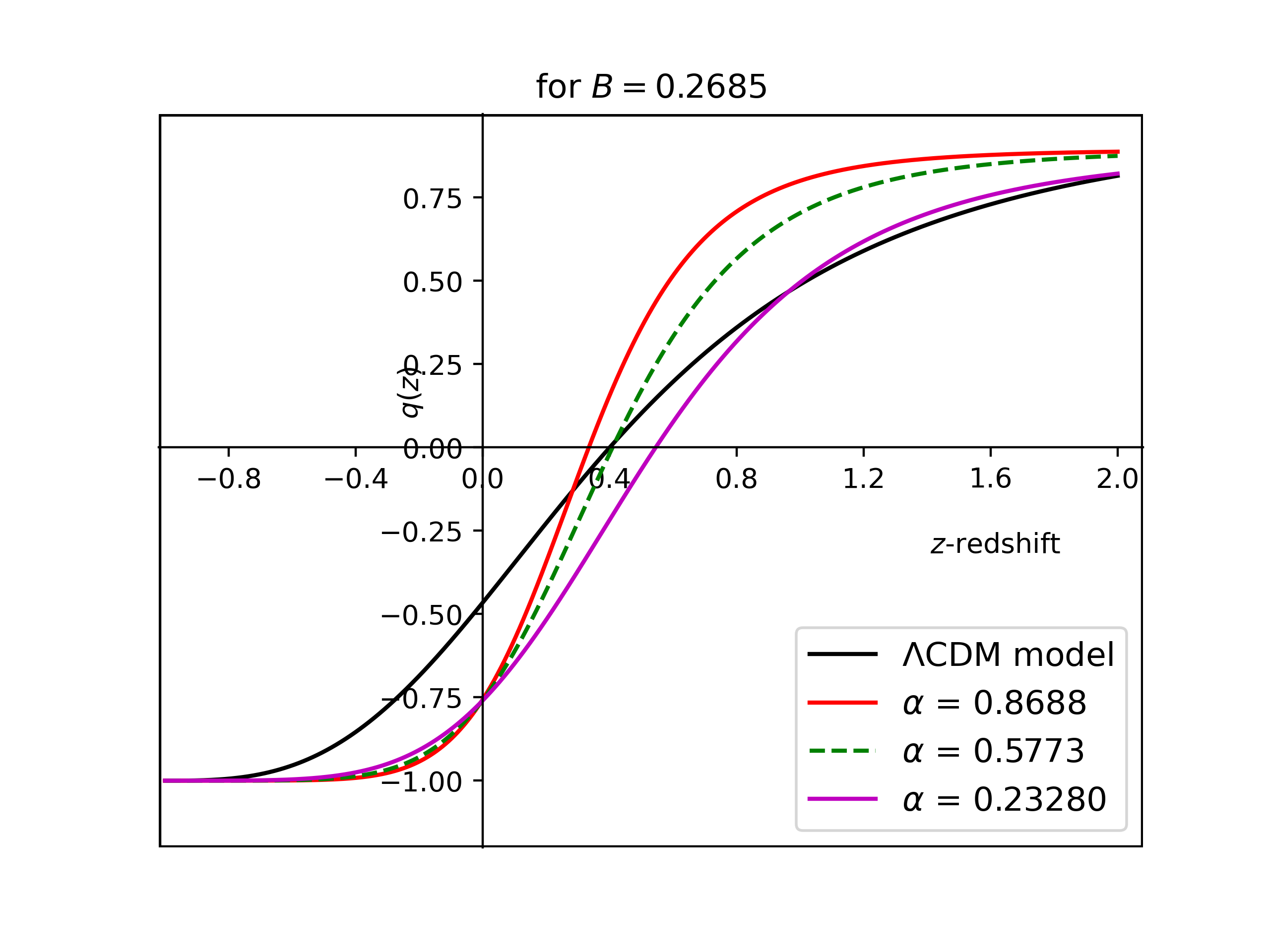}
			\caption{The deceleration parameter $q(z)$ versus $z$ for MCG mode Eq. \ref{decelerationMCG} for different values of $\alpha$ and B= 0.2685.}
			\label{fig:MCG4}
		\end{minipage}
	\end{figure}
	
	
	\subsection{The deceleration parameter in MGCG cosmology}
	
	A more generalised version of the MCG model with some interesting properties \cite{elmardi2016chaplygin, bouhmadi2015scalar, morais2015can} is the so-called modified generalised CG (MGCG) model described by the equation of state:
	\begin{eqnarray}
		p_{ch}=B\rho_{ch}-(1+B)\frac{A}{\rho_{ch}^\alpha}\;,
	\end{eqnarray}
	with A, B, and $\alpha$ as the constant parameters of the model. With the appropriate choice of these parameters, the model can be shown to have the equation of state of both the cosmological constant and the MCG as its limiting cases.
	
	The energy density of such a CG gas model is given by:
	\begin{eqnarray}
		\rho_{ch}(a)=\left[A+C_4a^{-3(\alpha+1)(B +1)}\right]^{\frac{1}{1+\alpha}}\;,
	\end{eqnarray}
	where $C_4$ is an integration constant. The normalized Hubble and deceleration parameters of the model are therefore:
	\begin{eqnarray}
		\rho_{ch}(a)=\left[A\Big(1+\mathcal{K}_2 a^{-3(\alpha+1)(B +1)}\Big) \right]^{\frac{1}{1+\alpha}}\;,
	\end{eqnarray}
	where $\mathcal{K}_{2} =\frac{C_4}{A}\;$. The normalised Hubble parameter yield is:
	\begin{eqnarray}\label{eq. MGCG Hubble parameter} 
		h^2(z)&=& \left[F_1\Big(1+\mathcal{K}_2(1+z)^{3(\alpha+1)(B+1)}\Big)\right]^{\frac{1}{1+\alpha}} +\Omega_{b0}(1+z)^{3}\;,
	\end{eqnarray}
	and the deceleration parameter becomes
	\begin{eqnarray}
		&&q(z)=\frac{(1+3B)\left[F_1\Big(1+\mathcal{K}_2(1+z)^{3(\alpha+1)(B +1)}\Big)\right]^{\frac{1}{1+\alpha}} }{{2\left[\left(F_1\Big(1+\mathcal{K}_2(1+z)^{3(\alpha+1)(B+1)}\Big)\right]^{\frac{1}{1+\alpha}} ++\Omega_{b0}(1+z)^{3}\right)}}\nonumber \\
		&&\quad \quad - \frac{3(1+B)A\left[F_3\Big(1+\mathcal{K}_2(1+z)^{3(\alpha+1)(B +1)}\Big)\right]^{-\frac{\alpha}{1+\alpha}}}{2\left[\left(F_1\Big(1+\mathcal{K}_2(1+z)^{3(\alpha+1)(B+1)}\Big)\right]^{\frac{1}{1+\alpha}} ++\Omega_{b0}(1+z)^{3}\right)} \\
		&&\quad \quad + \frac{\Omega_{b0}(1+z)^{3}}{2\left[\left(F_1\Big(1+\mathcal{K}_2(1+z)^{3(\alpha+1)(B+1)}\Big)\right)^{\frac{1}{1+\alpha}}+\Omega_{b0}(1+z)^{3}\right]}\;. \nonumber 
	\end{eqnarray}
	
	The numerical results of the deceleration parameter are presented in Figs. \ref{fig:GMCCG01} - \ref{fig:GMCCG002} with different values of $B$ and $\alpha$. 
	\begin{figure}[ht!]
		\begin{minipage}{0.45\linewidth}
			\includegraphics[width=1\textwidth]{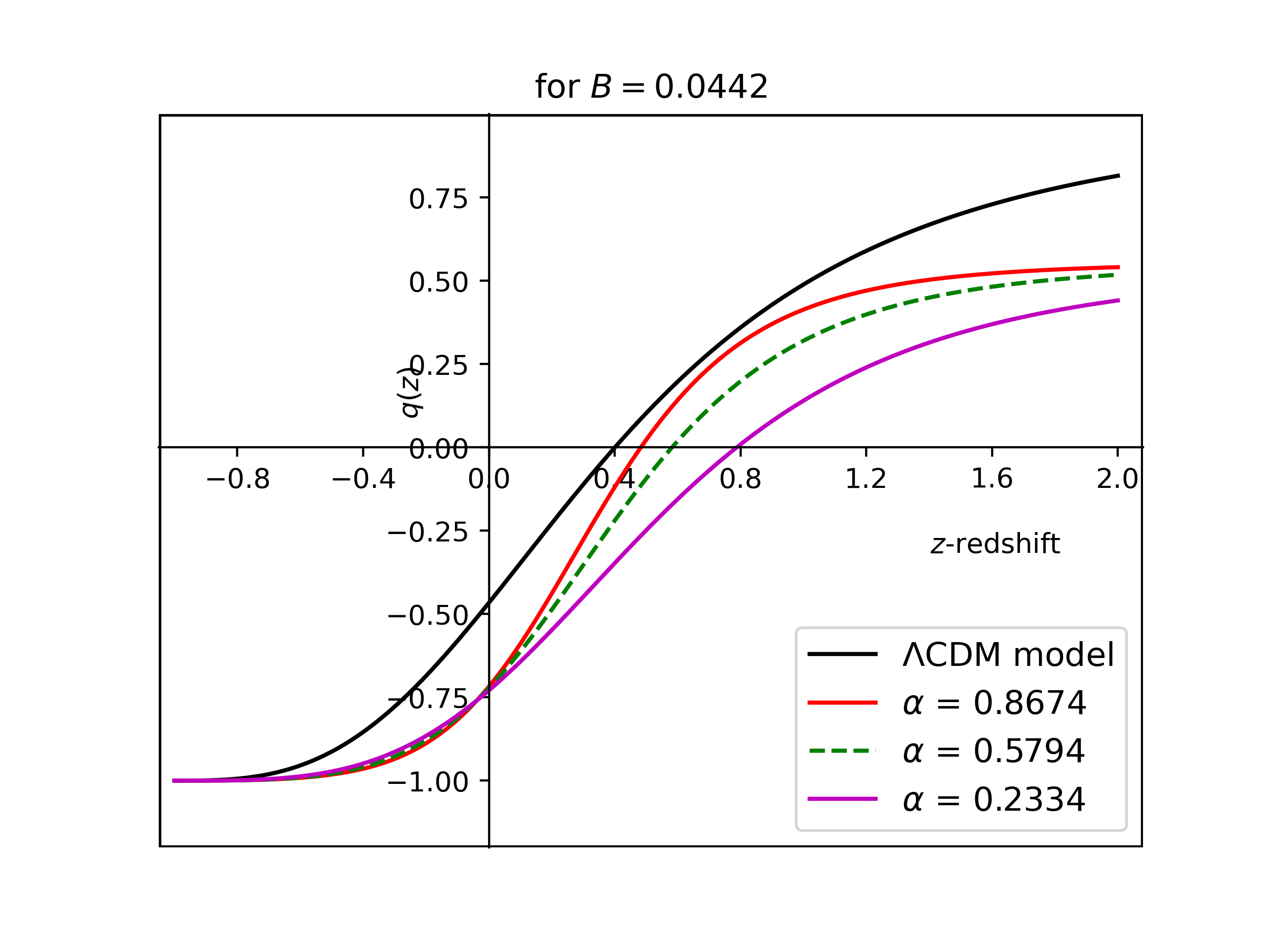}
			\caption{The decelerating parameter versus cosmological redshift $z$ for MGCG model for different values of $\alpha$ at B = 0.0442.}
			\label{fig:GMCCG01}
		\end{minipage}
		\qquad
		\begin{minipage}{0.45\linewidth}
			\includegraphics[width=1\textwidth]{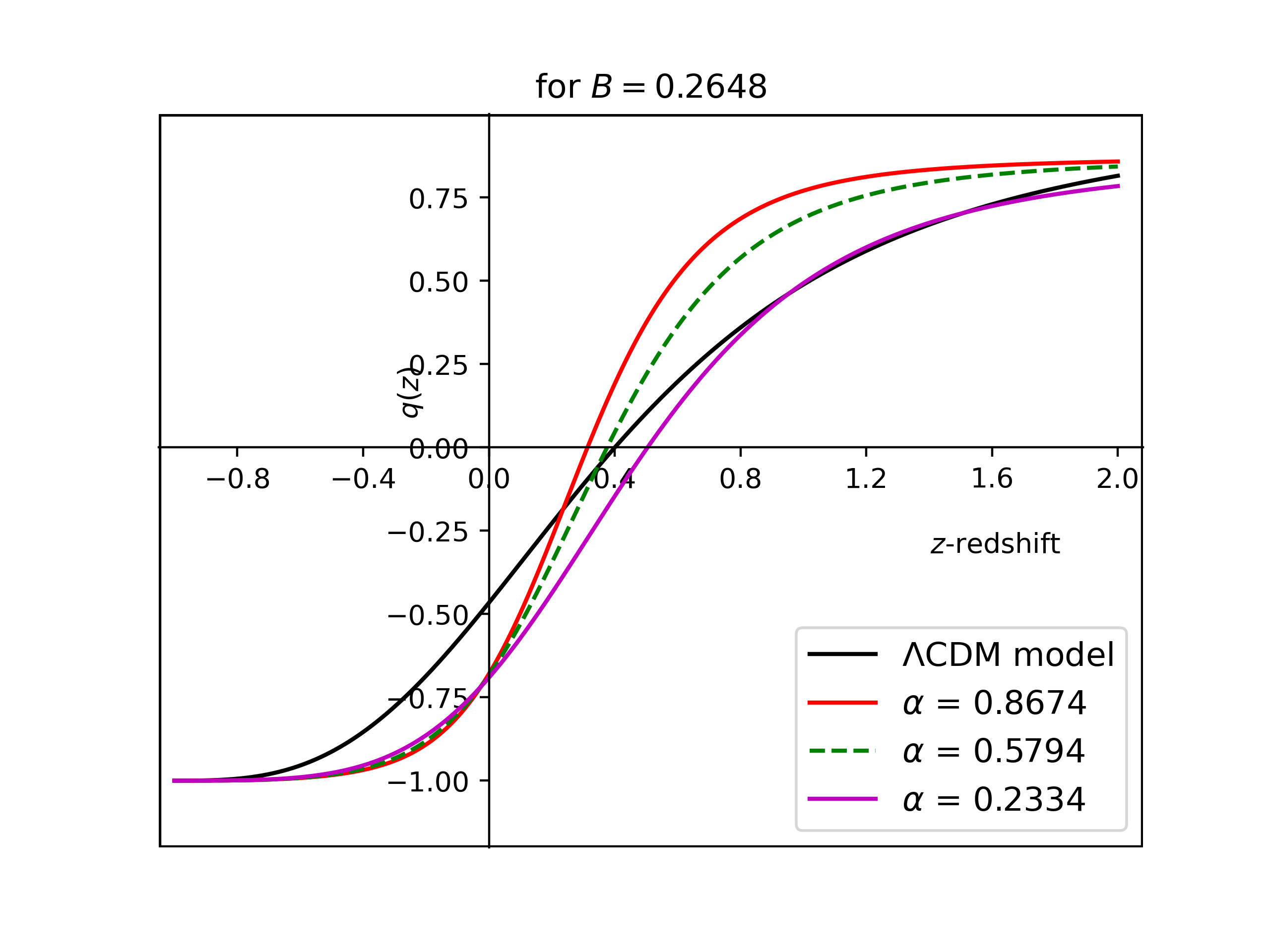}
			\caption{The decelerating parameter versus cosmological redshift $z$ for MGCG model for different values of $\alpha$ at B = 0.2648.}
			\label{fig:GMCCG002}
		\end{minipage}
	\end{figure}
	
	
	\subsection{The deceleration parameter in ECG Cosmology}\label{ECG}
	
	Here, we introduce the extended Chaplygin gas models (ECG) which has an equation of state of the form \cite{kahya2014observational, pourhassan2014extended}:
	\begin{eqnarray}\label{ECG}
		p_{ch}=\sum_{n =1}^\infty {A_n \rho^n_{ch}}-\frac{B}{\rho_{ch}^\alpha}\;,
	\end{eqnarray}
	
	This model is an extension of the MCG and MGCG models; it includes a quadratic barotropic equation of state of the form $ p=p_{0} +\omega_{1} \rho + \omega_{2} \rho^{2}$ \cite{kahya2014observational, pourhassan2014extended}. While the linear barotropic equation of state can be recovered for $p_{0}= \omega_{1}=0$ and $\omega_{2} =\omega$. By expanding the sigma notation in Eq. \ref{ECG} to the second notation, we get:
	\begin{equation}\label{ECG0}
		p_{ch}= A_0 \rho^0_{ch} +A_1 \rho_{ch}+ A_2 \rho^2_{ch}-\frac{B}{\rho_{ch}^\alpha}\;,
	\end{equation}
	and assuming that the parameter $A_{0}=0$, we get
	\begin{equation}\label{ECG2}
		p_{ch}= A_1 \rho_{ch}+ A_2 \rho^2_{ch}-\frac{B}{\rho_{ch}^\alpha}\;.
	\end{equation}
	This model reduces to the MCG model for $n=1$. The conservation equations for this model are given as:
	\begin{equation}
		\dot{\rho}_{ch} + 3H\Big(\rho_{ch}+A_1 \rho_{ch}+ A_2 \rho^2_{ch}-\frac{B}{\rho_{ch}^\alpha}\Big) = 0\;.
	\end{equation}
	
	Now we need to solve this conversation equation to get the energy density $\rho_{ch}$, we assume that $ A_{1}=2( A_{2} \rho_{ch} -1)$, and $B=A_{2} \rho^{\alpha+ 2}_{ch}$, the conservation equation for ECG model read as:
	\begin{equation}
		\dot{\rho}_{ch} + 3H\Big( 3A_2  \rho^{2}_{ch}- \rho_{ch}- \frac{A_{2} \rho^{(\alpha+2)}_{ch}}{\rho^{\alpha}_{ch}}\Big) = 0\;, 
	\end{equation}
	
	The energy density of the exotic fluids reads as:
	\begin{equation}
		\rho_{ch} (a)= \Big( 2A _{2} + D^{-1} a^{-3}\Big)^{-1}
	\end{equation}
	where $D= e^{C} $, and $C$ is a constant of integration, and the total energy
	\begin{equation}
		\rho_{tot} (a)= \Big( 2A _{2} + D^{-1} a^{-3}\Big)^{-1}+ \rho_{b,0} a^{-3}\;, 
	\end{equation}
	The Friedmann equation reads as:
	\begin{equation} \label{HECG}
		3H^{2} (z)=  \Big( 2A _{2} + D^{-1} (1+z)^{3}\Big)^{-1} + \rho_{b,0} (1+z)^{3}\;, 
	\end{equation}
	and the normalised Hubble parameter is given as:
	\begin{equation}
		h^{2} (z)= D_{1} \Big(2+  D_{2}(1+z)^{3} \Big)^{-1}  + \Omega_{b,0} (1+z)^{3}\;, 
		\label{eq. ECG hubble parameter}
	\end{equation}
	with $D_{1}=  \frac{1}{3 H^{2}_{0}A_{2}}$ and $D_{2}= \frac{1}{D A_{2}}$\;. The deceleration parameter:
	\begin{eqnarray}
		q=  \frac{1}{2}+  \frac{ 3 h^{2}(z)\Omega_{ch}\Big( \frac{h^{2}(z)} {D_{1}} 
			\Omega_{ch}-1\Big)}{  \Big[ \Big( \frac{2}{D_{1}} + \frac {(1+z)^{3} D_{2}}{D_{1}}\Big)^{-1}+ \Omega_{b,0} (1+z)^{3}\Big] } \;.
		\label{11q}
	\end{eqnarray}
	where $ \Omega_{ch} = \frac{\rho_{ch}}{3H^{2}}=\frac{ D_{1}\Big( 2 + D_{2} (1+z)^{3}\Big)^{-1}}{h^{2}(z)}$. 
	
	In Fig. \ref{fig:GMCCG011}, we present the numerical results of the deceleration parameter for the ECG model of Eq. \ref{11q}. From the numerical plots of its deceleration parameter, we see that the model has ruled out in the early universe and ruled in the late-time universe. 
	\begin{figure}[ht!]
		\centering
		\includegraphics[width=10cm,height= 6cm ]{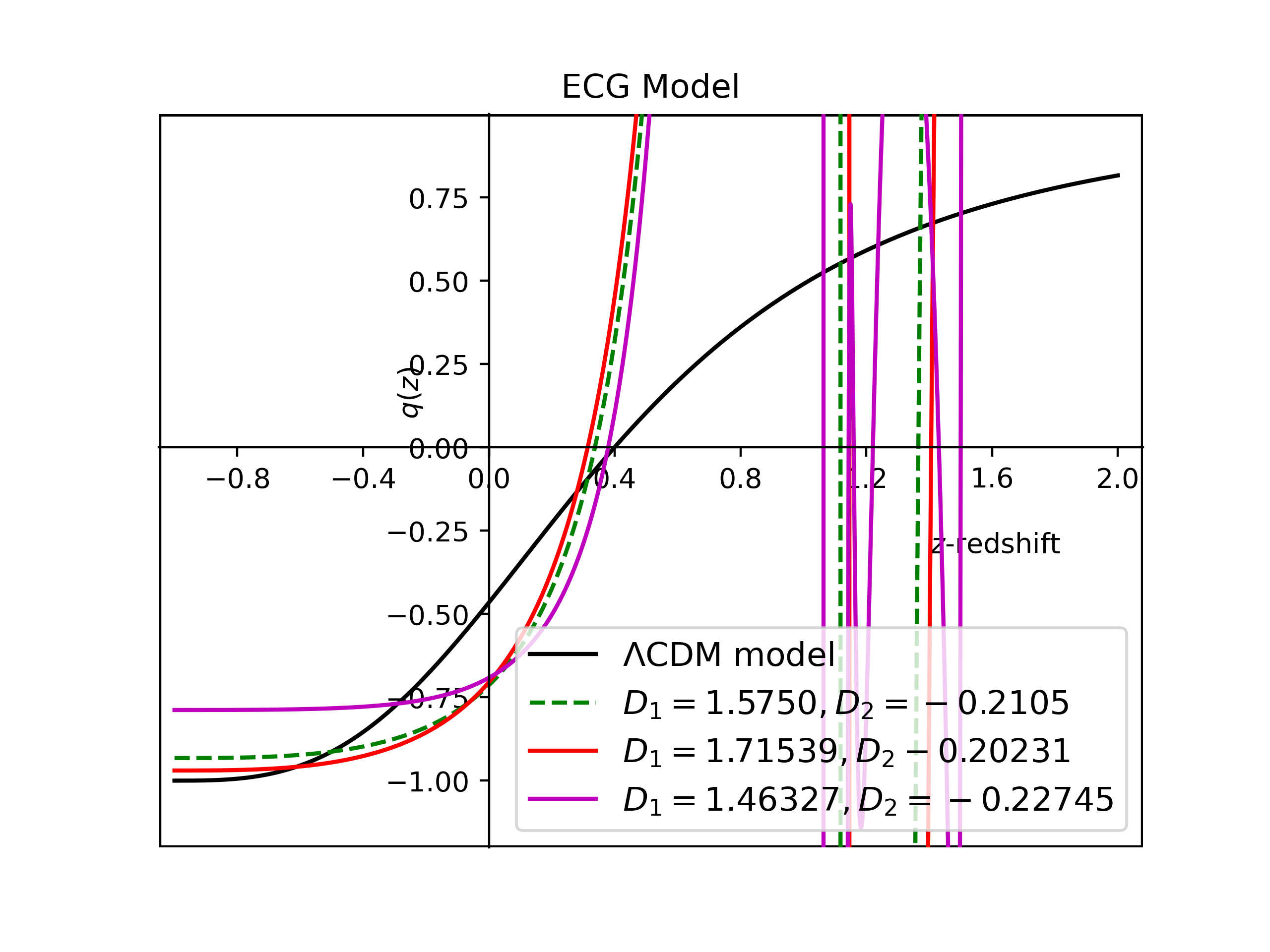}
		\caption{The decelerating parameter versus cosmological redshift $z$ for ECG model in different values of $D_1$ and $D_2$, from Eq. \ref{11q}}
		\label{fig:GMCCG011}
	\end{figure}
	
	
	\subsection{The deceleration parameter in GCCG cosmology}\label{OGCG2}
	
	In this section, we consider the generalised cosmic Chaplygin gas (GCCG) models to present the numerical results of the deceleration parameter to explain the cosmic acceleration. The equation state is given as \cite{gonzalez2003you, kahya2014observational}:
	\begin{eqnarray}
		p_{ch}=-\rho^{-\alpha}_{ch}\left[C+\left(\rho_{ch}^{1+\alpha}-C\right)^{-\omega}\right]\;,
	\end{eqnarray}
	where $C=\frac{A}{1+\omega}-1$, for $A$ a constant which can take on both negative and positive numbers, and $0>\omega>-l$ and $l$ here is a positive definite constant which can take on values larger than unity \cite{Elmardi:2018ryn}.
	
	The continuity equation of the exotic fluid reads as:
	\begin{eqnarray}
		&&\dot{\rho}_{ch}+3\frac{\dot{a}}{a}\left[\rho_{ch} -\rho^{-\alpha}_{ch}\left(C+(\rho_{ch}^{1+\alpha}-C)^{-\omega}\right)\right]=0\;.
	\end{eqnarray}
	
	From this equation, we have a total energy density and total pressure including the baryonic matter components can be written respectively as:
	\begin{eqnarray}
		\rho_{tot}&=&\left[C+\left(1+Ba^{-3(\alpha+1)(\omega +1)}\right)^{\frac{1}{1+\omega}}\right]^{\frac{1}{1+\alpha}}+\rho_{b0}a^{-3}
	\end{eqnarray}
	and the normalized Hubble parameter is given as:
	\begin{eqnarray}
		h^2(z)  \left(\frac{C}{(3H_0^{2})^{(1+\alpha)}}+\frac{\left[1+B(1+z)^{3(\alpha+1)(\omega +1)}\right]^{\frac{1}{1+\omega}}}{(3H_0^{2})^{(1+\alpha)}}\right)^{\frac{1}{1+\alpha}} +\Omega_{b0}(1+z)^{3}\;.
		\label{cGGG}
	\end{eqnarray}
	
	Then, the deceleration parameter for the GCCG model is:
	\begin{eqnarray}
		&&q(z)=-\frac{3\mathcal{N}(z)^{-\frac{\alpha}{1+\alpha}}\left[C+\left(1+B(1+z)^{3(\alpha+1)(\omega +1)}\right)^{-\frac{\omega}{1+\omega}}\right]}{2\mathcal{N}(z)^{\frac{1}{1+\alpha}}+2\rho_{b0}(1+z)^{3}} \nonumber\\
		&& \quad\quad+\frac{\mathcal{N}(z)^{\frac{1}{1+\alpha}}+\rho_{b0}(1+z)^{3}}{2\mathcal{N}(z)^{\frac{1}{1+\alpha}}+2\rho_{b0}(1+z)^{3}}\;
		\label{RR}
	\end{eqnarray}
	where $\mathcal{N}(z) = C+\left(1+B(1+z)^{3(\alpha+1)(\omega +1)}\right)^{\frac{1}{1+\omega}}$. 
	
	\begin{figure}[ht!]
		\begin{minipage}{0.45\linewidth}
			\includegraphics[width=1\textwidth]{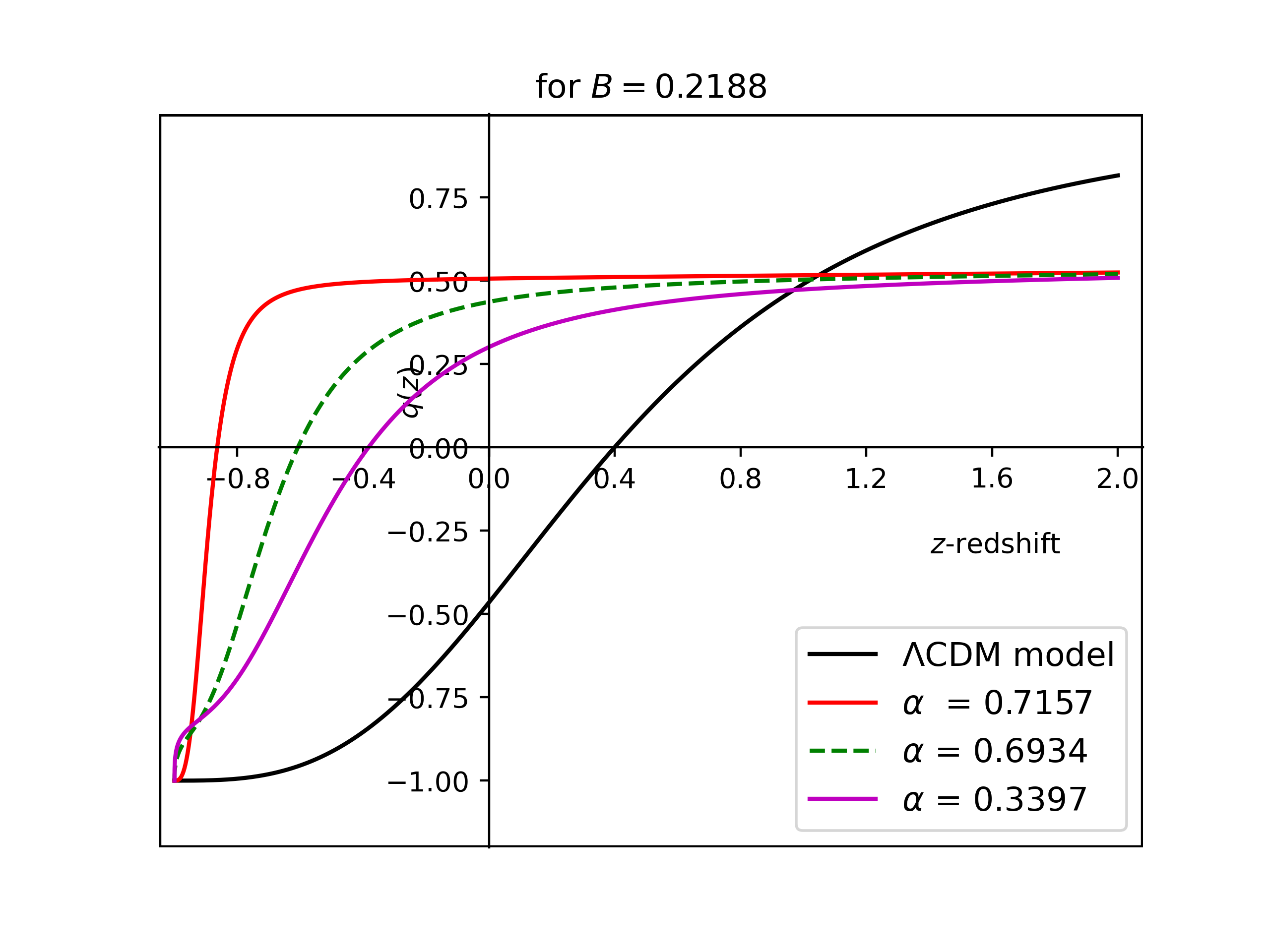}
			\caption{$q(z)$ versus cosmological redshift $z$ for GCCG model for $B = 0.2188$.}
			\label{fig:GMCCG}
		\end{minipage}
		\qquad
		\begin{minipage}{0.45\linewidth}
			\includegraphics[width=1\textwidth]{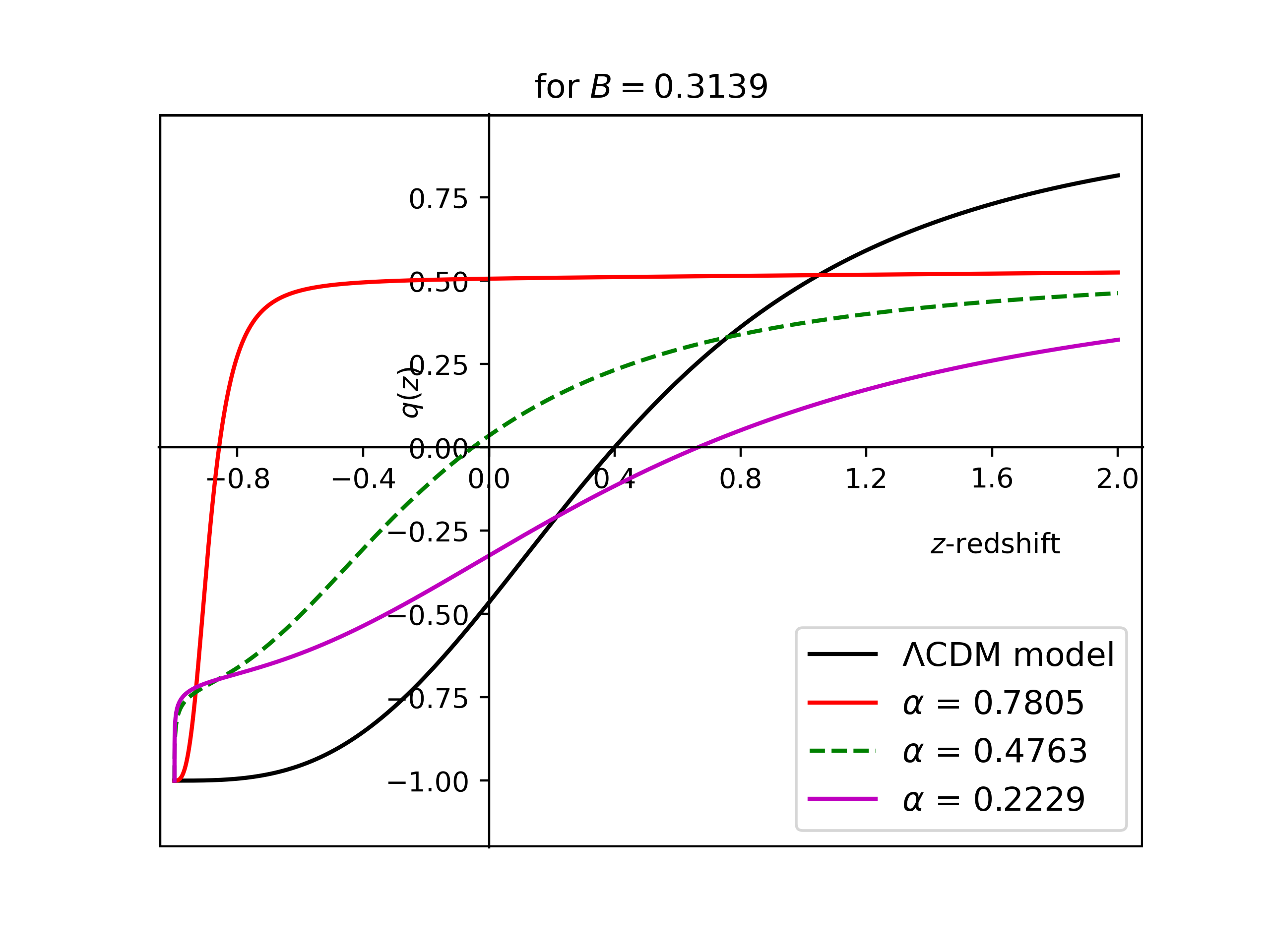}
			\caption{$q(z)$ versus cosmological red-shift $z$ for GCCG model for $B = 0.3139$.}
			\label{fig:GMCCG}
		\end{minipage}
	\end{figure}
	
	As we did in the previous sections, we present the numerical results of Eq. \ref{RR} in the following Figs. \ref{fig:GMCCG} for the deceleration parameter in the GCCG models. From the plots we observe that this model can explain the phantom phase of the universe as presented in \cite{gonzalez2003you} for small values of $\alpha$ and $B$ and which are less candidates than the other CG models to explain the accelerating expansion universe. We use the average values $\omega$ from Table \ref{tab: best-fitting parameter values} for numerical purposes. 
	
	From all the plots shown in Figs. \ref{qvsz} - \ref{fig:GMCCG}, we tried to show the accelerated expansion within the observationally known range of the cosmological redshift $z$. All results are consistent with what one would expect the expansion history of the Universe to be in the current concordance model; there should be a decelerating matter-dominated phase followed by a dark-energy-driven late-time acceleration phase except in the GCCG model. However, the numerical results of GCCG rule out this model, and its further statistical analysis is investigated in Sec. \ref{stastic}. 
	
	
	\section{Supernovae cosmology for CG models}\label{numerical}
	\subsection{The $\Lambda$CDM model}
	
	The distance modulus that can be obtained by combining the different cosmological distance definitions, found in \cite{deza2009encyclopedia}, is given by\footnote{This distance modulus is given in terms of \textit{Mpc}.}:
	\begin{equation}
		\mu = m-M = 25-5\times\log_{10}\left[3000\bar{h}^{-1}(1+z)\int^{z}_{0}\frac{dz^{\prime}}{h(z^{\prime})}\right]\;,
		\label{eq: Distance modulus}
	\end{equation}
	where $m$ and $M$ are the apparent and absolute magnitudes for a particular measured supernova. We also used the definition for the Hubble uncertainty parameter, which is given as $\bar{h} = \frac{H(z)}{100 \frac{km}{s.Mpc}}$, with $H(z)$ being the Hubble parameter. We obtained the normalised Friedmann equation $h(z)$, as a result of the definition for the line-of-sight co-moving distance \cite{deza2009encyclopedia}. 
	
	Now that we have a model to fit against the supernovae data, we will use 359 low- and intermediate-redshift data points obtained from the SDSS-II/SNLS3 Joint Light-curve Analysis (JLA). This will be sufficient to test the transition phase between the matter-dominated epoch (a decelerating expansion) and the late-time dark energy epoch (an accelerating expansion), which occurs at $z\approx 0.5$ \cite{linder2001understanding, capozziello2019model}. We will be using the absolute magnitudes for these supernovae in the B-filter that were calculated in the research papers \cite{conley2010supernova, Neill2009, Hicken2009}, and are represented by blue, green, and magenta coloured dots in the best-fit supernovae figures\footnote{These absolute magnitude values can also be found on NED.}. 
	
	A question that may arise now is: ``What about the degeneracy between the Hubble parameter and the absolute magnitude of the supernovae?". Generally speaking, since there exists a degeneracy between these two parameters and the absolute magnitude is usually unknown, we cannot constrain the parameters using the supernova data \cite{colgain2019hint}. Normally you would have to use either a fixed absolute magnitude value for the supernovae (as done by \cite{colgain2019hint}, since it gives a robust enough prediction that the other free parameters, excluding the Hubble constant, can still be constrained), or you will need to use Cepheid variable star data, as done by \cite{freedman2001final, riess2009redetermination, riess20113}, where they determined the error on the distance modulus created by the uncertainty of the absolute magnitude, and add this correction to the supernova distance modulus.
	
	However, in our case, it is not necessary, since we used the absolute magnitudes determined by \cite{Hicken2009, Neill2009, conley2010supernova}. In their work, they already incorporated the corrections to the absolute magnitudes when they calculated it. The corrections they made were specifically made to enable them to constrain the Hubble constant on Supernovae Type 1A data. Therefore, our distance modulus also incorporates these corrections. Furthermore, we use a Markov Chain Monte Carlo (MCMC) simulation that takes into account the error on our distance modulus, when calculating the best-fitting parameters. These errors include the systematic errors that were obtained when acquiring the data (the error on the apparent magnitude found in the JLA dataset) and the errors on the absolute magnitudes as given by \cite{Hicken2009, Neill2009, conley2010supernova}. Therefore, we can use our calculated distance modulus as is. 
	
	There is a downside to using these absolute magnitude values. Limits the amount of supernovae Type 1a data that we can use. For instance, the full JLA data set already includes 740 supernovae data, which is then succeeded by the Pantheon Analysis, which itself is then succeeded by the most recent data set Pantheon+, which contains 1048 supernovae Type Ia data \cite{Scolnic2021}. However, as in the case of the JLA dataset, they estimate the distance modulus by calculating a covariance matrix \cite{Betoule2014} (see Table F.2) in \cite{Betoule2014}. This means that they do not supply the absolute magnitude values for these supernovae (see Table F.3). For our MCMC simulation, we need the absolute magnitude, so we can only use data for particular supernovae where the absolute magnitudes were calculated and calibrated against Cepheid variable stars.
	
	Now that we have a model and the accompanying dataset, we want to compare how the different CG models perform against the concordance model, namely the $\Lambda$CDM model. We will start with the best-fitting model for the $\Lambda$CDM model, to get an idea of what our ``true model" will look like. We are assuming $\Omega_{k}=0$ (that is, a flat universe as required by the setup of the Chaplygin gas solutions) and $\Omega_{r} = 0$. The reason for not including radiation was that we found that the resolution of our MCMC simulation is not strong enough to solve the radiation density and we would end up with a uniform distribution over the entire search space. Another reason why the MCMC simulation is struggling to constrain the radiation density is due to observations. Radiation density is only a dominant factor in the early universe epoch. Whereas the supernovae observations are observed between the matter dominate epoch and the late-time dark energy accelerated expansion epoch, as mentioned in the data collection paragraph. Therefore, supernovae data are not suitable to constrain the radiation density parameter. We will also be as making the substitution $\Omega_{\Lambda}= 1-\Omega_{m0}$, we find the normalized Friedmann equation for the $\Lambda$CDM model as:
	\begin{equation}
		h(z) = \sqrt{\Omega_{m0}\left(1+z\right)^{3}+1-\Omega_{m0}}\;.
		\label{eq: LCDM normalized Friedmann}
	\end{equation}
	
	We can then use our MCMC simulation as mentioned, which was developed in \cite{hough2020constraining} and later used in \cite{hough2019constraining, hough2020viability}, specifically for the testing of modified gravity models on supernovae data. This MCMC simulation uses the Metropolis-Hasting algorithm and, for simplicity, uses a Gaussian probability distribution to calculate the likelihood of each best-fitting parameter value depending on the distance modulus. This code incorporates the built-in Python package \textit{EMCEE Hammer} \cite{foreman2013emcee} to perform the MCMC simulation. This build-in package uses different random walkers, each executing the MH algorithm and all starting at the same initial parameter values, following their path to determine the most probable parameter values. This creates a Gaussian distribution based on each random walker's ending parameter values. Using the average values for each probability distribution for each parameter, on average we will obtain the best-fitting parameter values and its $1\sigma$-deviation \cite{hough2020constraining, hough2020viability}.
	
	Executing this MCMC simulation for the $\Lambda$CDM model, by combining Eqs. \ref{eq: Distance modulus} and \ref{eq: LCDM normalized Friedmann}, we find on average that the best-fitting parameter value for each free parameter is $\bar{h} = 0.6967^{+0.0048}_{-0.0047}$ (constrained) for the Hubble uncertainty parameter and $\Omega_{m0}=0.2674^{+0.0249}_{-0.0239}$ (constrained) for the matter density parameter. The results of the MCMC simulation are shown in Fig. \ref{fig: MCMC LCDM results}.
	\begin{figure}[ht!]
		\centering
		\includegraphics[width=8cm,height= 5cm ]{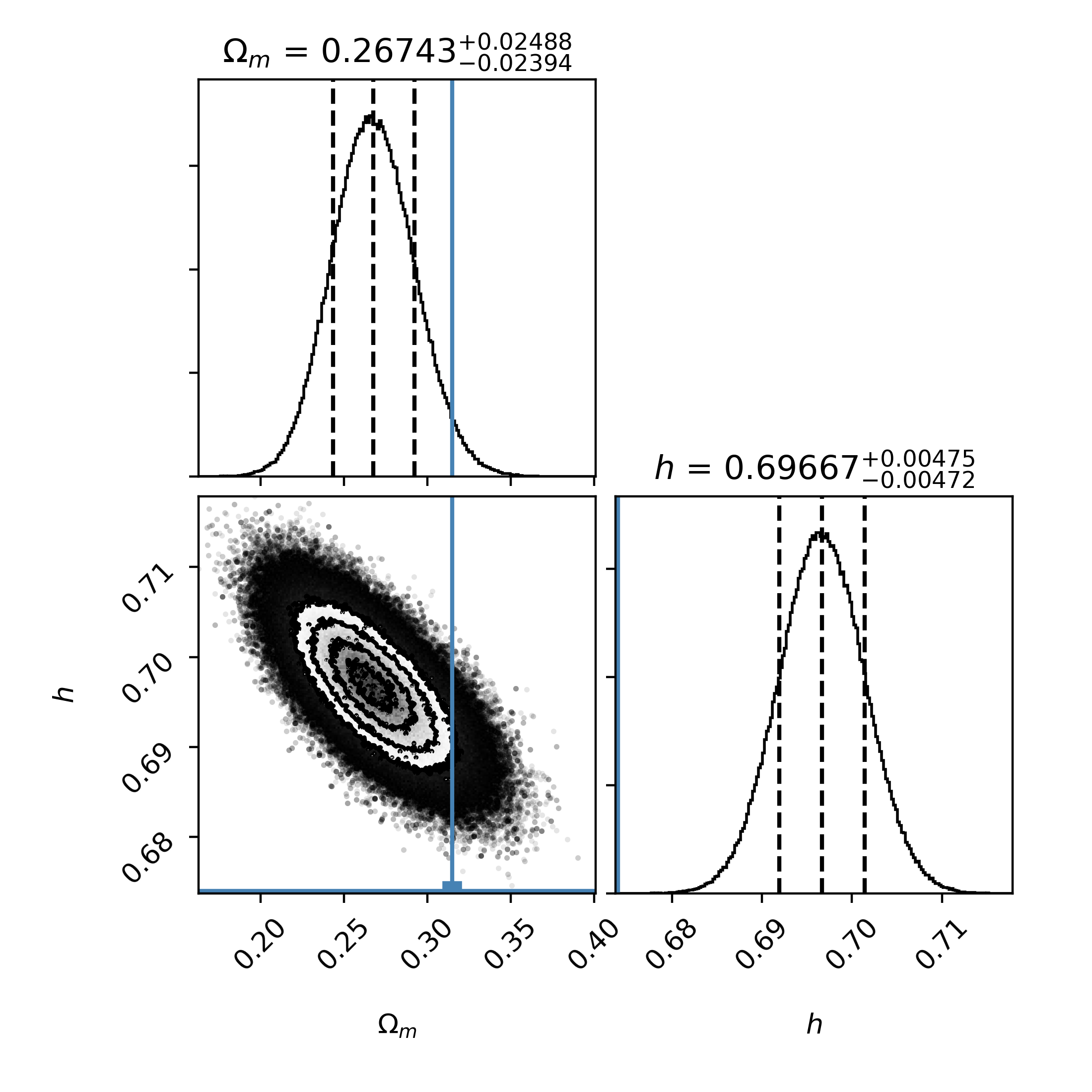}
		\caption{The MCMC simulation results for the $\Lambda$CDM model's (Eq. \ref{eq: LCDM normalized Friedmann}) cosmological free parameters ($\Omega_{m0}$ and $\bar{h}$), with the ``true" values (blue lines: $\Omega_{m0} = 0.315$ and $\bar{h} =0.674$) provided by the Planck2018 collaboration data release \cite{aghanim2020planck}. We used 100 random walkers and 10 000 iterations.} 
		\label{fig: MCMC LCDM results}
	\end{figure}
	
	Note that not all parameters are always fully constrained. This can be due to the amount of data or the actual simulation running into the numerical errors that occur when approximating the values. There are mainly 3 ways to define how they are constrained. The first is that the parameter is fully constrained. This occurs when a parameter was able to create a full Gaussian probability distribution in the available parameter search space. The second is that the parameter is unreliably constrained. This occurs when only one side of the Gaussian distribution was found in the search space. This means that there is a value that is more desirable than the one it received, since the average is slightly shifted from the peak of the distribution. Lastly, the parameter is unconstrained, which occurs when a uniform distribution is obtained for the particular parameter. This means that according to the MCMC simulation, this parameter does not play a role in the best fit (at least in this numerical resolution) for a particular model, such as the radiation density. Using this knowledge and the results from the MCMC simulation for the $\Lambda$CDM model, we can plot the best fit model on the data and obtain Fig. \ref{fig:LCDM model graphs}.
	\begin{figure}[ht!]
		\centering
		\begin{minipage}{0.45\textwidth}
			\centering
			\includegraphics[scale = 0.4]{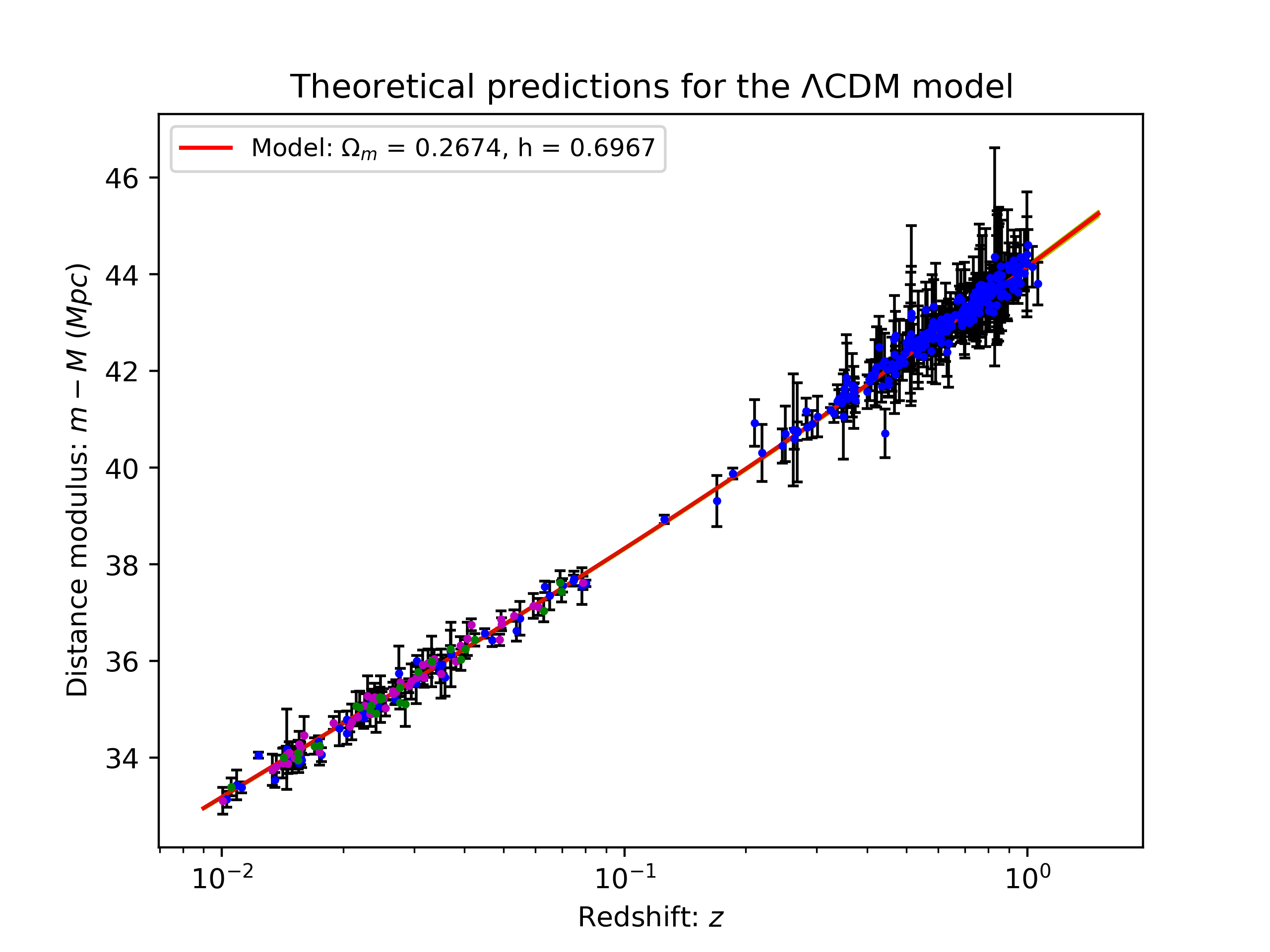}		
		\end{minipage}
		\hfill
		\begin{minipage}{0.45\textwidth}
			\centering
			\includegraphics[scale=0.4]{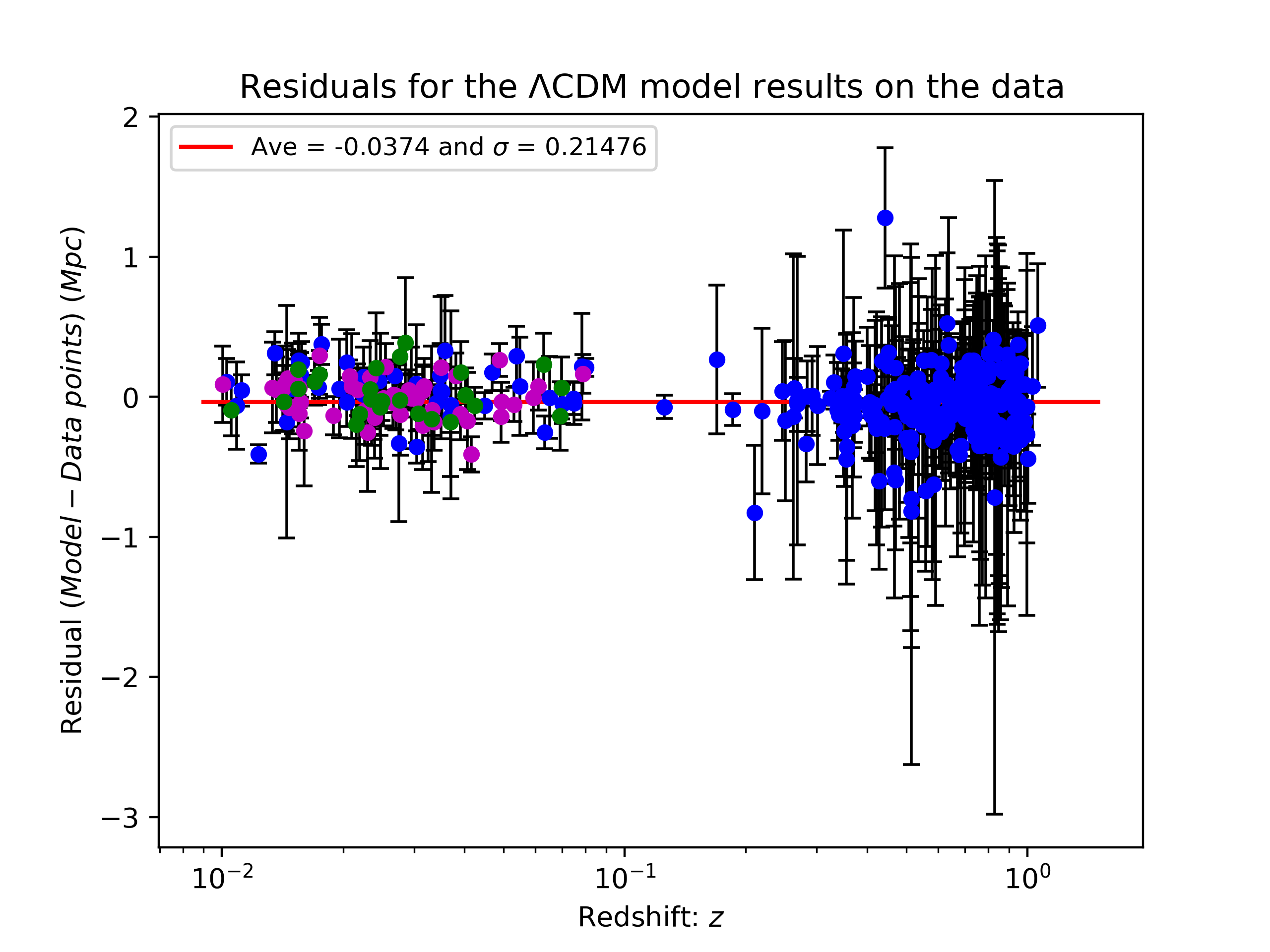}
		\end{minipage}
		\caption{The $\Lambda$CDM model's Eq. \ref{eq: LCDM normalized Friedmann} best-fitting free parameters for the Supernovae Type 1A data with cosmological parameter values as $\bar{h} = 0.6967^{+0.0048}_{-0.0047}$ (constrained) and $\Omega_{m0}=0.2674^{+0.0249}_{-0.0239}$ (constrained) calculated by the MCMC simulation (L.H.S. panel). The R.H.S. panel shows the residual distance in \textit{Mpc} between the predicted model values and the data points.}
		\label{fig:LCDM model graphs}
	\end{figure}
	
	From Fig. \ref{fig:LCDM model graphs} on the top panel, as expected, we see that the $\Lambda$CDM model fits the data extremely well. Even the $1\sigma$-deviation does not have an impact on the full range predicted by the $\Lambda$CDM model. Furthermore, on the bottom panel, where the residuals are shown, we can see that at no point does the model over-or under-estimate the resulting distance modulus for each supernova. We also note that the average offset the model has, compared to the data, is $\bar{x}_{res} = -0.0374$ Mpc with a standard deviation of $\sigma_{res} =0.21476$, showing that this is a thigh relation between the model and the data points.
	
	Regarding comparing these results with other results to determine whether the model and simulation actually work, we saw that our results are not within $1\sigma$ of the Hubble constant parameter value determined by Planck2018 collaboration \cite{aghanim2020planck}. However, this is expected, since it is known that there exists a discrepancy between the predicted Hubble constant values, due to studies using the CMB data, and the studies using the supernova data. It is noted in the research paper \cite{riess20162}, that supernovae-related studies tend to find a higher predicted value than CMB-related studies. It is still uncertain why this discrepancy exists. When we compare our results with other supernova studies, we see that we are within $1\sigma$ of their results \cite{riess20162}. This indicates that our MCMC simulation is indeed working as intended.
	
	
	\subsection{The various CG model results}
	
	Now that we have a way of testing the various CG models against observational data, we can go ahead and use each different CG model's normalised Friedmann equation in our MCMC simulation. Since all of the MCMC simulation results look relatively the same and take up a lot of space when plotting, we will only provide the necessary results, such as the most probable best-fitting parameter values and whether or not the parameter was constrained. We start with the original CG model. The only difference that the original CG model has compared to the general CG model is $\alpha =1.0$ in Eq. \ref{eq:chaplygin friedmann model}. We can then insert equation \ref{eq:chaplygin friedmann model} into the distance modulus function, with $\alpha =1.0$, and execute the MCMC simulation. We plot the best-fitting model on the data in Fig. \ref{fig:OCG model graphs}.
	\begin{figure}[ht!]
		\centering
		\begin{minipage}{0.45\textwidth}
			\centering
			\includegraphics[scale = 0.4]{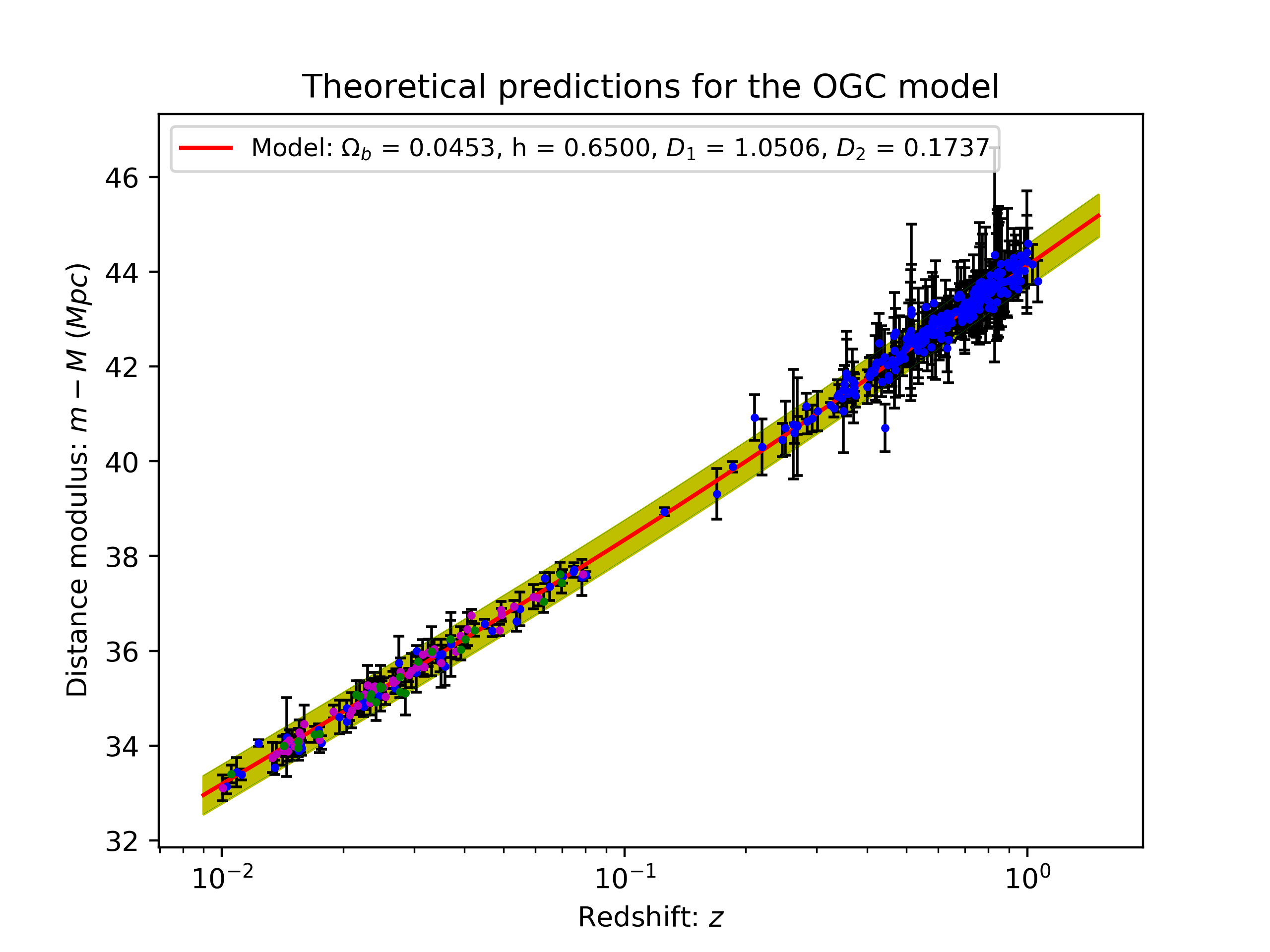}				
		\end{minipage}
		\hfill
		\begin{minipage}{0.45\textwidth}
			\centering
			\includegraphics[scale=0.4]{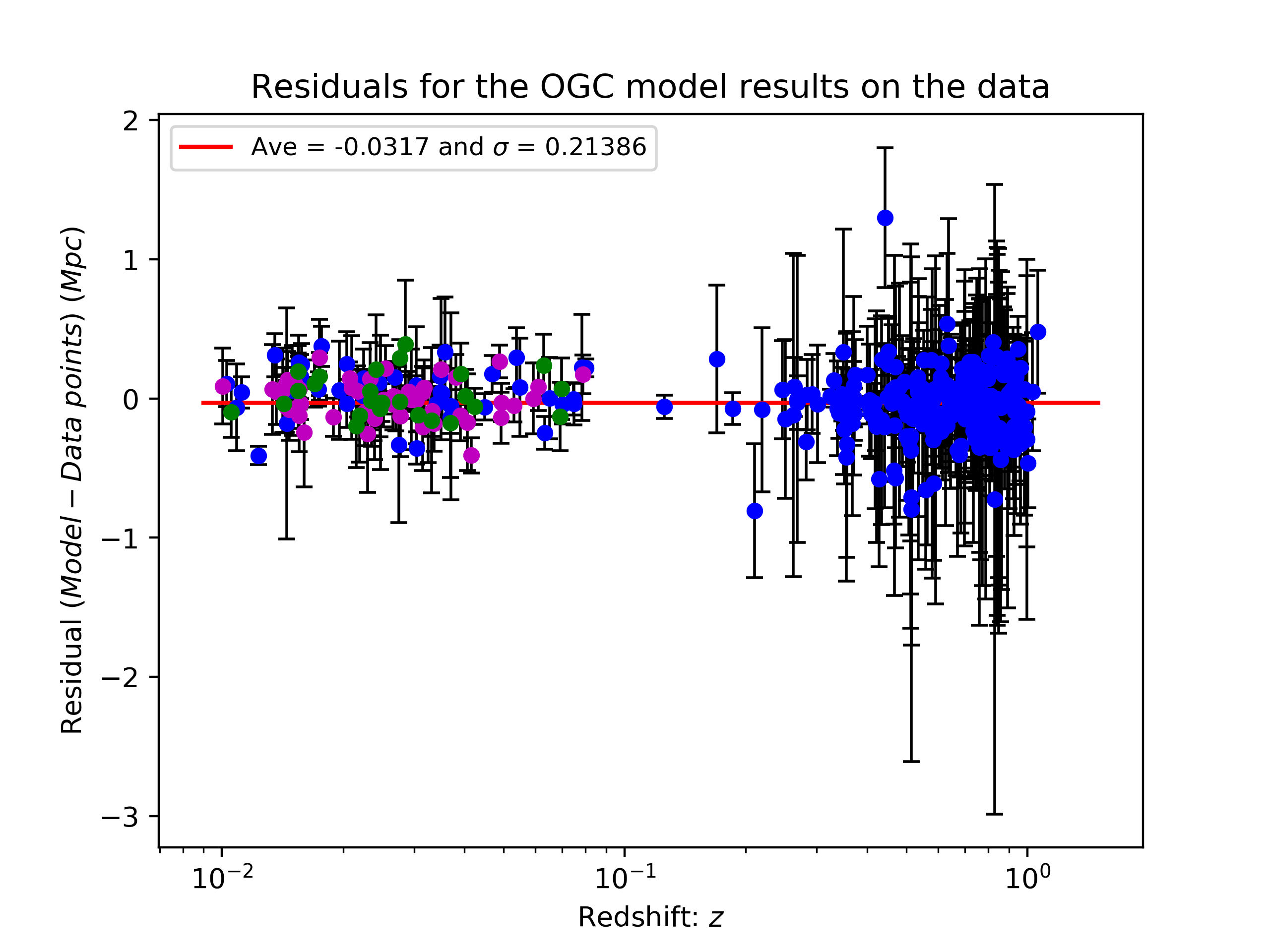}
		\end{minipage}
		\caption{The original CG model's Eq. \ref{eq:chaplygin friedmann model}, with $\alpha=1.0$, best-fitting free parameters for the Supernovae Type 1A data with cosmological parameter values obtained as $\bar{h} = 0.6500^{+0.0743}_{-0.0375}$ (unreliably constrained), $\Omega_{b0}=0.0453^{+0.0356}_{-0.0308}$ (unreliably constrained), $D_{1} = 1.0506^{+0.2936}_{-0.3785}$ (constrained) and $D_{2} = 0.1737^{+0.0675}_{-0.0693}$ (constrained) calculated by the MCMC simulation (L.H.S. panel). The R.H.S. panel shows the residual distance in \textit{Mpc} between the predicted model values and the data points.}
		\label{fig:OCG model graphs}
	\end{figure}
	
	From Fig. \ref{fig:OCG model graphs}, we see that the original CG model does explain the supernovae data. This is evident on the bottom panel since the model does not over-or under-estimate the supernovae at the low or intermediate redshift with the average offset being $\bar{x}_{res} = -0.0317$ Mpc, with a standard deviation on this average of $\sigma_{res} = 0.21386$ (which is even smaller than the $\Lambda$CDM model). However, even though the matter-density and Hubble uncertainty parameters gave realistic values that are within $1\sigma$ of their expected values, they are not fully constrained (leading to the large error region). 
	
	Now that we have seen that the OCG model does indeed show promise in being able to explain the supernovae data, we can try to fit the general CG model, where $\alpha$ is no longer fixed, but is a free parameter with boundary values of $0 < \alpha \leq 1.0$. We can then once again insert Eq. \ref{eq:chaplygin friedmann model} into our distance modulus model, execute the MCMC simulation and plot the results on the supernovae data in Fig. \ref{fig: CG model graphs}.
	\begin{figure}[ht!]
		\centering
		\begin{minipage}{0.45\textwidth}
			\centering
			\includegraphics[scale = 0.4]{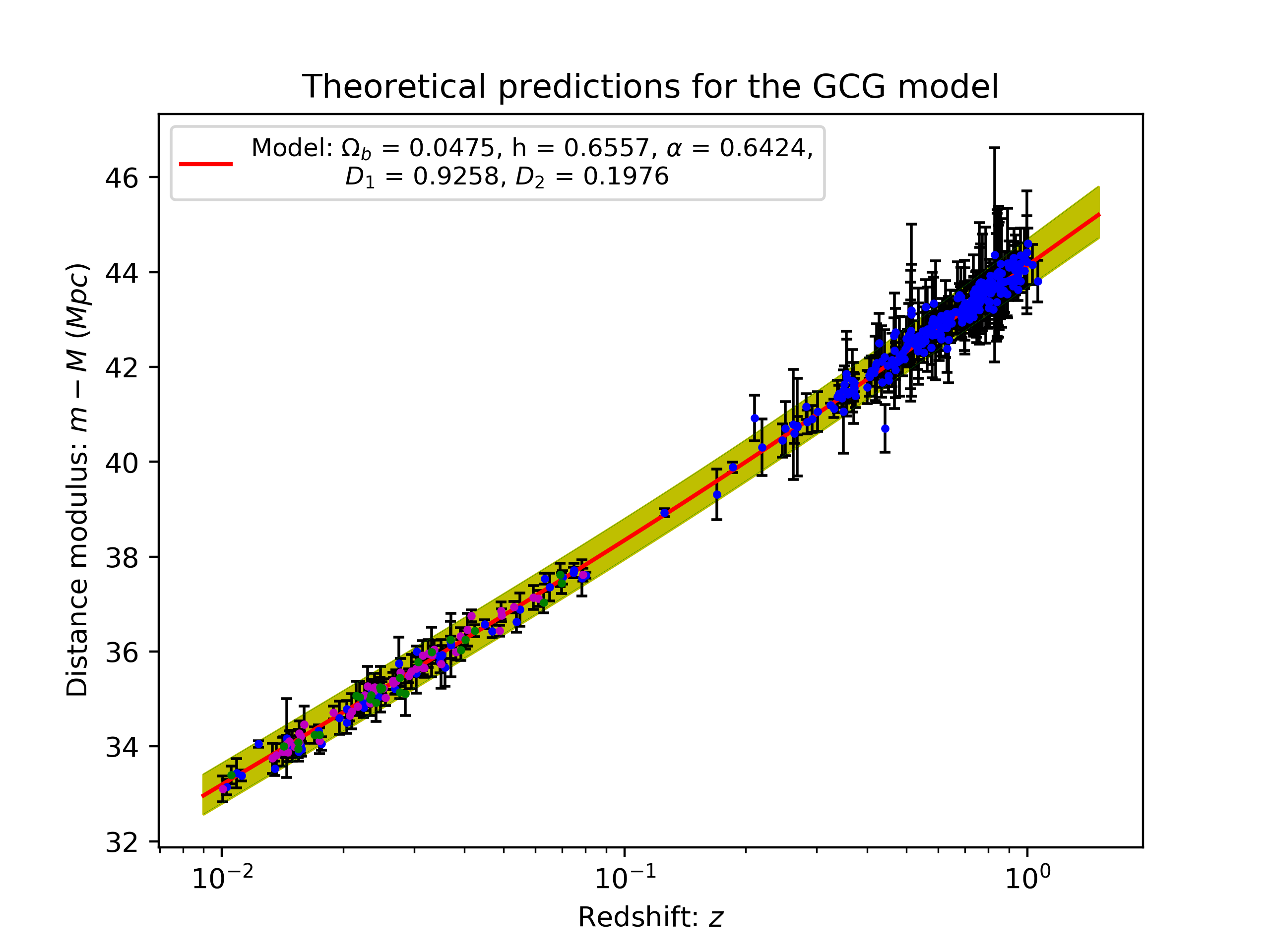}				
		\end{minipage}
		\hfill
		\begin{minipage}{0.45\textwidth}
			\centering
			\includegraphics[scale=0.4]{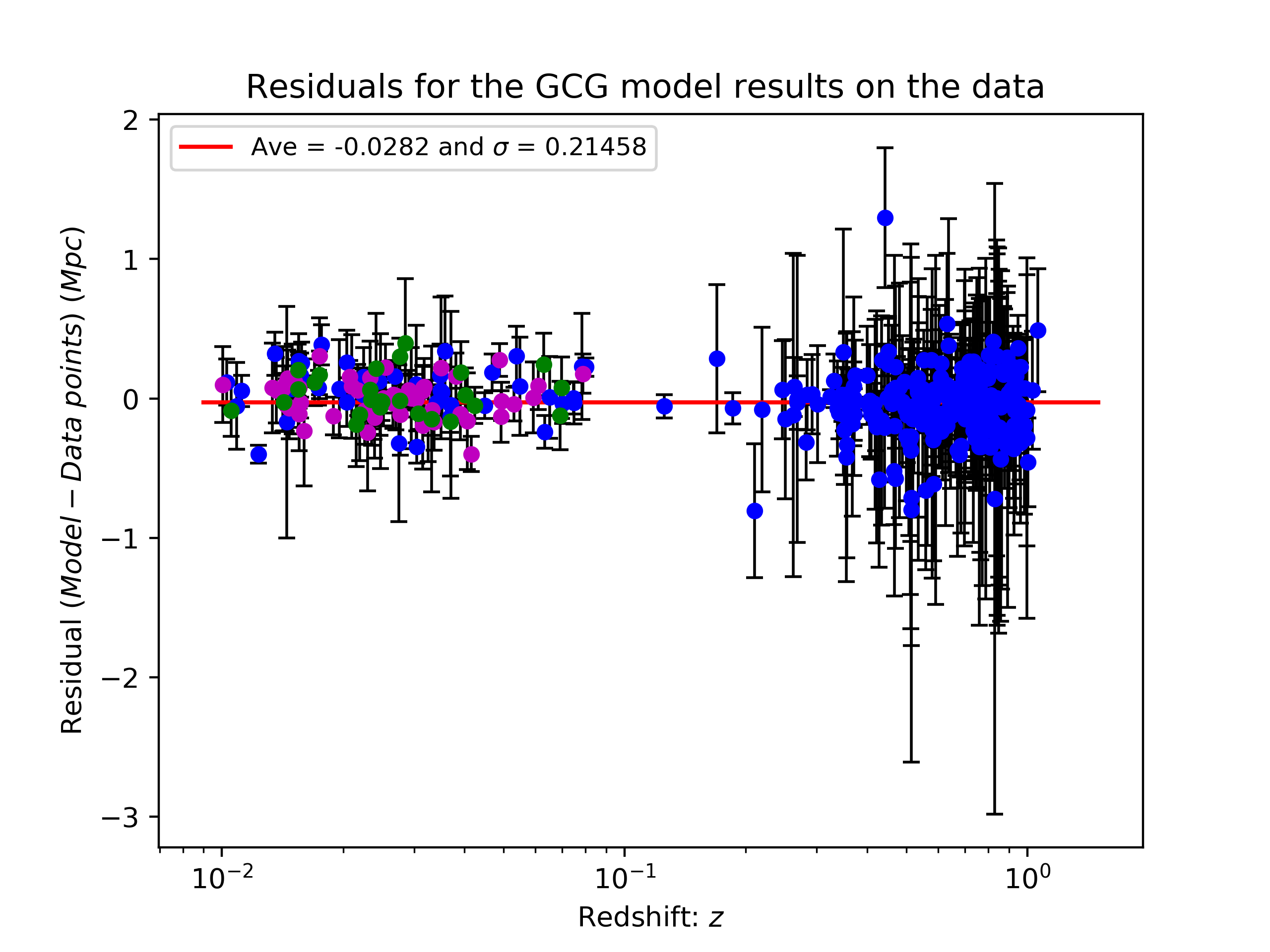}
		\end{minipage}
		\caption{The general CG model's Eq. \ref{eq:chaplygin friedmann model} best-fitting free parameters for the Supernovae Type 1A data with cosmological parameter values obtained to be $\bar{h} = 0.6557^{+0.0741}_{-0.0408}$ (unreliably constrained), $\Omega_{b0}=0.0475^{+0.0346}_{-0.0320}$ (unconstrained), $\alpha = 0.6424^{+0.2516}_{-0.3594}$ (unreliably constrained), $D_{1} = 0.9258^{+0.2476}_{-0.2669}$ (constrained), and $D_{2} = 0.1976^{+0.0675}_{-0.0675}$ (constrained) calculated by the MCMC simulation (L.H.S. panel). The R.H.S. panel shows the residual distance in \textit{Mpc} between the predicted model values and the data points.}
		\label{fig: CG model graphs}
	\end{figure}
	
	Fig. \ref{fig: CG model graphs} once again show that the CG models can explain the supernovae data, with realistic parameter values for the cosmological parameter, albeit unreliably constrained, as well as explaining the supernovae data without any over- or under-estimation on the low- or intermediate-redshift data. Even the average underestimation on the model on the data, namely $\bar{x}_{res} = -0.0282$ Mpc, with a standard deviation on this average of $\sigma_{res} = 0.21458$, which is on average closer to the actual data point values than even the OCG model (which were closer than the $\Lambda$CDM model), although the deviation for this model on the data is more in-line with the $\Lambda$CDM model's results (with the OCG model faring better in this case). However, as with the OCG model, we need to be careful when using these unreliable constraints. At least in the OCG model's case, both are unreliably constrained, which shows that it is closer to the actual best-fitting parameter value, than being unconstrained, such as with the GCG model.
	
	However, it is interesting that the values of the free parameters $D_{1}$ and $D_{2}$ are the parameters that the MCMC simulation gave as the dominant parameters when trying to fit the model. Therefore, it would seem as if $D_{1}$ and $D_{2}$ can up to a point determine the shape of the function, with the other parameters only being used to ``fine tune" the best-fitting model. This raises another question: since $D_{1}$ and $D_{2}$ are related to the free parameters $A$, $C_1$, and $H_{0}$, are the two free parameters the dominant parameters or is it $H_{0}$. In the case of the second, would it then not result in a constrained parameter value for $\bar{h}$, or is this parameter not needed due to using the Hubble parameter in $D_{1}$ and $D_{2}$, therefore, with them already constrained, is the Hubble parameter more reliable than the MCMC simulation predicts? 
	
	Moving on to the modified CG model (MCG), we can find the best-fitting parameters using Eq. \ref{eq: MCG model}. We use the boundary conditions $0< B\leq 1/3$ for the parameter $B$. The MCMC results for the MCG model are plotted on the supernovae data and are shown in Fig. \ref{fig: MCG model graphs}. 
	\begin{figure}[ht!]
		\centering
		\begin{minipage}{0.45\textwidth}
			\centering
			\includegraphics[scale = 0.4]{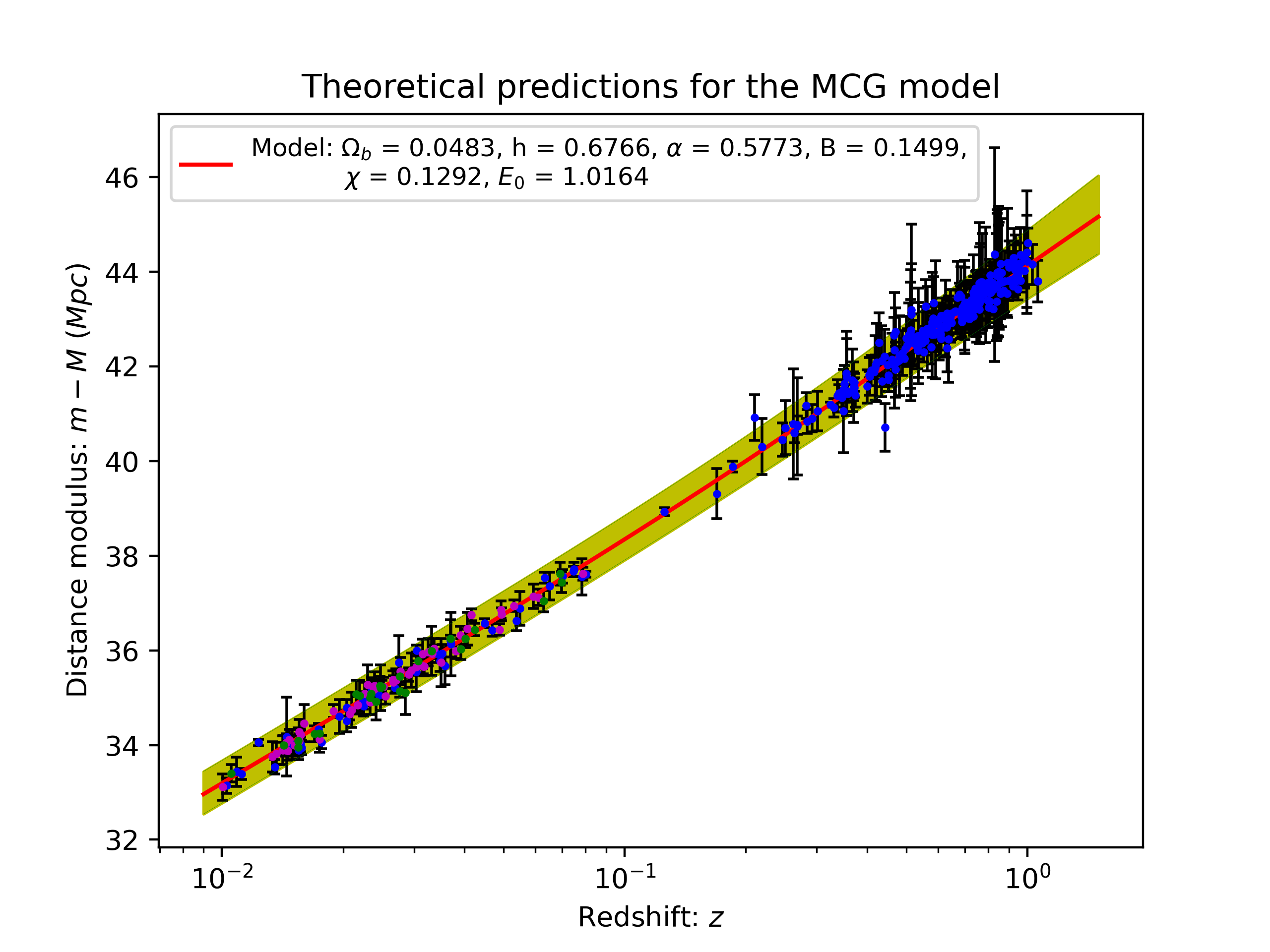}				
		\end{minipage}
		\hfill
		\begin{minipage}{0.45\textwidth}
			\centering
			\includegraphics[scale=0.4]{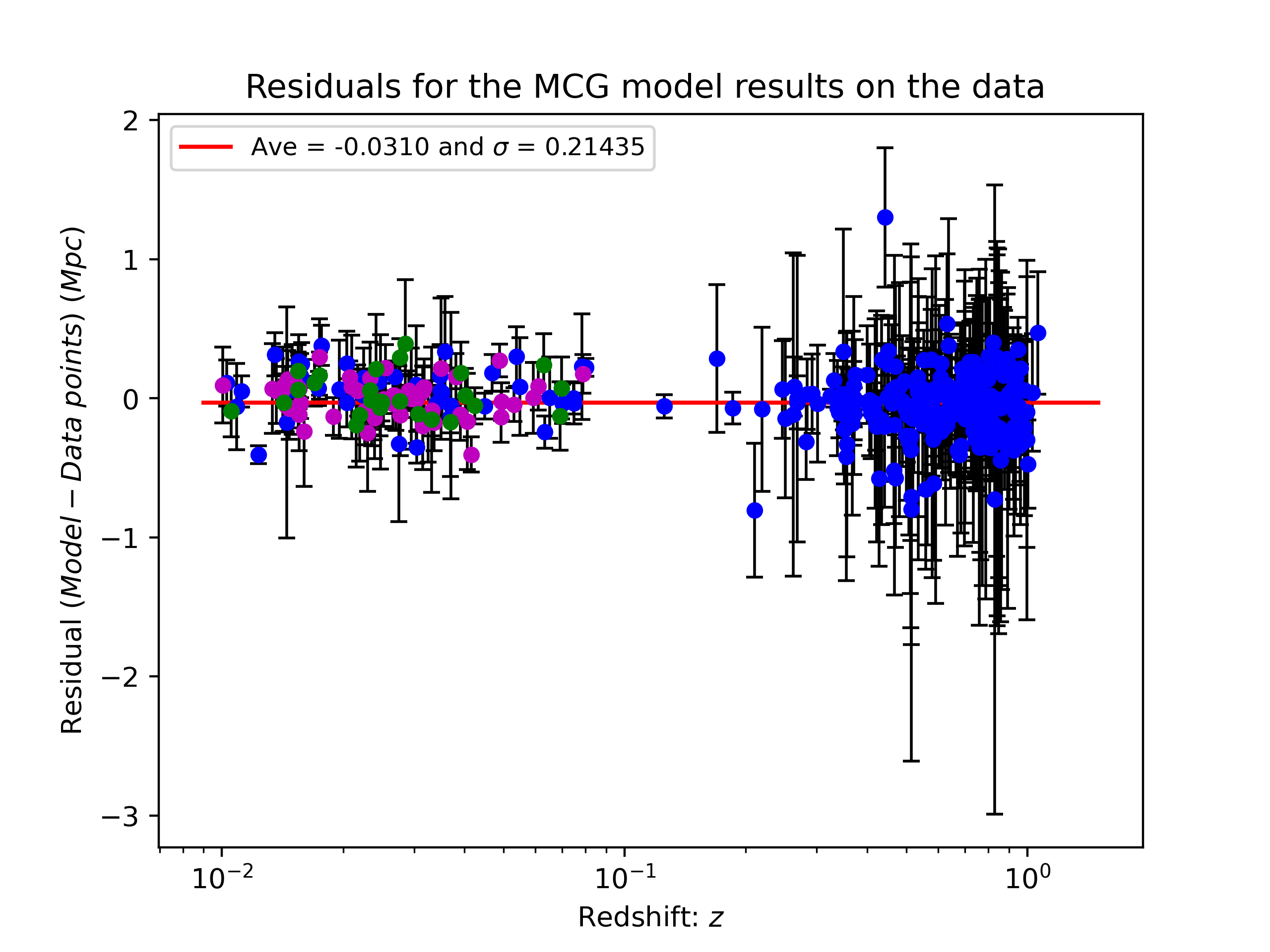}
		\end{minipage}
		\caption{The modified CG model's Eq .\ref{eq: MCG model} best-fitting free parameters for the Supernovae Type 1A data with cosmological parameter values obtained to be $\bar{h} = 0.6766^{+0.0738}_{-0.0564}$ (unreliably constrained), $\Omega_{b0}=0.0483^{+0.0350}_{-0.0333}$ (unconstrained), $\alpha = 0.5773^{+0.2915}_{-0.3445}$ (unreliably constrained), $B = 0.1499^{+0.1186}_{-0.1047}$ (unreliably constrained), $E_{0} = 1.0164^{+0.3670}_{-0.2904}$ (constrained), and $\mathcal{K} = 0.1292^{+0.0652}_{-0.0466}$ (constrained) calculated by the MCMC simulation (L.H.S. panel). The R.H.S. panel shows the residual distance in \textit{Mpc} between the predicted model values and the data points.}
		\label{fig: MCG model graphs}
	\end{figure}
	
	From Fig. \ref{fig: MCG model graphs}, we can see that the modifications to the general CG model hold up since it was able to find an average underestimate of the residual between the model and the supernovae data points of only $\bar{x}_{res} = -0.0310$ Mpc, which has a difference of $0.0028$ Mpc further away relative to the GCG model, but has a tighter relationship with the data, with a standard deviation of $\sigma_{res} = 0.21435$. We also once again note that this model has no under- or overestimation of the data at any red-shift. It is also worth noting that the only parameters that were fully constrained were the free parameters, $E_{0}$ and $\mathcal{K}$, enhancing our suspicion that the free parameters determine the general shape of the function, while the other parameters are only there for fine adjustments to make it more accurate.
	
	Moving on to the next CG model, namely the modified generalised CG model (MGCG) given by eq. \ref{eq. MGCG Hubble parameter}. The MCMC results for the MGCG model are plotted on the supernova data and are shown in Fig. \ref{fig: MGCG model graphs}. 
	\begin{figure}[ht!]
		\centering
		\begin{minipage}{0.45\textwidth}
			\centering
			\includegraphics[scale = 0.4]{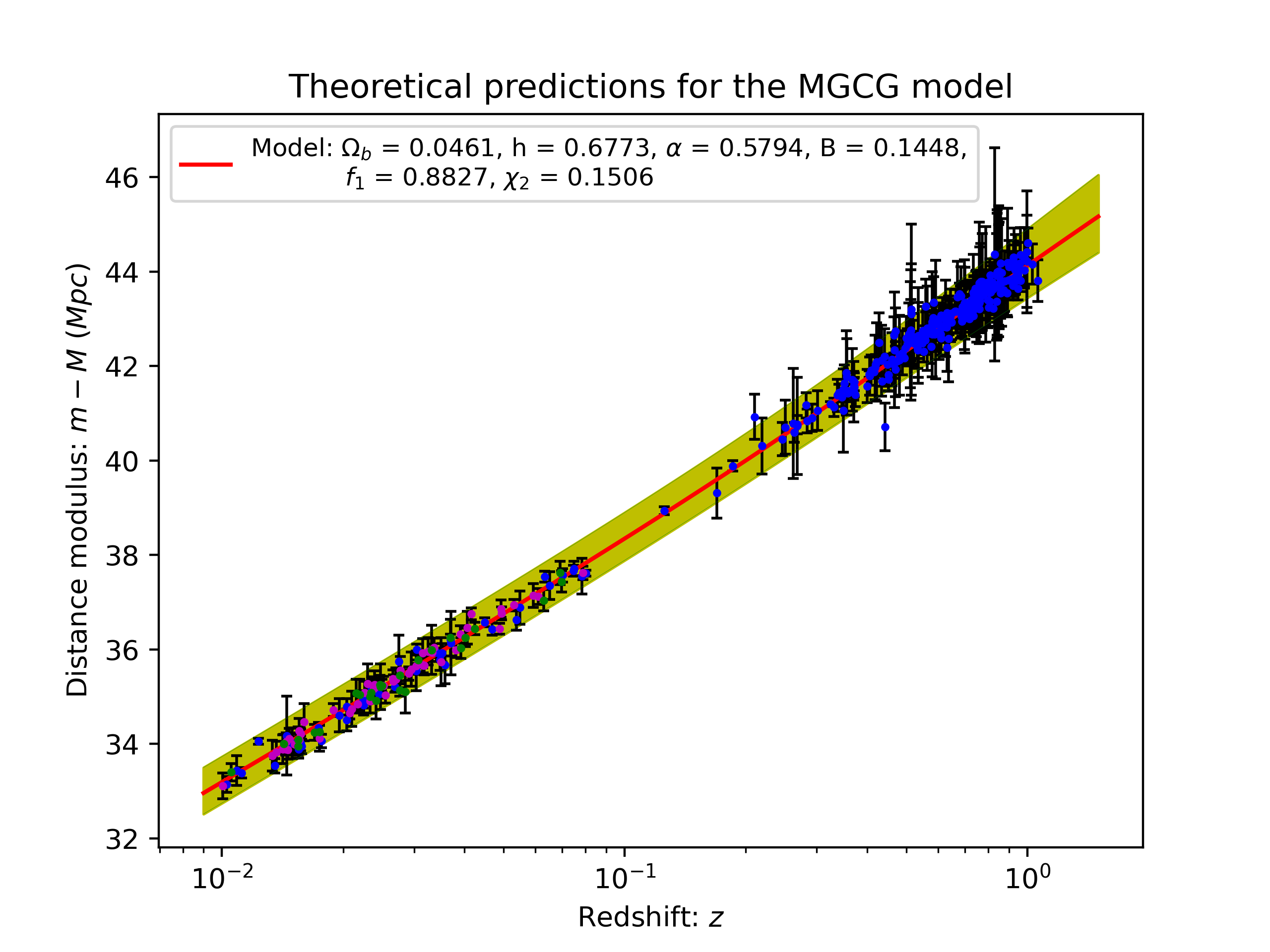}			
		\end{minipage}
		\hfill
		\begin{minipage}{0.45\textwidth}
			\centering
			\includegraphics[scale=0.4]{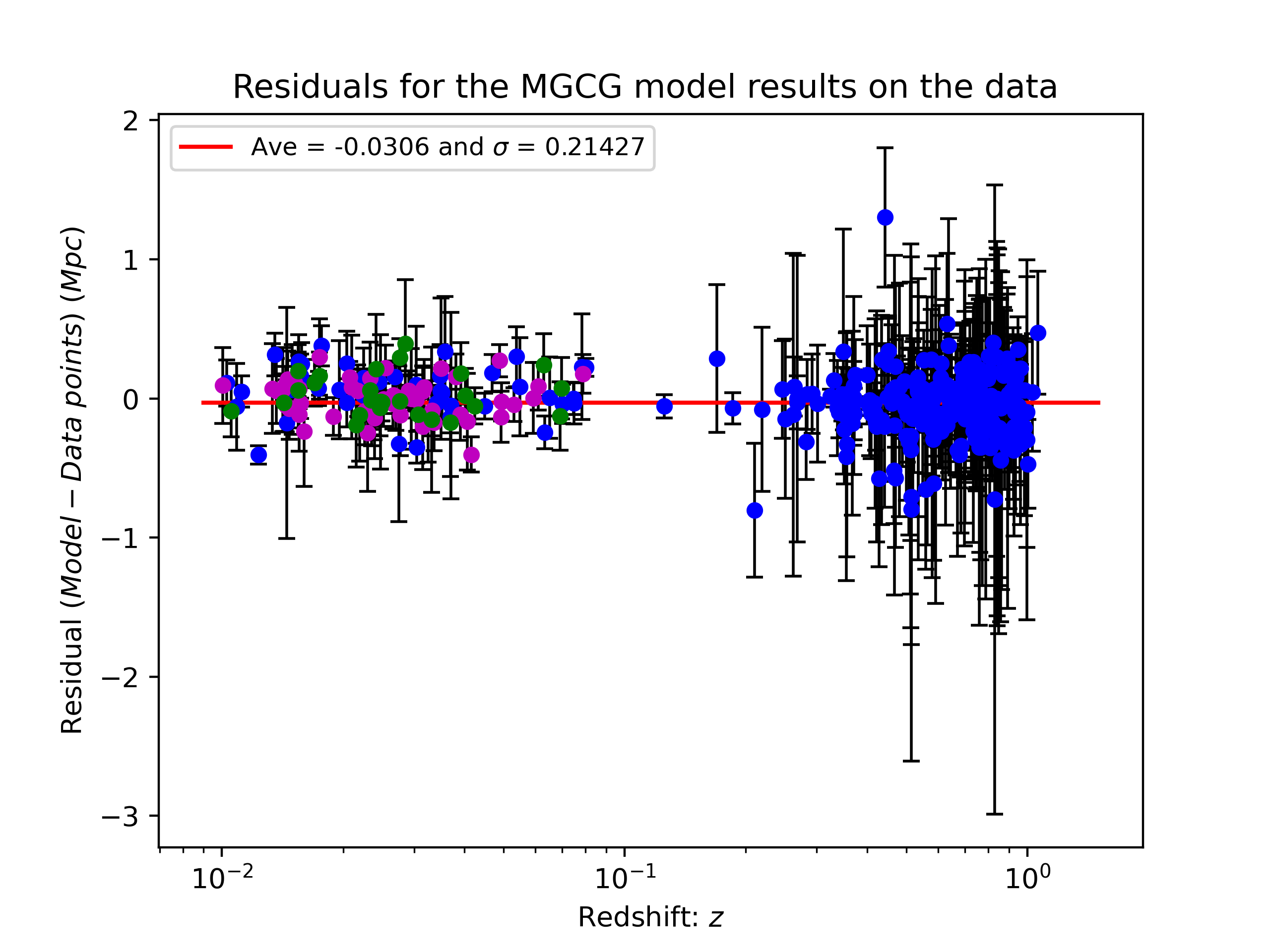}
		\end{minipage}
		\caption{The modified generalised CG model's Eq. \ref{eq. MGCG Hubble parameter} best-fitting free parameters for the Supernovae Type 1A data with cosmological parameter values obtained to be $\bar{h} = 0.6773^{+0.0751}_{-0.0557}$ (unreliably constrained), $\Omega_{b0}=0.0461^{+0.0363}_{-0.0320}$ (unconstrained), $\alpha = 0.5794^{+0.2880}_{-0.3460}$ (unreliably constrained), $B = 0.1448^{+0.1200}_{-0.1006}$ (unreliably constrained), $ F_{1} = 0.8827^{+0.2847}_{-0.2464}$ (constrained) and $\mathcal{K}_{2} = 0.1506^{+0.0600}_{-0.0469}$ (constrained) as calculated by the MCMC simulation (L.H.S. panel). The R.H.S. panel shows the residual distance in \textit{Mpc} between the predicted model values and the data points.}
		\label{fig: MGCG model graphs}
	\end{figure}
	
	From Fig. \ref{fig: MGCG model graphs}, it is clear that the MGCG model fits the supernova data as we have seen with the other CG models. Furthermore, we once again see that the cosmological parameters do not play a big role in determining the outcome of the model, which is not useful for determining the physical attributes of the Universe. We also see that there is no over-or-underestimation of the observational data at low- or intermediate-redshifts with the average offset being $\bar{x}_{res} = -0.0306$ Mpc, with a standard deviation of $\sigma_{res} = 0.21427$ (which is in line with the results from the MCG model), therefore confirming that this is the best possible fit with our MCMC simulation resolution.
	
	Getting to our second to last model, namely the ECG is given by Eq. \ref{eq. ECG hubble parameter}. We also have to remember that this is only one solution and not the complete model, therefore, we will not be able to fully constrain this model or make conclusions about the viability of this model, but we will get insight into how the model works. The MCMC results for this model are plotted on the supernovae data and are shown in Fig. \ref{fig: ECG model graphs}.
	\begin{figure}[ht!]
		\centering
		\begin{minipage}{0.45\textwidth}
			\centering
			\includegraphics[scale = 0.4]{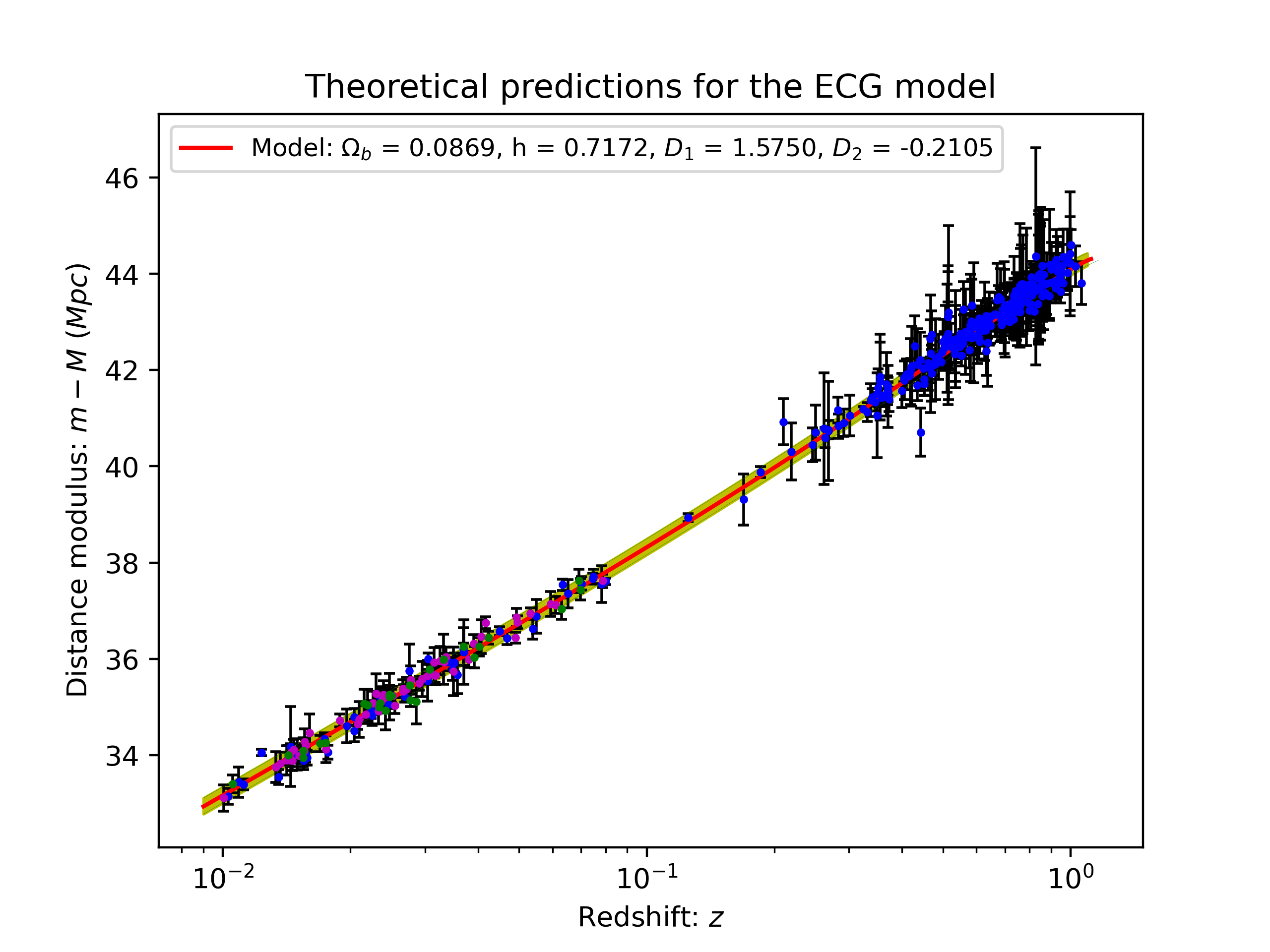}				
		\end{minipage}
		\hfill
		\begin{minipage}{0.45\textwidth}
			\centering
			\includegraphics[scale=0.4]{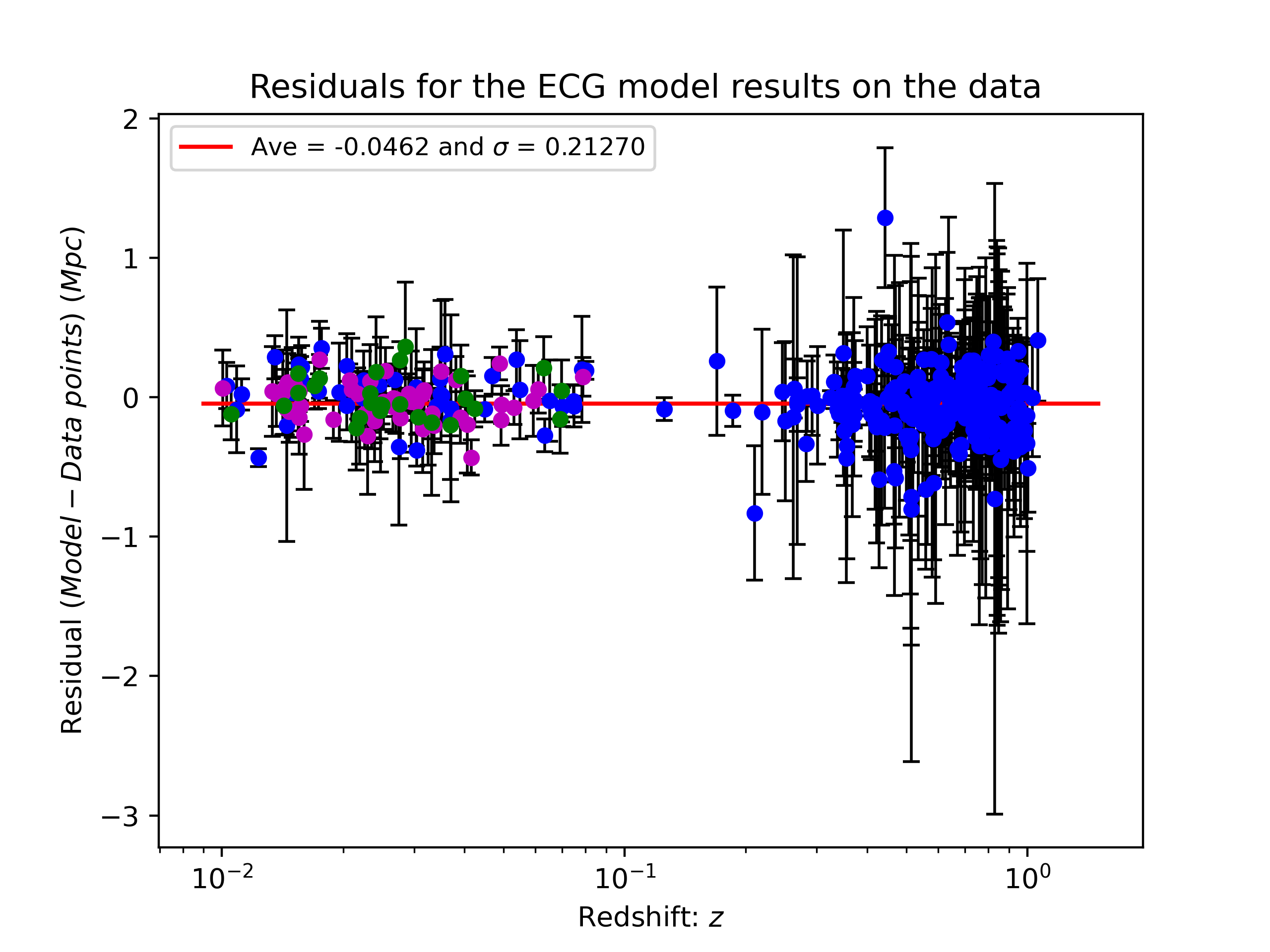}
		\end{minipage}
		\caption{The extended CG model's Eq. \ref{eq. ECG hubble parameter} best-fitting free parameters for the Supernovae Type 1A data with cosmological parameter values obtained to be $\bar{h} = 0.7172^{+0.0231}_{-0.0276}$ (unreliably constrained), $\Omega_{b0}=0.0869^{+0.0095}_{-0.0170}$ (unreliably constrained), $D_{1} = 1.5750^{+0.1404}_{-0.1117}$ (constrained) and $D_{2} = -0.2105^{+0.0128}_{-0.0069}$ (unreliably constrained) as calculated by the MCMC simulation (L.H.S. panel). The R.H.S. panel shows the residual distance in \textit{Mpc} between the predicted model values and the data points}
		\label{fig: ECG model graphs}
	\end{figure}
	
	From Fig. \ref{fig: ECG model graphs}, we see that this particular solution to the ECG model is quite stable, having a smaller error region around the best fit than any of the previous CG models. We see that there is no real over or underestimation of the observational data at low- or intermediate-redshifts, but it has the highest average offset of any of the CG models with a value of $\bar{x}_{res} = -0.0462$ Mpc, but in turn has the smallest standard deviation of $\sigma_{res} = 0.21270$, confirming that this is the best possible fit with our MCMC resolution. One thing that is quite worrying about this model is the very high baryonic matter density. The search space allowed that the MCMC simulation uses for matter density is less than $0.1$ since we know from other observational results that it is about 5\%. The MCMC simulation found a matter density value close to the high end of the search space and it is unreliably constrained. Meaning that the best-fit value is even higher. This is not possible and goes against other experimental results. 
	
	We can now go over to our last model's Hubble parameter equation, namely the generalised cosmic CG (GCCG) model, given by Eq. \ref{cGGG}. The MCMC simulation results (with these manually constrained parameter values) are shown in Fig. \ref{fig: GCCG model graphs}.
	\begin{figure}[ht!]
		\centering
		\begin{minipage}{0.45\textwidth}
			\centering
			\includegraphics[scale = 0.4]{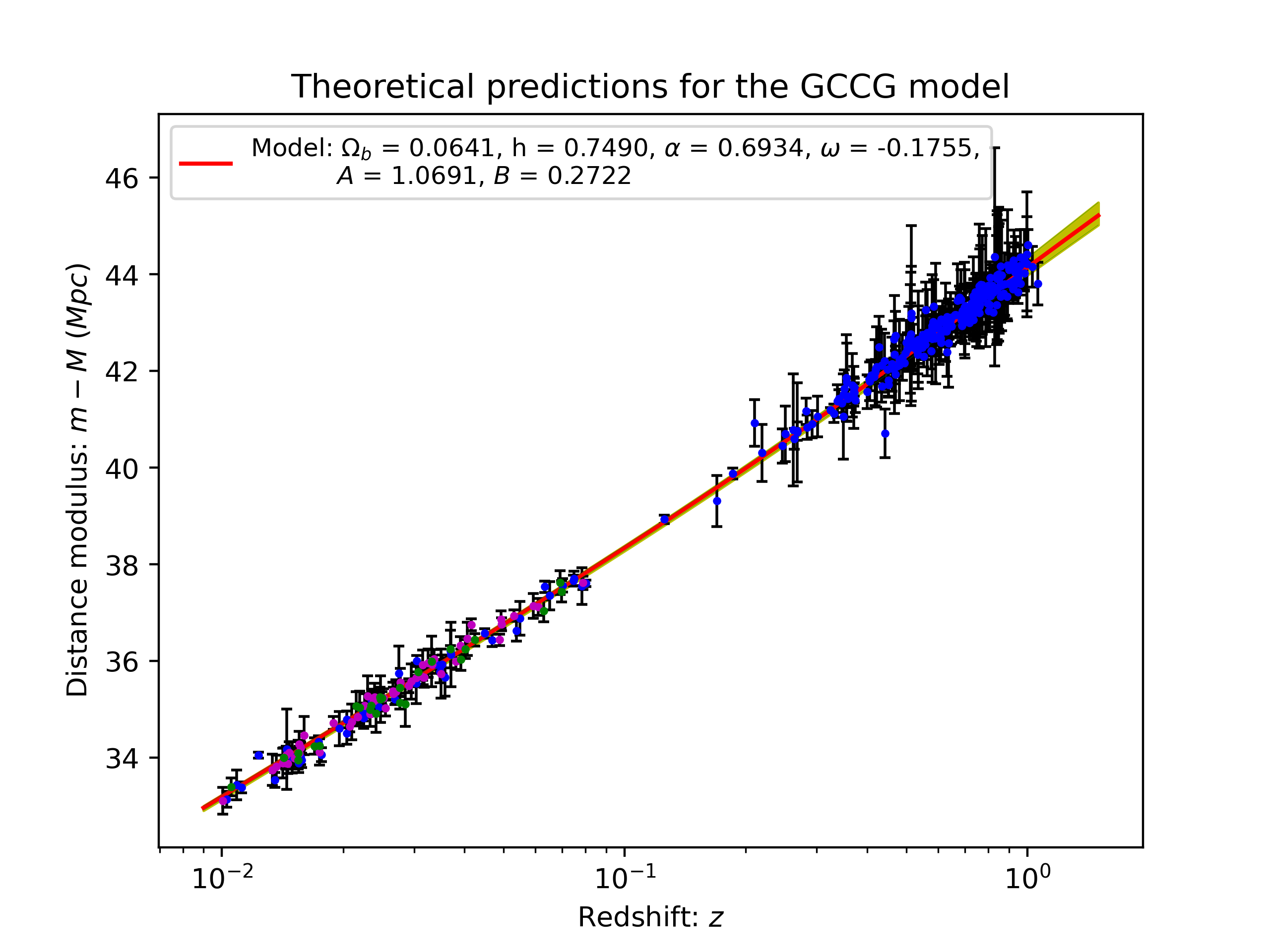}
		\end{minipage}
		\hfill
		\begin{minipage}{0.45\textwidth}
			\centering
			\includegraphics[scale=0.4]{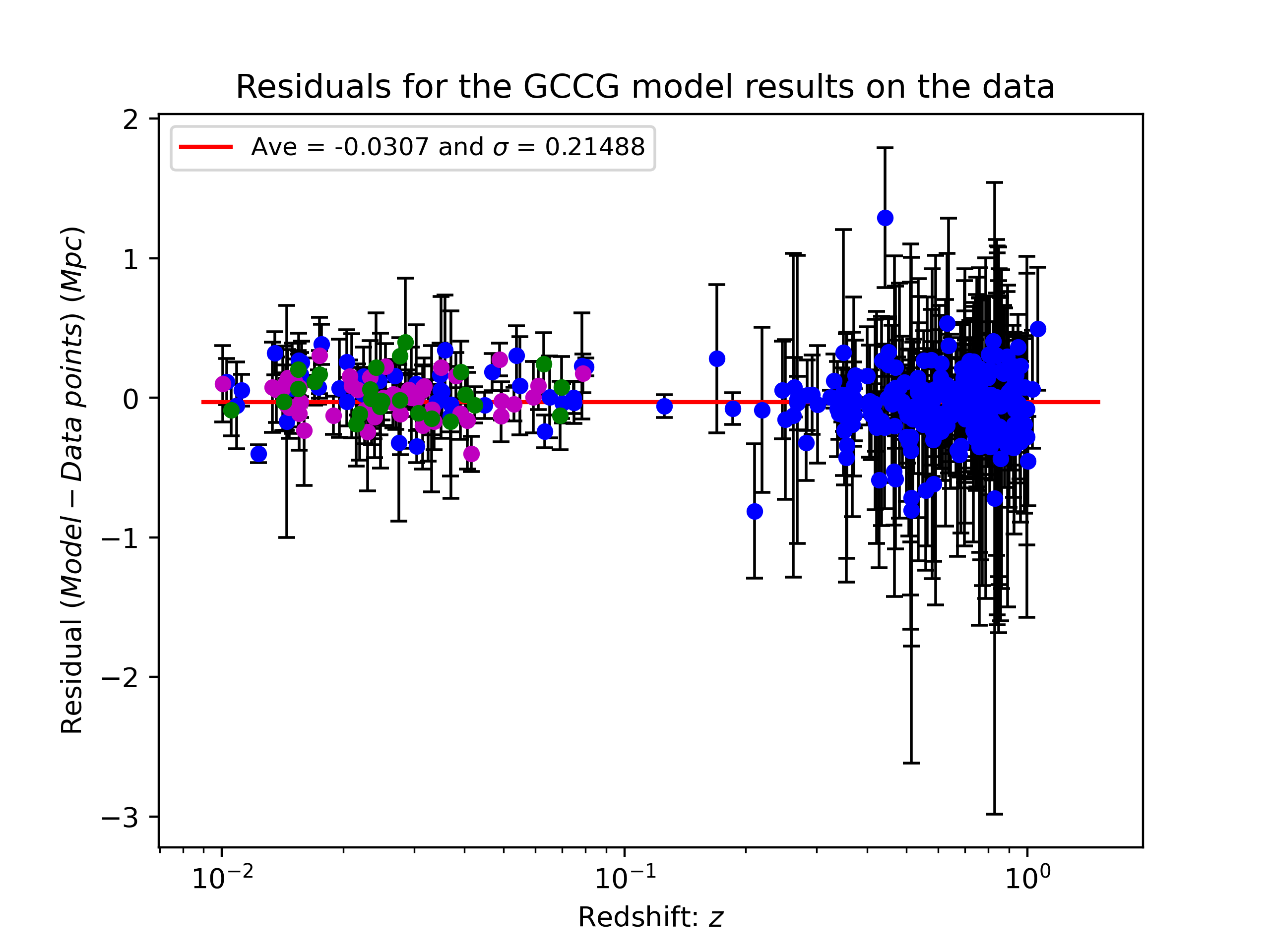}
		\end{minipage}
		\caption{The generalised cosmic CG model's Eq. \ref{cGGG} best-fitting free parameters for the Supernovae Type 1A data with cosmological parameter values obtained to be $\bar{h} = 0.7490^{+0.0689}_{-0.0729}$ (unconstrained), $\Omega_{b0}=0.0641^{+0.0243}_{-0.0274}$ (unreliably constrained), $\alpha = 0.6934^{+0.2223}_{-0.3537}$ (unreliably constrained), $\omega = -0.1755^{+0.1282}_{-0.2224}$ (unreliably constrained), $ A = 1.0691^{+0.2961}_{-0.4024}$ (constrained), and $B = 0.2722^{+0.0417}_{-0.0535}$ (unreliably constrained) as calculated by the MCMC simulation (L.H.S. panel). The R.H.S. panel shows the residual distance in \textit{Mpc} between the predicted model values and the data points.}
		\label{fig: GCCG model graphs}
	\end{figure}
	
	From Fig. \ref{fig: GCCG model graphs} we get some interesting points. First, this is the Chaplygin gas model that is the most stable according to the MCMC simulation, just being beat out by the $\Lambda$CDM model in terms of its error regime. At very low redshift, the errors, as in the case with the $\Lambda$CDM, are almost not even visible, but the error regime starts to increase after the transition phase between the matter-dominated epoch and the late-time accelerating epoch. The second thing to note is that even though the baryonic matter density ($\Omega_{b0}$) and the free parameter $B$ are classified as unreliably constrained, their positive and negative errors are very much the same. This is the first time this happened. The reason for it is that the distribution plot obtains one side of the Gaussian distribution (therefore unreliably constrained), but it is not at the edge of the search space. It then gets to the peak of "most probable value" and then obtains a plateau. Therefore, the most probable value is a range of values for this parameter. We have a uniform description (which would have made it unconstrained) for a region in the search space, but that search space is also at the most probable value (therefore actually constraining the parameters).
	
	The following thing to note (and probably the most important thing) is that the Chaplygin gas model parameters, namely $\alpha$, $A$ and $B$ are very similar to the other Chaplygin gas model parameter values that we found, but the observable parameters, such as the baryonic matter density $\Omega_{b0}$, the Hubble constant $\bar{h}$ and the equation of state parameter $\omega$, are different from their counterpart parameter values. Not only the other Chaplygin gas model solutions, but also the literature results. Take, for example, the Plank2018 results \cite{Planck2018}. The values for these parameters are(depending on which combination of models and observations you use: $\Omega_{b0}=0.0493$, $\bar{h}=0.6736$, and $\omega = 1.04^{+0.01}_{-0.10}$. As you can see for the first two, we are much higher than the CMB values, while the equation-of-state parameter value is not even close. Now you may recall that the supernovae results are known to have a higher Hubble parameter, but even their results are lower than this, with $\bar{h}= 0.7490^{+0.0689}_{-0.0729}$. They also agree with Planck on the equation of state parameter and the values of the matter density parameter \cite{Abbott2019}. Therefore, this is not a very favourable result for this model. It does not agree at all with the supernovae or the Planck2018 results. While the other Chaplygin gas models agree with either the Planck2018 results, even though they are on supernovae data.
	
	From the residuals plot in Fig. \ref{fig: GCCG model graphs}, we do not obtain any over- or underestimation on a specific section of the data, therefore matching the transition phase going from the matter-dominated epoch, to an accelerating Universe epoch with an average distance estimation to the supernovae of $\bar{x}_{res} = -0.0307$ Mpc. The standard deviation on this average best fit over the entire redshift range is $\sigma_{res} = 0.21488$, which is comparable to the previous models. We summarise all of the different parameter values for each model in Tab. \ref{tab: best-fitting parameter values}.
	\begin{table}
		\scalebox{0.7}{
			\begin{tabular}{lrrrrrrr}
				\hline\noalign{\smallskip}
				\multicolumn{1}{c}{\textbf{Model}} & \multicolumn{1}{c}{\textbf{$\bar{h}$}} & \multicolumn{1}{c}{\textbf{$\Omega_{b0} / \Omega_{m0}$}} & \multicolumn{1}{c}{\textbf{$D_{1} / E_{0}/ F_{1}/ A$}}& \multicolumn{1}{c}{\textbf{$D_{2} / \mathcal{K}/ \mathcal{K}_{2}$}}& \multicolumn{1}{c}{\textbf{$\alpha$}} & \multicolumn{1}{c}{\textbf{$B$}}& \multicolumn{1}{c}{\textbf{$\omega$}}\\
				\noalign{\smallskip}\hline\noalign{\smallskip}
				$\Lambda$CDM & $0.6967^{+0.0048}_{-0.0047}$ & $0.2674^{+0.0249}_{-0.0239}$ & & & & &\\
				\noalign{\smallskip}
				Original CG & $0.6500^{+0.0743}_{-0.0375}$ & $0.0453^{+0.0356}_{-0.0308}$ & $1.0506^{+0.2936}_{-0.3785}$ & $0.1737^{+0.0675}_{-0.0693}$ & & &\\
				\noalign{\smallskip}
				General CG & $0.6557^{+0.0741}_{-0.0408}$ & $0.0475^{+0.0346}_{-0.0320}$ & $0.9258^{+0.2476}_{-0.2669}$& $0.1976^{+0.0675}_{-0.0675}$ & $0.6424^{+0.2516}_{-0.3594}$ & &\\
				\noalign{\smallskip}
				Modified CG & $0.6766^{+0.0738}_{-0.0564}$  & $0.0483^{+0.0350}_{-0.0333}$ & $1.0164^{+0.3670}_{-0.2904}$ & $0.1292^{+0.0652}_{-0.0466}$  & $0.5773^{+0.2915}_{-0.3445}$ & $0.1499^{+0.1186}_{-0.1047}$ &\\
				\noalign{\smallskip}
				Modified GCG & $0.6773^{+0.0751}_{-0.0557}$  & $0.0461^{+0.0363}_{-0.0320}$ & $0.8827^{+0.2847}_{-0.2464}$ & $0.1506^{+0.0600}_{-0.0469}$ & $0.5794^{+0.2880}_{-0.3460}$ & $0.1448^{+0.1200}_{-0.1006}$ &\\
				\noalign{\smallskip}
				Extended CG & $0.7172^{+0.0231}_{-0.0276}$ & $0.0869^{+0.0095}_{-0.0170}$ & $1.5750^{+0.1404}_{-0.1117}$ & $-0.2105^{+0.0128}_{-0.0069}$ & & &\\
				\noalign{\smallskip}
				GCCG & $0.7490^{+0.0689}_{-0.0729}$ & $0.0641^{+0.0243}_{-0.0274}$ & $1.0691^{+0.2961}_{-0.4024}$ & & $0.6934^{+0.2223}_{-0.3537}$ & $0.2722^{+0.0417}_{-0.0535}$ & $-0.1755^{+0.1282}_{-0.2224}$\\
				\noalign{\smallskip}\hline
		\end{tabular}}
		\caption{\it{The MCMC simulation best fitting parameters values for each tested model, including the $\Lambda$CDM model. The models are listed in the order they were tested. NB! Due to limitations, the following groupings are done: 1) The $\Lambda$CDM's $\Omega_{m0}$ parameter is grouped in the $\Omega_{b0}$ column. 2) Since there are similar free parameters, $D_{1}$, $E_{0}$, $F_{1}$ and $A$ are grouped, while $D_{2}$, $\mathcal{K}$ and $\mathcal{K}_2$ are grouped. 3) The B-parameters for the MCG and the MGCG models are limited to a maximum of $B=\frac{1}{3}$ due to the definitions of these models, while the B-parameter of the GCCG model is only an integration constant; therefore, even though their names are the same, they are inherently different.}}
		\label{tab: best-fitting parameter values} 
	\end{table}
	
	
	\subsection{Comparing the residuals of the various CG models to the $\Lambda$CDM's}
	
	Now that we have all the models that we are going to work with, we can compare each of their best-fitting residual plots to the $\Lambda$CDM model, to understand how each of them handles the transition phase between the matter-dominated epoch ($z > 0.5$) and the accelerating epoch that we are currently experiencing ($z<0.5$) \cite{linder2001understanding, capozziello2019model}. We visualise this comparison in Fig. \ref{fig: Theoretical residual plots}
	\begin{figure}[ht!]
		\centering
		\hspace{-6mm}
		\includegraphics[width=8cm,height= 5cm ]{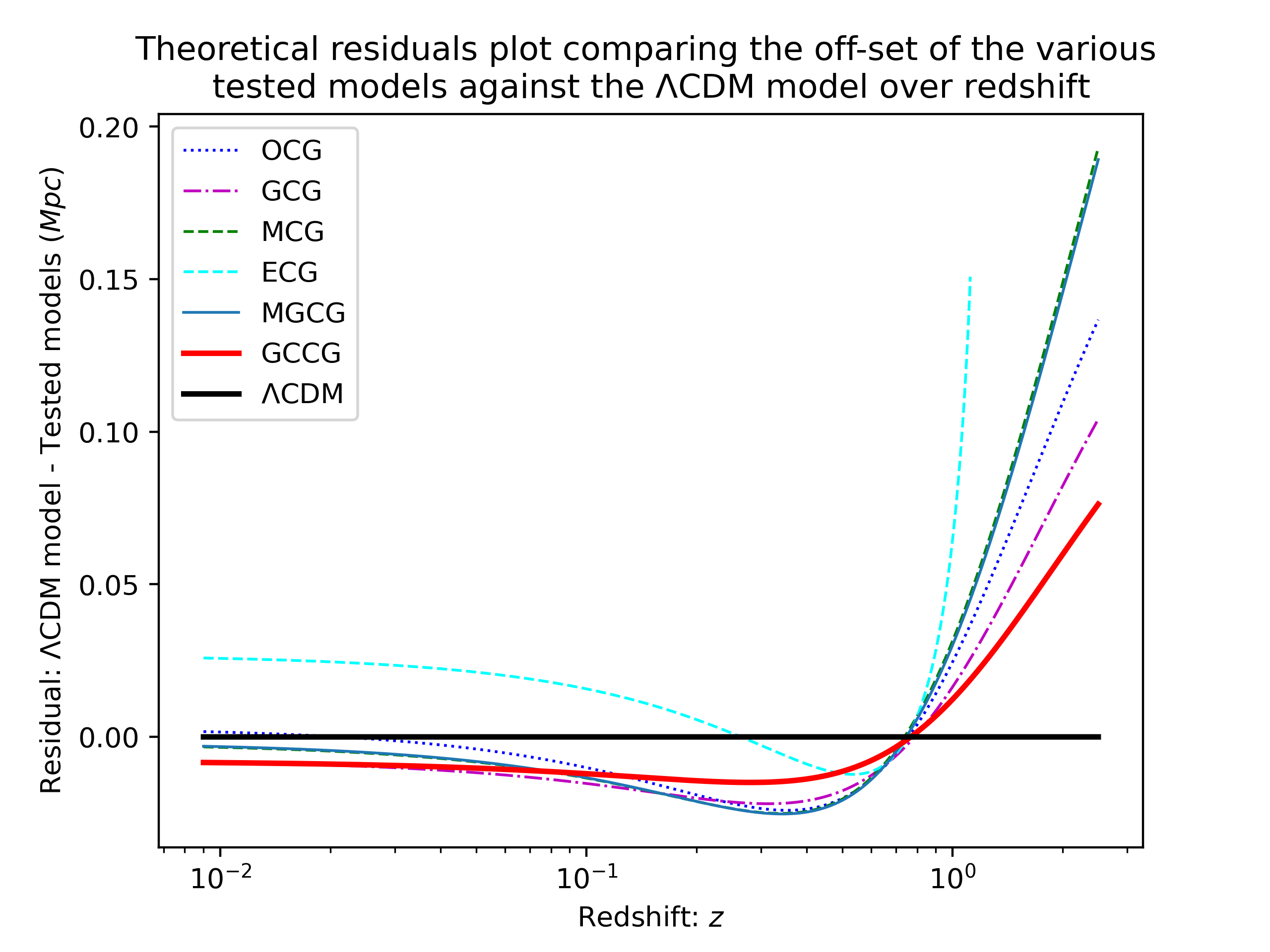}			
		\caption{Theoretical residuals comparing the different tested models against the $\Lambda$CDM model.}
		\label{fig: Theoretical residual plots}
	\end{figure}
	
	From Fig. \ref{fig: Theoretical residual plots}, we learn a few interesting details about the behaviour of the various Chaplygin gas models. First, all Chaplygin gas models have a very similar residual plot compared to the $\Lambda$CDM model. Second, we notice that at the transition point between the matter-dominated epoch transitioning to an accelerating epoch, all six CG models change their relative behaviour and start matching the trend produced by the $\Lambda$CDM model with the GCCG model handling the transition phase the best, while the specific ECG model solution was the worst. This is quite a distinction for these models, since they can explain an accelerating universe with the best-fitting parameter values that were obtained from observational data. Therefore, these models have theoretical backing to explain an accelerating Universe and have some support from observations. 
	
	Third, it is interesting to note that as time progresses, the CG models converge onto the $\Lambda$CDM model, with the OCG matching the very late time-distance predictions of the $\Lambda$CDM model almost exactly. All of these models end up within a distance modulus prediction of approximately $\Delta\mu \sim 0.001$ Mpc (the ECG model being the only exception). This is relatively small on the scale of these observations. To put this in perspective, the distance between the Sun and the centre of the Milky Way is estimated at $\sim 0.008$ Mpc \cite{reid1993distance}. Therefore, these models can mostly predict the distance to supernovae happening in various galaxies, to a distance less than the distance to the centre of our galaxy relative to the $\Lambda$CDM model's prediction. This is quite an admirable result.
	
	Lastly, although these models do quite well at late times, none of them succeeds in following the trend of matter-dominated epochs set by the $\Lambda$CDM model. This is worrisome, since a model that can explain both of the epochs, as well as the transition phase between the two, will be more useful in explaining the various cosmological phenomena. For instance, if we were to use higher redshift supernovae data, which is very difficult to obtain due to the incoming flux decreasing over distance \cite{leibundgut2001cosmological}, these models will struggle more and more relative to the $\Lambda$CDM model. The reason for this is that they are already diverging away from the $\Lambda$CDM model at the redshifts that we were able to obtain. The specific solution to the ECG model is probably the biggest culprit for not matching earlier times, such as the matter-dominated epoch, since it experiences a hard cutoff at $z=1$. This is a sign that this particular model is not valid at $z=1$. On the basis of this alone, we can already rule out this solution to the model ECG model. However, we will continue with the other tests, as we knew that this is only a single solution and not the whole model, so we can still learn important details about the ECG model.
	
	As an additional interesting test that we did, we saw that all the models' relative difference distance modulus begins to plateau at about $z \sim 10$ (except for the ECG model which has its hard cutoff at $z=1$), and starts to match the trend set by the $\Lambda$CDM model once again albeit with a difference ranging anywhere between $\Delta\mu \sim 0.15$ Mpc for the GCCG model to $\Delta\mu \sim 0.7$ Mpc for the MCG and MGCG models, with the GCCG model having the smallest relative difference between the $\Lambda$CDM model. Therefore, even though these models differ from the values predicted by the $\Lambda$CDM for earlier times, they still follow the same trend after $z\sim 10$ and find predictions with a relative distance modulus difference of less than $\pm 1$ Mpc.
	
	
	\subsection{Comparing the deceleration parameters of the various CG models to the $\Lambda$CDM's}\label{comparasion}
	
	Similarly to comparing the best-fitting residual plots, we can compare the behaviours of the deceleration parameters of each of the different models tested against the $\Lambda$ CDM plot. As shown in Sec. \ref{Sec: Background}, different combinations of the parameter values can lead to different ways in which the Universe can be shaped. This is important since the rate at which the Universe is changing cannot only tell us about the history of the shape of the Universe, but it can also enable us to make predictions about the future fate of the Universe. Therefore, using the best-fitting parameter obtained from the supernovae observations, we will be able to compare how each model predicts this rate of change compared to the accepted shape, predicted by the $\Lambda$CDM model, also using the same dataset and MCMC simulation. This is shown in Fig. \ref{fig: Best-fitting deceleration parameters}.
	\begin{figure}[ht!]
		\centering
		\hspace{-6mm}
		\includegraphics[width=8cm,height= 5cm]{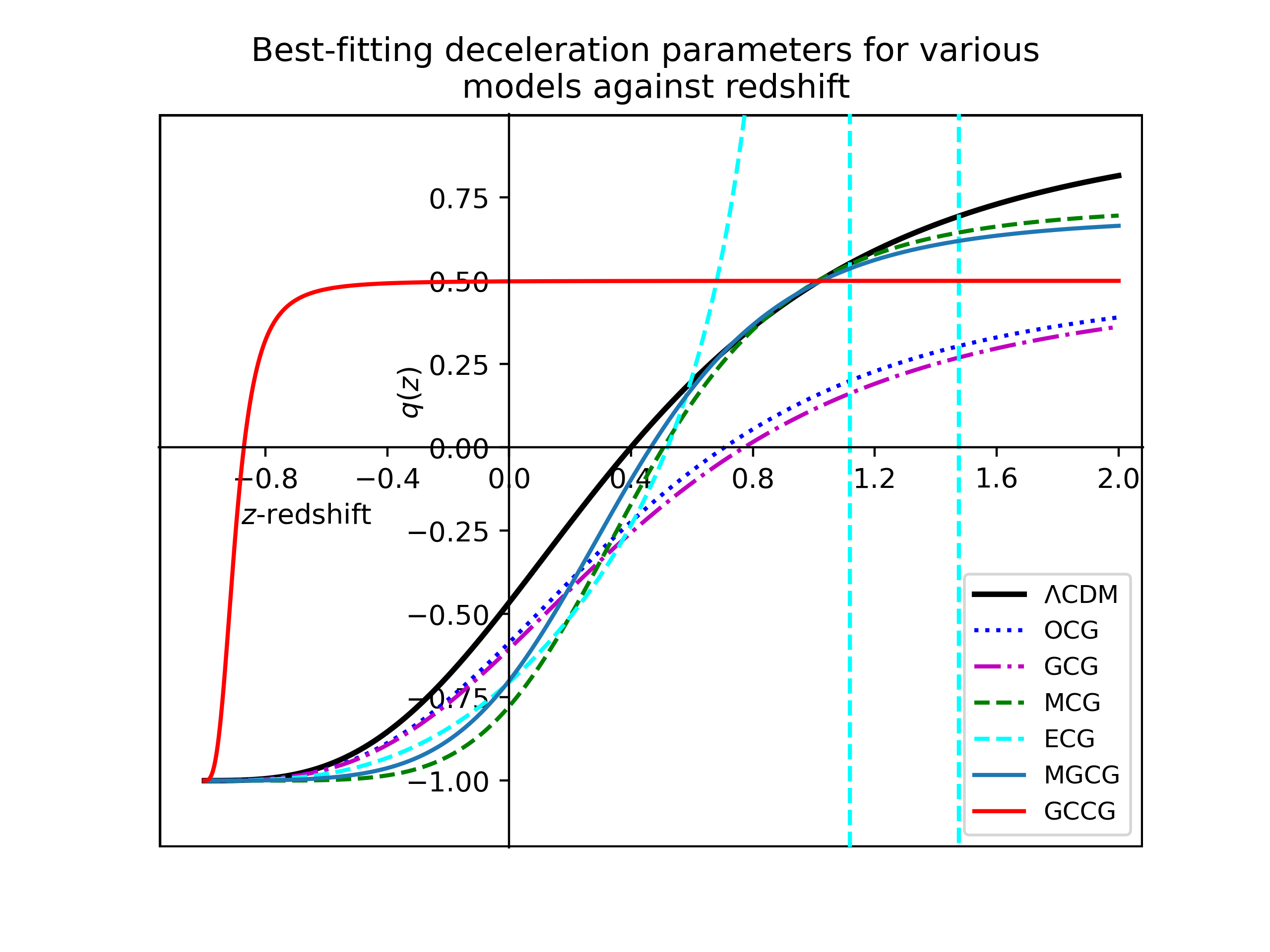}	
		\caption{The deceleration parameters from each of the different tested models compared to the $\Lambda$CDM model based on the best-fitting parameter values as predicted by the MCMC simulation on the supernovae data.}
		\label{fig: Best-fitting deceleration parameters}
	\end{figure}
	
	From Fig. \ref{fig: Best-fitting deceleration parameters}, we can again note a few important points about the behaviour of these CG models based on their best-fitting parameter values. First, we can see that the first four Chaplygin gas models obtained deceleration parameter revolutions that are very similar to one another, in particular the two pairings of the OCG/GCG models and the MCG/MGCG models. The only models that failed in this regard were the specific solution to the ECG model and the GCCG. In the graphs, we can once again see the hard cut-off that occurs at $z=1$ for this particular solution to the ECG, but excluding the cut-off $z=1$, we see that it follows the late-time deceleration parameter set by the $\Lambda$CDM model quite well. While for the GCCG model, things are more complicated. According to the graph. The Universe only had a constant deceleration to its expansion for a late time (which goes against observations), and will only start to accelerate in the future.
	
	Second, while focussing on the other more successful models, we see that the MGCG model matches on average the $\Lambda$CDM model's trend the best; however, it must be noted that to the present day ($z=0$), the OCG and GCG models were the closest to the rate at which the size of the universe expands as predicted by the $\Lambda$CDM model, while the MCG and MGCG models predicted that the Universe is expanding at a rate of about $\sim 25\%$ faster than the $\Lambda$CDM model.
	
	Third, we note that these four models, the specific solution to the ECG model and the $\Lambda$CDM model predict the transition period between the matter-dominated epoch and the accelerating epoch, is between $0.4 < z <0.8$, the ECG model being the closest to the value $z\sim 0.5$ found in the literature \cite{linder2001understanding, capozziello2019model}, just being outdone by the MCG model. Remember that this is based on our results from the MCMC simulations just to test the viability of these models, which can differ from the major experimental papers that constrains these parameters by combining various datasets. This is just to observe whether or not these models follow the accepted trends when tested with observational data.
	
	\section{Statistical analysis}\label{stastic}
	
	From a statistical analysis point of view, we can determine the small details that could have been missed in the visual comparisons to the $\Lambda$CDM model. The statistical analysis test that we will be using is the Akaike information criterion (AIC) and Bayesian/Schwarz information criterion (BIC) selections. These information criteria evaluate the plausibility of an alternative model explaining the data compared to an ``accepted/true model". In our case, the $\Lambda$CDM model will be considered as the ``true model". These selections take into account the best-fitting likelihood function values of the various models, but then also punish the model based on the number of free parameters it uses. The reason for this is that, with more free parameters, it is easier to manipulate a model into any shape you need, thus making it easier to fit the model to any dataset. This means that a model that can explain the data with the same accuracy with fewer free parameters that can be manipulated would be more likely to be statistically accepted than models with a lot of free parameters. The values for each model's AIC and BIC selections are calculated using the following equations:
	\begin{equation}
		\begin{split}
			\bullet\quad AIC &= \chi ^{2} +2K,\\
			\bullet\quad BIC &= \chi ^{2} +K\log(n),
		\end{split}
	\end{equation}
	where $\chi^{2}$ is calculated using the model's Gaussian likelihood function $\mathcal{L}(\hat{\theta} |data)$ value, $K$ is the amount of free parameters for the particular model, while $n$ is the amount of data points in our dataset. Since the value of $\chi^{2}$ can be any positive value based on the number of data points, the calculated values for the AIC and the BIC can be very random. Therefore, we will have to use the difference in the AIC and BIC values of each model compared to the ``true model's" AIC and BIC values. This statistical test is also used in other similar research papers \cite{nunes2017new}. To make sense of the information-criterion difference values between the CG models (aka the alternative models) and the $\Lambda$CDM model, we will use Jeffrey's scale to make conclusions about the viability of the various CG models. It should be noted that this is not an exclusive scale and should be handled with care \cite{nesseris2013jeffreys}. The Jeffrey's scale ranges are:
	\begin{itemize}
		\item  $\Delta IC \leq 2$ -\textrm{substantial observational support},\\
		\item  $ 4 \leq \Delta IC \leq 7$ - \textrm{less observational support},\\ 
		\item  $\Delta IC \geq 10$ -\textrm{no observational support}.\\
	\end{itemize}
	
	Lastly, since we already have the calculated $\chi^{2}$-values, we will also be able to calculate a reduced $\chi^{2}_{red.}$-value for each tested model. This will allow us to mathematically compare the overall best fits, namely the goodness-of-fit, for each model on the supernovae data. 
	
	
	\subsection{Using SNIa}\label{stastic1}
	
	All statistical analysis values calculated for each model, including the $\Lambda$ CDM model, are shown in the Tab.
	\ref{tab:statistical analysis}.
	\begin{table}
		\scalebox{0.7}{
			\begin{tabular}{llllllll}
				\hline\noalign{\smallskip}
				\textbf{Model} & \textbf{$\mathcal{L}(\hat{\theta}|data)$} & \textbf{$\chi ^{2}$} & \textbf{Red. $\chi ^{2}$} & \textbf{$AIC$} &\textbf{$|\Delta AIC|$} & \textbf{$BIC$} & \textbf{$|\Delta BIC|$}\\
				\noalign{\smallskip}\hline\noalign{\smallskip}
				Original CG & -120.1494 & 240.2987 & 0.6769 & 248.2987 & 2.8978 & 263.8320 & 10.6645\\
				\noalign{\smallskip}
				Modified CG & -120.2718 & 240.5435 & 0.6814 & 252.5435 & 7.1426 & 275.8434 & 22.6759\\
				\noalign{\smallskip}
				Modified Generalised CG & -120.2832 & 240.5665 & 0.6815 & 252.5665 & 7.1656 & 275.8664 & 22.6989\\
				\noalign{\smallskip}
				Extended CG & -120.5708 & 241.1415 & 0.6793 & 249.1415 & 3.7406 & 264.6748 & 11.5073\\
				\noalign{\smallskip}
				Generalized Cosmic CG & -120.6540 & 241.3079 & 0.6836 & 253.3079 & 7.9071 & 276.6079 & 23.4404\\
				\noalign{\smallskip}
				General CG & -120.6823 & 241.3645 & 0.6818 & 251.3645 & 5.9636 & 270.7811 & 17.6136\\
				\noalign{\smallskip}
				$\Lambda$CDM & -120.7004 & 241.4009 & 0.6762 & 245.4009 & 0 & 253.1675 & 0\\
				\noalign{\smallskip}\hline
		\end{tabular}}
		\caption{The best fit for each tested model, including the $\Lambda$CDM model. The models are listed in order from the highest value of the likelihood function $\mathcal{L}(\hat{\theta}|data)$ to the lowest likelihood of being viable. The reduced $\chi^{2}$-values are given as an indication of the goodness of fit for a particular model. The AIC and BIC values are shown, as well as the $\Delta IC$ for each information criterion. The $\Lambda$CDM model is chosen as the ``true model".}
		\label{tab:statistical analysis} 
	\end{table}
	
	From Tab. \ref{tab:statistical analysis}, we can take note of some interesting information. First, we found that all different CG models obtained a better likelihood function value than the $\Lambda$CDM model based on a Gaussian probability distribution, with the OCG model obtaining the largest likelihood function value. However, in the reduced $\chi^{2}$ values, where the number of parameters is taken into account when determining the goodness of fit, the $\Lambda$CDM model has the best value with only the OCG and to some extent the particular solution to the ECG model that manages to match this precision. The MCG, GCG, MGCG, and GCCG models have a better fit to the data than the $\Lambda$CDM model, but due to their increasing amount of free parameters, they did not obtain a goodness of fit that matches the results of the $\Lambda$CDM and OCG model.
	
	This increase in parameters is especially evident with the GCCG model. As with the MCG and MGCG model, it has 6 free parameters and, just as they, the GCCG has a better likelihood function value than both the $\Lambda$CDM and the GCG. However, due to the number of free parameters, it is the weakest statistically performing model based on the reduced $\chi^{2}$ that specifically tests the goodness of fit. Move towards the information criteria to determine their viability of being an alternative model to explain the accelerated expansion of the Universe. In terms of the AIC test, we see that the OCG, the GCG and the specific ECG model solution obtained some observational support, with the OCG model just missing out on the substantial observational support category, but is still higher than the boundary for less observational support. Therefore, it can be concluded that the OCG model has some observational support according to the AIC criterion, while the GCG and the ECG model has less observational support. The MCG, MGCG, and GCCG models miss out on the less observational support but are not yet in the no observational support, so we cannot rule it out. 
	
	In terms of the BIC criteria, we did not obtain one model that had some observational support category, but the OCG model was the closest to being in one of the categories while the GCCG model is the worst-performing model. Therefore, statistically, based on likelihood, goodness of fit, AIC, and BIC criteria, the OCG model is the most likely to be an alternative model to the $\Lambda$CDM model, with the GCG, MCG, MGCG and the specific ECG model solution not being ruled out, but will have to be tested on other data sets before being accepted or rejected. The GCCG model has the least statistical backing since it is the worst performer in the goodness-of-fit test, and in both the AIC and BIC test. of the other four boundary case models, statistically the ECG model has the best chance of being a viable option, however, since we know from the other tests this solution is already ruled out, which means the MGCG model has the best chance of being a viable option due to it following the trends produced by the $\Lambda$CDM model on average the closest, as well as obtaining almost the best likelihood- and $\chi ^{2}$ values between all these three different models.
	
	Lastly, the GCCG model was statistically not viable to be an alternative model. This conclusion is further enhanced by the fact that the cosmological parameter values ($\Omega_{b0}$, $\bar{h}$ and $\omega$) are so different from other experiments and results found in the literature. We can also eliminate this particular solution of the ECG model, but due to the statistical analysis showing some promising results for the ECG model as a whole, we cannot exclude the ECG model, just this particular solution. Therefore, we can continue with it in the perturbations section. Furthermore, because the MCMC model obtains a best-fitting OCG model that can be considered to be an alternative model, pending further testing, we know that the MCMC simulation was indeed calibrated correctly and that these differences in the parameter values are strictly related to the model itself and not the MCMC simulation.
	
	This concludes the viability test of these different Chaplygin gas models using supernova cosmology. As mentioned above, after finding the viable models, we can proceed to different tests on those particular viable models. This includes the perturbation test that will be performed in the next section. It also includes different tests that can be studied in future work. We will not go into a lot of detail as this is out of the scope of this paper, but different tests at different redshifts can lead to new information about different epochs. For instance, since CMB cosmology occurs on a very high redshift, you will be able to add constraints on the radiation density. Cepheid variable stars that were used to calibrate the supernovae data can themselves be used for a very low-redshift (present-day) study.
	
	
	\subsection{Using OHD data}
	{We consider the Hubble parameter data (OHD) at redshifts of 0.57 and 2.34 from the recent BAO observations provided by Sloan Digital Sky Survey (SDSS), DR9, and DR11 as presented in \cite{sahni2014model}. After we restricted different observational parameters, the statistical analysis values of $\Lambda$CDM, OCG, GCG, MCG and ECG are calculated using OHD for further consideration, while the MGCG and GCCG models are ruled out in the previous section. As presented in Sec. \ref{stastic1}, the values of $\chi^2$, reduced Red.$\chi^2$, $AIC$, $\Delta AIC$,$BIC$, $\Delta BIC$ are calculated to compare the best-fit CG models with $\Lambda$CDM and available in Table \ref{tab:OHD1} using OHD data.}
	\begin{table}
		\centering
		\scalebox{0.7}{
			\begin{tabular}{llllllll}
				\hline\noalign{\smallskip}
				\textbf{Model} & \textbf{$\mathcal{L}(\hat{\theta}|data)$} & \textbf{$\chi ^{2}$} & \textbf{Red. $\chi ^{2}$} & \textbf{$AIC$} &\textbf{$|\Delta AIC|$} & \textbf{$BIC$} & \textbf{$|\Delta BIC|$}\\
				\noalign{\smallskip}\hline\noalign{\smallskip}
				$\Lambda$CDM & -16.0794 & 33.8814 & 0.7059 &39.8814 & 0 & 45.6769 & 0\\
				\noalign{\smallskip}
				Original CG &-15.6627 & 31.3254 & 0.6664 & 39.3255 & 0.5559 & 47.0527& 2.6242\\
				\noalign{\smallskip}
				General CG & -15.9842 & 31.9885 & 0.6801 & 39.9685 & 0.0871 & 47.6958 & 2.0189\\
				\noalign{\smallskip}
				MCG & -14.7626 & 29.5253 & 0.6284 & 37.5252 & 0.3562 & 45.2526 & 0.4243\\
				\noalign{\smallskip}
				ECG & -20.8592& 41.7184 & 0.9270 & 53.7184 & 8.4658 & 65.3094 &19.6325 \\
				\noalign{\smallskip}\hline
		\end{tabular}}
		\caption{The best fit for $\Lambda$CMD model, OCG, GCG, MCG and ECG models.}
		\label{tab:OHD1}
	\end{table}
	
	\begin{figure}[ht!]
		\centering
		\includegraphics[width=12cm,height= 8cm]{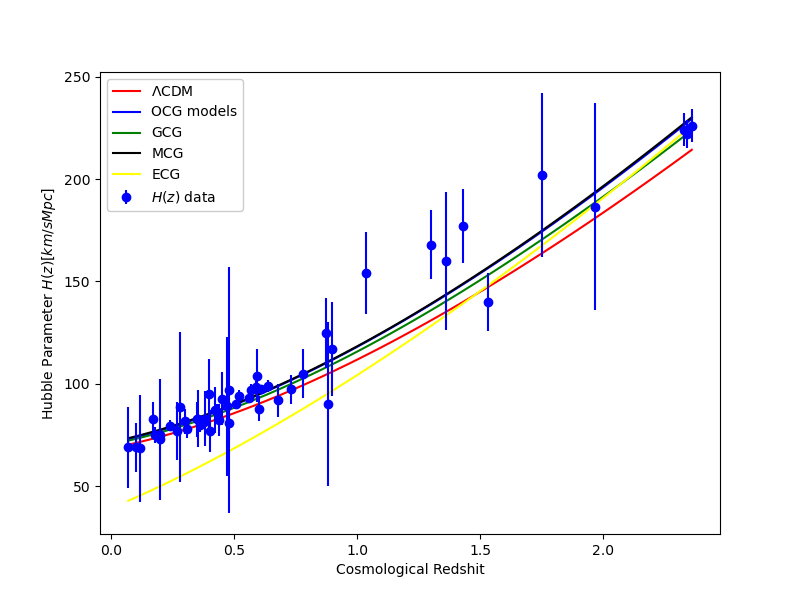}	
		\caption{The Hubble parameter $H(z)$ diagram versus redshift with various CG models and $\Lambda$CDM.}
		\label{fig:OHD1}
	\end{figure}

	{From ranges of Jeffrey’s scale \cite{nesseris2013jeffreys} and our statistical result in Table \ref{tab:OHD1}, we notice that the $\Delta$IC values of OCG, GCG and MCG indicate that these three models have substantial observational support in the background level. However, the calculated values of $\Delta$IC, ECG model indicate that the model has less observational support. As presented in Sec. \ref{stastic1}, MCG has failed in the statistical analysis in Sec. \ref{stastic} but it has substantial observational support in this analysis, and we consider it as a perturbation level in Secs. \ref{perturbation} and \ref{sigma} for further analysis We also present the Hubble parameter diagram in Fig. \ref{fig:OHD1} and we observe that all three models are alternative models to explain the accelerating expansion of the universe.}
	
	{After we explore the background cosmology of the CG models using SNIa and OHD data, here we also extended our knowledge at a linear cosmological perturbation level to see the contributions of CG to growth matter density fluctuations. At the perturbation level, we also consider the growth structure data $f_{\sigma 8}$ for the detailed analysis in Sec. \ref{sigma}. Based on Jeffrey’s scale ranges of our statistical results \cite{nesseris2013jeffreys}, we consider OCG and GCG that obtained substantial / less observational support. The density contrast is studied using the $\Lambda$CDM and baryonic-Chaplygin gas approaches. Based on the statistical analysis presented in \ref{stastic}, the MGCG and GCCG models are not preferred by observation and are not considered in this section to study density contrasts. However, MCG is considered to study at least this model fails in the background and see how it behaves at the perturbation level.
	}
	
	
	\section{Linear perturbations in Chaplygin gas cosmology}\label{perturbation}
	
	To do this, we consider the non-interacting non-relativistic fluid $\rho_b$ and the relativistic fluid $\rho_{\rm{r}}$, $\rho_m \equiv \rho_r+\rho_b$ with exotic fluid $\rho_{ch}$ in our Universe. In the background FLRW cosmological, we define spatial gradients of gauge-invariant variables such as:
	\begin{eqnarray}
		&& D^m_a \equiv \frac{a}{\rho_m}\tilde{\nabla}_a\rho_m\;,\quad Z_a \equiv a \tilde{\nabla}_a \theta\;,\\&& D^{ch}_a \equiv \frac{a}{\rho_{ch}}\tilde{\nabla}_a\rho_{ch}\;,\quad D^t_a \equiv \frac{a}{\rho_{t}}\tilde{\nabla}_a\rho_{t}\;,\nonumber
		\label{variables}
	\end{eqnarray}
	where the energy density $D^m_a$, volume expansion of the fluid $Z_a$ as follows \cite{dunsby1992covariant, abebe2012covariant, carloni2006gauge, ananda2008detailed}, $D^{ch}_a$ is the energy density for CG and $D^t_a$ is the spatial energy density for total fluids [Matter fluid +CG]. These four gradient variables are key to examining the evolution equation for matter density fluctuations. 
	
	For a perfect multi-fluid system, the following conservation equations, considered in \cite{dunsby1991gauge, dunsby1992covariant} hold:
	\begin{eqnarray}
		&&\dot{\rho}_i = -\theta(\rho_i+p_i)=-\theta(1+w_i)\rho_i\;, \\
		&&\dot{\rho}_t = -\theta(\rho_t+p_t)=-\theta(1+w_t)\rho_t\;, \\ 
		&&(\rho_t+p_t)\dot{u}_a +\tilde{\nabla}_a p_t=0 \;.
	\end{eqnarray}
	
	From these conservation equations, the 4-acceleration in the energy frame of the total fluid can be given as:
	\begin{eqnarray}
		\dot{u}_a &&= -\frac{\tilde{\nabla}_ap_t}{\rho_t+p_t}\\ 
		&&= -\frac{1}{a\left(1+w_t\right)\rho_t}\left[w_m\rho_m D^m_a+w_{ch}\rho_{ch}D^{ch}+a\rho_{ch}\tilde{\nabla}_a w_{ch}\right]\nonumber\;,
	\end{eqnarray} 
	with $w_{m}$ and $w_{ch}$ the equation of state parameters for matter fluid and Chaplygin gas, and $w_t$ is the effective equation of state parameter for the total fluid. Another key equation for a general fluid is the so-called Raychaudhuri equation and it can be expressed as:
	\begin{equation}
		\dot{\theta} = -\frac{1}{3}\theta^2 - \frac{1}{2}(\rho_t +3p_t)+\tilde{\nabla}^a\dot{u}_a\;.
		\label{Raychaudhuri}
	\end{equation}
	
	The above-defined spatial gradient variables evolve according to:
	\begin{eqnarray}
		&&\dot{D}^m_a-\left(\frac{1+w_{m}}{1+w_t}\right)\frac{w_{m}\rho_{m}}{\rho_t}\theta D^{m}_a -\left(\frac{1+w_{m}}{1+w_t}\right)\frac{w_{ch}\rho_{ch}}{\rho_t}\theta D^{ch}_a\nonumber\\ 
		&& \quad \quad-\left(\frac{1+w_m}{1+w_t}\right)\frac{\rho_{ch}}{\rho_t}a\theta\tilde\nabla_aw_{ch}+(1+w_{m}) Z_{a}=0 \;,\label{basic2}\\ 
		&& \dot{D}^{ch}_a-\left(\frac{1+w_{ch}}{1+w_t}\right)\frac{w_{ch}\rho_{ch}}{\rho_t}\theta D^{ch}_a-\left(\frac{1+w_{ch}}{1+w_t}\right)\frac{w_m\rho_m}{\rho_t}\theta D^{m}_a\nonumber\\ 
		&& \quad \quad+\left(\frac{1+w_m}{1+w_t}\right)\frac{\rho_m}{\rho_t}a\theta\tilde\nabla_aw_{ch}+(1+w_{ch}) Z_{a} =0 \;,\\
		&& \dot{D}^t_a-w_t\theta D^t_a+(1+w_t) Z_{a}=0 \;,\\
		&& \dot{Z}_a +\frac{2}{3} \theta Z_{a}+\Big(\frac{1}{2}\left(1+3w_m\right)\rho_m-\frac{w_m\rho_m}{(1+w_t)\rho_t}\left(\frac{1}{3}\theta^2+\frac{1}{2}(1+3w_t)\rho_t\right) \nonumber\\ 
		&&\quad \quad+\frac{w_m\rho_m}{(1+w_t)\rho_t}\tilde{\nabla}^2\Big)D^m_a +\Big(\frac{1}{2}\left(1+3w_{ch}\right)\rho_{ch}-\frac{w_{ch}\rho_{ch}}{(1+w_t)\rho_t}\left(\frac{1}{3}\theta^2+\frac{1}{2}(1+3w_t)\rho_t\right) \nonumber\\ 
		&&\quad \quad+\frac{w_{ch}\rho_{ch}}{(1+w_t)\rho_t}\tilde{\nabla}^2\Big)D^{ch}_a +\frac{\rho_{ch}}{(1+w_t)\rho_t}\Big(\rho_t-\frac{1}{3}\theta^2+\tilde{\nabla}^2\Big)a\tilde{\nabla}_a w_{ch}=0\;.
		\label{basic1}
	\end{eqnarray}
	
	Note that the above equations \ref{basic2} - \ref{basic1} are the generalised evolution equations of the perturbations for matter-Chaplygin gas models. 
	To simplify these evolution equations, we will consider the OCG, GCG, and MCG models by specifying the gradient of their equation of state parameters as $a\tilde{\nabla}_{a} w_{ch}$, and present the evolution equations for each model accordingly to analyse the growth of density contrasts with cosmic time. 
	
	
	\subsection{The evolution for the GCG model}\label{GCGq}
	From the GCG equation of state form \ref{CGpressure}, we have
	\begin{eqnarray}
		a\tilde{\nabla}_{a} w_{ch}= -w_{ch} (\alpha+1) D^{ch}_{a}\;.
	\end{eqnarray}
	
	Then, the evolution equations for GCG models are obtained as:
	\begin{eqnarray}
		&&\dot{D}^{m}_{a} +(1+w_{m}) Z_{a} - \frac{\theta w_{m} (1+ w_{m}) \rho_{m}}{\rho_{t}(1 + w_{t})} D^{m}_{a} + \frac{\theta \alpha w_{ch} (1+ w_{m}) \rho_{ch}}{\rho_{t} (1+ w_{t}) } D^{ch}_{a}=0\;,\\ 
		&&\dot{D}^{ch}_{a} +(1+w_{ch}) Z_{a} - \frac{\theta w_{m} (1+ w_{ch}) \rho_{m}}{\rho_{t}( 1+ w_{t})} D^{m}_{a} \nonumber\\
		&& \quad \quad -\frac{\theta w_{ch}}{ (1+w_{t}) \rho_{t}} \Big((1+ w_{m} )(1+\alpha)\rho_{m} + (1+ w_{ch}) \rho_{ch}\Big) D^{ch}_{a}=0\;,\\
		&&\dot{D}^{t}_{a} +(1+w_{t}) Z_{a} - \theta w_{t} D^{t}_{a}=0\; ,\\
		&&\dot{Z}_{a}+ \frac{2}{3} \theta Z_{a} +\Big\lbrace \frac{1}{2} (1+3w_{m})\rho_{m}+ \frac{w_{m} \rho_{m}}{\rho_{t}(1+ w_{t})} \Big(-\frac{1}{3} \theta^{2} -\frac{1}{2} \rho_{t}(1+3 w_{t}) + \tilde{\nabla}^{2}\Big)\Big\rbrace D^{m}_{a}\nonumber\\ 
		&&\quad \quad+\Big\lbrace \frac{1}{2} \rho_{ch}(1+3w_{ch})+\frac{\rho_{ch} w_{ch}\alpha}{(1+w_t)\rho_t}\left(\frac{1}{3} \theta^{2}- \tilde{\nabla}^{2}\right)\nonumber\\ 
		&&\quad \quad- \frac{\rho_{ch} w_{ch}}{(1+w_{t})} \Big( \frac{1}{2} (1+3w_{t}) +(\alpha+1)\Big) \Big\rbrace D^{ch}_{a}=0\;.
	\end{eqnarray}
	Applying the scalar and harmonic decomposition of the vector gradient variables defined in Eq. \ref{variables}, as presented in \cite{abebe2012covariant, sahlu2020scalar}, the evolution of the $kth$ harmonic of the scalar perturbations:
	
	\begin{eqnarray}
		&&\dot{\Delta}^{k}_{m} +(1+w_{m}) Z_{k} - \frac{\theta w_{m} (1+ w_{m}) \rho_{m}}{\rho_{t}(1 + w_{t})} \Delta^{k}_{m} + \frac{\theta \alpha w_{ch} (1+ w_{m}) \rho_{ch}}{\rho_{t} (1+ w_{t}) } \Delta^{k}_{ch}=0\;,\\ 
		&&\dot{\Delta}^{k}_{ch} +(1+w_{ch}) Z_{k} - \frac{\theta w_{m} (1+ w_{ch}) \rho_{m}}{\rho_{t}( 1+ w_{t})} \Delta^{k}_{m} \nonumber\\ 
		&&\quad \quad -\frac{\theta w_{ch}}{ (1+w_{t}) \rho_{t}} \Big[(1+ w_{m} )(1+\alpha)\rho_{m} + (1+ w_{ch}) \rho_{ch}\Big] \Delta^{k}_{ch}=0\;,\\ 
		&&\dot{\Delta}^{k}_{t} +(1+w_{t}) Z _{k}- \theta w_{t} \Delta^{k}_{t}=0\; ,\\
		&&\dot{Z}_{k}+ \frac{2}{3} \theta Z_{k} +\Big\lbrace \frac{1}{2} (1+3w_{m})\rho_{m} + \frac{w_{m} \rho_{m}}{\rho_{t}(1+ w_{t})} \Big[- \frac{1}{3} \theta^{2} -\frac{1}{2} \rho_{t}(1+ 3w_{t}) -\frac{k^2}{a^2}\Big]\Big\rbrace \Delta^{k}_{m}\nonumber\\
		&&\quad \quad+\Big\lbrace \frac{1}{2} \rho_{ch}(1+3w_{ch})+\frac{\rho_{ch} w_{ch}\alpha}{(1+w_t)\rho_t}\left(\frac{1}{3} \theta^{2}+\frac{k^{2}}{a^{2}}\right)\nonumber\\ 
		&&\quad \quad- \frac{\rho_{ch} w_{ch}}{(1+w_{t})} \Big( \frac{1}{2} (1+3w_{t}) +(\alpha+1)\Big) \Big\rbrace\Delta^{k}_{ch}=0\;.
	\end{eqnarray}
	
	In redshift space as presented in \cite{sahlu2020scalar, ntahompagaze2020multifluid, sami2021covariant}, the above equations can be recast as\footnote{From here on-wards, we drop the index $k$ indicating the harmonic to avoid overcrowding of notations.}:
	\begin{eqnarray}
		&&\Delta'_{m} + \frac{3 w_{m} (1+ w_{m}) \Omega_{m}}{\Omega_{t}(1 + w_{t})(1+z)} \Delta_{m} - \frac{3 \alpha w_{ch} (1+ w_{m}) \Omega_{ch}}{\Omega_{t} (1+ w_{t}) (1+z)} \Delta_{ch} -\frac{(1+w_{m})}{H(1+z)} Z =0\;,\\ 
		&& \Delta'_{ch} +\frac{3 w_{m} (1+ w_{ch}) \Omega_{m}}{\Omega_{t}( 1+ w_{t})(1+z)} \Delta_{m} +\frac{3 w_{ch}}{ (1+w_{t}) \Omega_{t}(1+z)} \Big[(1+ w_{m} )(1+\alpha)\Omega_{m} \nonumber\\ 
		&& \quad \quad+ (1+ w_{ch}) \Omega_{ch}\Big] \Delta_{ch}-\frac{(1+w_{ch})}{H(1+z)} Z =0\;,\\
		&&\Delta'_{t} +\frac{ 3 w_{t}}{(1+z)} \Delta_{t}-\frac{(1+w_{t})}{H(1+z)} Z=0\; ,\\&& Z'-\frac{2}{(1+z)} Z -\frac{3H}{(1+z)}\Big\lbrace \frac{1}{2} (1+3w_{m})\Omega_{m} \nonumber\\ 
		&& \quad \quad+ \frac{w_{m} \Omega_{m}}{\Omega_{t}(1+ w_{t})} \Big[ -1 -\frac{1}{2} \Omega_{t}(1+ 3w_{t}) -\frac{k^2}{3H^{2}a^2}\Big]\Big\rbrace \Delta_{m} \nonumber\\ 
		&&\quad \quad-\frac{3H}{(1+z)}\Big\lbrace \frac{1}{2} \Omega_{ch}(1+3w_{ch})+\frac{\Omega_{ch} w_{ch}\alpha}{(1+w_t)\Omega_t}\left(1+\frac{k^{2}}{3H^{2} a^{2}}\right)\nonumber\\ 
		&&\quad \quad- \frac{\Omega_{ch} w_{ch}}{(1+w_{t})} \Big( \frac{1}{2} (1+3w_{t}) +(\alpha+1)\Big) \Big\rbrace\Delta^{ch}=0\;.
	\end{eqnarray}
	To rewrite the equations in a dimensional uniform way, we define 
	\begin{equation}
		h\equiv H/H_0\;, {\mathcal{Z}}\equiv Z/H_0\;, \kappa\equiv k/H_0,
		\label{Newdefination}
	\end{equation}
	so that our new evolution equations for the GCG model are given as:
	\begin{eqnarray}
		&&\Delta'_{m} + \frac{3 w_{m} (1+ w_{m}) \Omega_{m}}{\Omega_{t}(1 + w_{t})(1+z)} \Delta_{m} - \frac{3 \alpha w_{ch} (1+ w_{m}) \Omega_{ch}}{\Omega_{t} (1+ w_{t}) (1+z)} \Delta_{ch} - \frac{(1+w_{m})}{(1+z)h}\mathcal{Z} =0\;,\label{dd}\\ 
		&&\Delta'_{ch} +\frac{3 w_{ch}}{(1+z) (1+w_{t}) \Omega_{t}} \Big((1+ w_{m} )(1+\alpha)\Omega_{m} + (1+ w_{ch}) \Omega_{ch}\Big) \Delta_{ch} \nonumber\\
		&& \quad \quad+\frac{3 w_{m} (1+ w_{ch}) \Omega_{m}}{\Omega_{t}( 1+ w_{t})(1+z)} \Delta_{m}-\frac{(1+w_{ch})}{(1+z)h} \mathcal{Z} =0\;,\label{evolution1}\\
		&&{\Delta}'_t +3\left(\frac{1}{1+z}\right) w_t\Delta_t-\left(\frac{1}{1+z}\right)\frac{(1+w_t)}{h}\mathcal{Z}=0 \;,\\
		&&\mathcal{Z}' -\frac{2}{(1+z)} \mathcal{Z} \nonumber\\ 
		&&\quad \quad -\frac{3h}{(1+z)}\left[\frac{1}{2}\left(1+3w_m\right)\Omega_m-\frac{w_m\Omega_m}{(1+w_t)\Omega_t}\left(1+\frac{1}{2}(1+3w_t)\Omega_t+\frac{k^{2} (1+z)^{2}}{3h^{2}} \right)\right]\Delta^m \nonumber\\
		&& \quad \quad- \frac{3h}{(1+z)}\Bigg[\frac{1}{2} \Omega_{ch}(1+3w_{ch})+\frac{\Omega_{ch} w_{ch}\alpha}{(1+w_t)\Omega_t}\left(1+\frac{k^{2} (1+z)^{2}}{3h^{2}}\right)\nonumber\\
		&& \quad \quad- \frac{\Omega_{ch} w_{ch}}{(1+w_{t})} \Big( \frac{1}{2} (1+3w_{t}) +(\alpha+1)\Big) \Bigg]\Delta^{ch}=0\;.\label{evolution5}
	\end{eqnarray}
	
	For the case of $\alpha = 1$, the evolution equations of OCG can be found from these equations \ref{dd} - \ref{evolution5}. The numerical results of density contrast in OCG and GCG models will be presented in Sec. \ref{final1}, to investigate the contribution of the Chaplygin gas-like fluid to large-scale formation.
	
	
	\subsection{The evolution of the perturbations for MCG model}\label{GCGq}
	
	With the equation of state \ref{eqstat} for the MCG model, the term $a\tilde{\nabla}_a w_{ch}$ becomes:
	\begin{eqnarray}
		a\tilde{\nabla}_a w_{ch} = (\alpha+1)(B-w_{ch})D^{ch}_{a}\;.
	\end{eqnarray}
	
	By applying the same manners as presented in the GCG model the evolution equation for MCG read as:
	\begin{eqnarray} 
		&&{\Delta}'_m+\left(\frac{3(1+w_{m})}{(1+w_t)}\right)\frac{w_{m} \Omega_{m}}{(1+z) \Omega_{t}} \Delta_{m}\nonumber\\ 
		&&\quad \quad+\left(\frac{3(1+w_{m})}{(1+w_t)}\right)\frac{ \Omega_{ch}}{(1+z) \Omega_{t}}\bigg(B+ \alpha(B-w_{ch})\bigg)\Delta^{ch} -\frac{(1+w_{m})}{h(1+z)}\mathcal{Z} =0 \;,\\
		&&{\Delta}'_{ch}+\frac{3 }{ (1+w_{t}) (1+z)\Omega_{t}} \Big( w_{ch} (1+ w_{ch}) \Omega_{ch}-(1+ w_{m} )(1+\alpha) (B-w_{ch})\Omega_{m}\Big) \Delta^{ch} \nonumber\\
		&& \quad \quad+\frac{3(1+w_{ch})w_m\Omega_m}{(1+z)(1+w_t) \Omega_{t}}\Delta^{m}-\frac{(1+w_{ch})}{h(1+z)} \mathcal{Z} =0 \;,\\
		&&{\Delta}'_t +3\left(\frac{1}{1+z}\right) w_t\Delta_t-\left(\frac{1}{1+z}\right)\frac{(1+w_t)}{h}\mathcal{Z}=0 \;,\\ \label{e12}
		&&\label{ZMCG} \mathcal{Z}' -\frac{2}{(1+z)} \mathcal{Z} \nonumber\\ 
		&&\quad \quad -\frac{3h}{(1+z)}\left[\frac{1}{2}\left(1+3w_m\right)\Omega_m-\frac{w_m\Omega_m}{(1+w_t)\Omega_t}\left(1+\frac{1}{2}(1+3w_t)\Omega_t+\frac{k^{2} (1+z)^{2}}{3h^{2}} \right)\right]\Delta^m \nonumber\\ 
		&&\quad \quad- \frac{3h}{(1+z)}\Bigg[\frac{1}{2} \Omega_{ch}(1+3w_{ch})+\frac{\Omega_{ch} \Big(w_{ch} + (\alpha+1) (B-w_{ch})\Big)}{(1+w_t)\Omega_t}\left(-1-\frac{k^{2} (1+z)^{2}}{3h^{2}}\right)\nonumber\\
		&& \quad \quad- \frac{\Omega_{ch}}{(1+w_{t})} \Big( \frac{1}{2} w_{ch} (1+3w_{t}) -(\alpha+1) (B-w_{ch})\Big) \Bigg]\Delta^{ch}=0\;.
		\label{e11}
	\end{eqnarray}
	
	Note that for the case of $B =0 $ above equations \ref{e11} - \ref{e12} are reduced to Eqs \ref{dd} - \ref{evolution5}. 
	
	
	\subsection{The evolution of perturbations for the ECG model}\label{ECGq}
	
	From the ECG equation of state form \ref{ECG0}, we can write
	\begin{equation}
		a\tilde{\nabla}_{a} w_{ch}= (\alpha+2) A_{2} \tilde{\nabla}_{a} \rho_{ch}\;.
	\end{equation}
	
	The evolution equations of the perturbations for this model can be given by:
	\begin{eqnarray}
		&&\Delta'_m-\frac{(1+w_{m}) }{h(1+z)} \mathcal{Z}+ \frac{3 w_{m} (1+ w_{m}) \Omega_{m}}{\Omega_{t}(1 + w_{t})(1+z)}\Delta_m \nonumber\\ 
		&& \quad \quad+ \frac{3 (1+ w_{m}) \Omega_{ch}}{\Omega_{t} (1+ w_{t})(1+z) } \Big( w_{ch} +\frac{ \Omega_{ch} (\alpha+2)h^{2} }{D_{1}}\Big) \Delta_{ch}=0\;,\\
		&&\Delta’_{ch}-\frac{(1+w_{m}) }{h(1+z)} \mathcal{Z}+ \frac{3 w_{m} (1+ w_{ch}) \Omega_{m}}{\Omega_{t}( 1+ w_{t}) (1+z)} \Delta_{m} \nonumber\\
		&& \quad \quad +\frac{3\Omega_{ch}}{ (1+w_{t} )(1+z) \Omega_{t}} \Big(w_{ch} (1+w_{ch})- \frac{ (\alpha+2) (1+w_{m})\Omega_{m} h^{2}}{D_{1}}\Big) \Delta_{ch}=0\;,\\
		&&\Delta'_{t} +\frac{ 3 w_{t}}{(1+z)} \Delta_{t}-\frac{(1+w_{t})}{h(1+z)} Z=0\; ,\\
		&&\mathcal{Z}’- \frac{2}{(1+z)} \mathcal{Z}  \nonumber\\
		&& \quad \quad-\frac{3h}{(1+z)} \Big[ \frac{1}{2} (1+3w_{m}) \Omega_{m} - \frac{w_{m} \Omega_{m}}{\Omega_{t} (1+w_{t})} \big(1 + \frac{1}{2} (1+3w_{t}) \Omega_{t} +\frac{\kappa^{2} (1+z)^{2}}{3h^{2}}\big)\Big] \Delta_{m} \nonumber\\
		&& \quad \quad -\frac{3h}{(1+z)} \Big[ \frac{1}{2} (1+3w_{ch}) \Omega_{ch} + \frac{\Omega_{ch}}{\Omega_{t} (1+w_{t})} \big( w_{ch} + \frac{(\alpha +2) h^{2} \Omega_{ch}}{D_{1}}\big)\big( -1-\frac{\kappa^{2}(1+z)^{2}}{3h^{2}} \big) \nonumber\\
		&& \quad \quad + \frac{\Omega_{ch}}{ (1+w_{t})}\big( -\frac{1}{2} w_{ch} (1+3w_{t}) +\frac{ (\alpha+ 2) h^{2} \Omega_{ch}}{D_{1}}\big) \Big] \Delta_{ch}=0\;.
	\end{eqnarray}
	The solutions to these equations are highly unstable to any choice of initial conditions, and the model did poorly in the background analysis. Therefore, we think that it is not worth pursuing this particular model with the particular choice of the defining parameters that mimic Ref. \cite{kahya2014observational}, but slightly adapted for dimensional reasons, any further.
	
	
	\subsection{The growth of the normalised density contrast in the GR and $\Lambda$CDM limits}\label{final}
	
	From Eqs. \ref{evolution1}, in the absence of Chaplygin gas-like fluids the linear evolution equation is reduced to the well-known results in GR:
	\begin{eqnarray}
		&&\Delta'_{m} + \frac{3  w_{m}}{(1+z)} \Delta_{m} -\frac{(1+w_{m})}{(1+z)h} \mathcal{Z} =0\;,\\
		&&\mathcal{Z}'-\frac{2}{(1+z)}\mathcal{Z} -\frac{3h}{(1+z)(1+w_{m})}\Big\lbrace \frac{1}{2} (1+3w_{m})\Omega_{m}\\
		&& \quad \quad+ w_{m} \Big[1 +\frac{\kappa^2}{h^2} (1+z)^2\Big]\Big\rbrace \Delta_{m}=0\;.\nonumber
		\label{evolutionxs11}
	\end{eqnarray} 
	
	The normalized energy density  can be defined as presented in \cite{sahlu2020scalar}:
	\begin{equation}
		\delta_{[m,ch]}(z)=\frac{\Delta _{[m,ch]}(z)}{\Delta (z_{in})} 
		\label{normalizeddensity} \;,
	\end{equation}
	where $\Delta_{in}$ is the initial value of $ \Delta_{[m,ch]}(z)$ at $z_{in} = 20$. The numerical results of equation \ref{evolutionxs11} are presented in Fig. \ref{fig:GR}. From the plot we see the contribution of baryonic matter to the formation of large-scale structures, which is expected theoretically and observationally.
	\begin{figure}[ht!]
		\includegraphics[width=6cm,height= 5cm ]{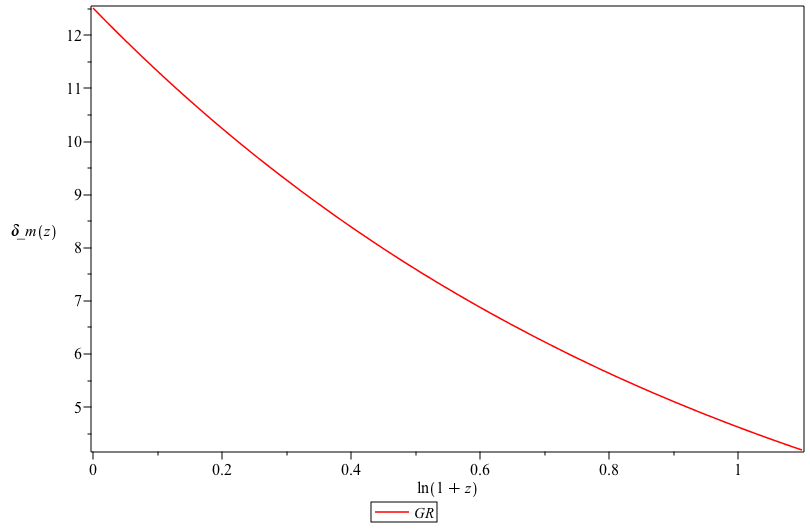}
		\includegraphics[width=6cm,height= 5cm ]{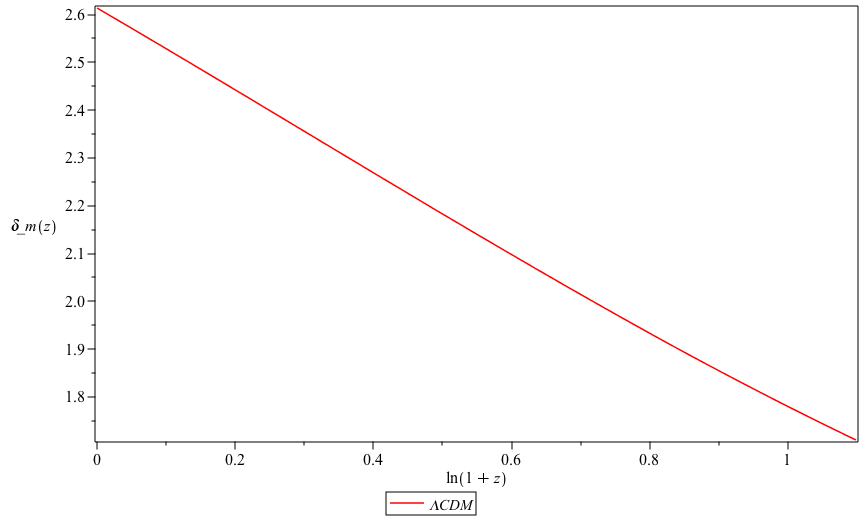}
		\caption{Left: Density contrasts of baryonic matter with the redshift $z$ in standard GR (without cosmological constant). Right: The density contrast of baryonic matter with the redshift $z$ in $\Lambda$ CDM.}
		\label{fig:GR}
	\end{figure}
	
	For the $\Lambda$CDM case, the linear evolution equations are given as:
	\begin{eqnarray}
		&&\Delta'_{m} + \frac{3 w_{m} \Omega_{m} }{\Omega_{t}(1+z)} \Delta_{m} -\frac{(1+w_{m})}{(1+z)h} \mathcal{Z} =0\;,\\
		&&\mathcal{Z}' - \frac{2}{(1+z)} \mathcal{Z} -\frac{3h}{(1+z)}\Big[ \frac{1}{2} \Omega _{m} (1+3 \omega_{m}) \nonumber \\
		&& \quad \quad -\frac{ w_m\Omega_m}{ \left(1+w_t\right)\Omega_t} \Big(1 +\frac{1}{2}\Omega_{m} (1+3w_m )- \frac{1}{2}\Omega_{\Lambda} +\frac{(1+z)^{2}k^2}{3h^{2} }\Big)\Big] \Delta^m=0 \;,
		\label{LCDM}
	\end{eqnarray}
	and the solution for the density contrast is also presented in Fig. \ref{fig:GR}. As expected, one observes a lower structure growth amplitude in an accelerated background than in the pure GR limit. In the following, we show how the addition of the Chaplygin gas in the CG models considered affects the growth rate of structure formation, compared to the pure limits of GR and $\Lambda$ CDM.
	
	
	\subsection{The growth of perturbations in baryonic matter-Chaplygin gas mixture}\label{final1}
	
	In this subsection, we explain the contribution of these coupling systems (i.e., baryonic and Chaplygin gas)  for the formation of large-scale structures. We assume the contribution of relativistic matter (radiation) is negligible for the contribution of the formation of large-scale structures today. The long-wavelength mode, where $\kappa \ll 1 $, and the short-wavelength mode within the horizon, where $\kappa \gg 1$ are considered in the whole system to present the numerical results of density contrasts in OCG, GCG, and MCG. From Tab. \ref{tab: best-fitting parameter values}, we use the values of the best-fitting parameters for each model.
	In the following, Figs. \ref{fig:baryonic} - \ref{fig:baryonic1} we present the behaviour of the density contrast in short- and long-wavelength modes for baryonic fluid, total and Chaplygin gas, or exotic fluids with cosmological red-shift. From the results, we observe that the amplitude of density contrast is  decaying and oscillating through cosmological redshift in the short wavelength mode. However, in the long-wavelength mode, the amplitude of density contrast is growing through cosmological redshift and shows the significance of baryonic matter-Chaplygin gas for the formation of large scale. We consider the long-wavelength mode for further investigation in Sec. \ref{sigma}
	\begin{figure}[ht!]
		\includegraphics[width=6cm,height= 5cm ]{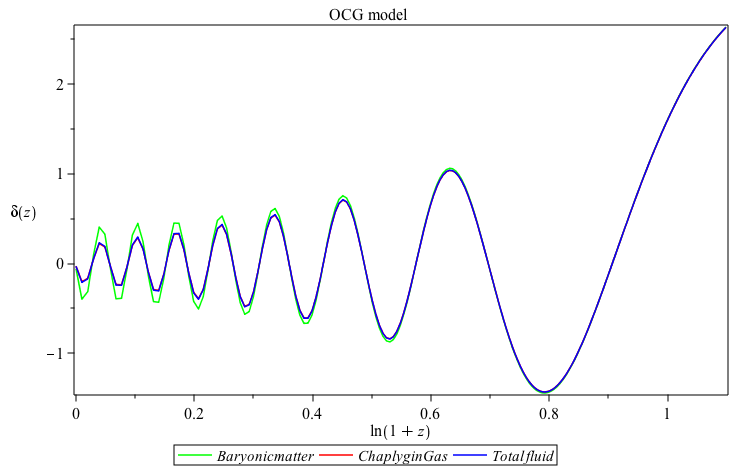}
		\includegraphics[width=6cm,height= 5cm ]{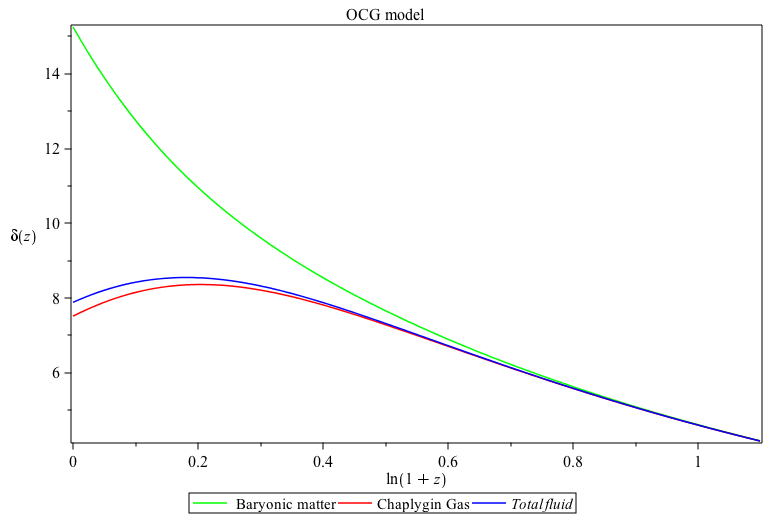}
		\caption{Left: The density contrasts of the baryonic-Chaplygin gas with $z$ for the short-wavelength mode at $\kappa =100$ for the OCG model. Right: The density contrasts of baryonic matter-Chaplygin gas with $z$ for long-wavelength mode at $\kappa =0$ for OCG model.} 
		\label{fig:baryonic}
	\end{figure} 
	
	\begin{figure}[ht!]
		\includegraphics[width=6cm,height= 5cm ]{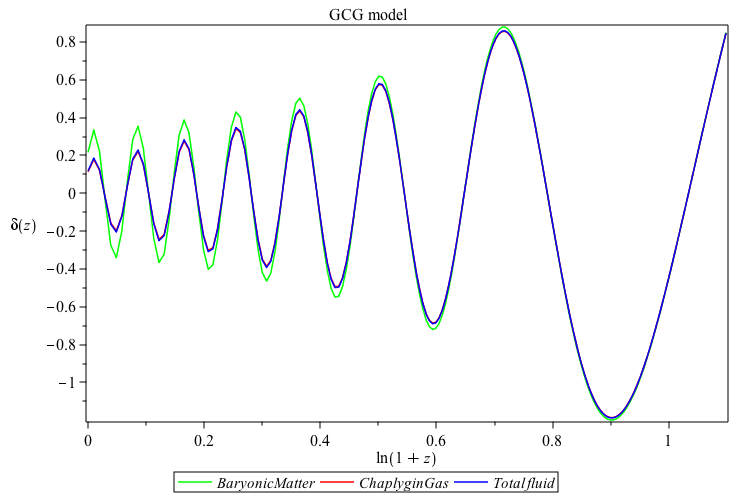}
		\includegraphics[width=6cm,height= 5cm ]{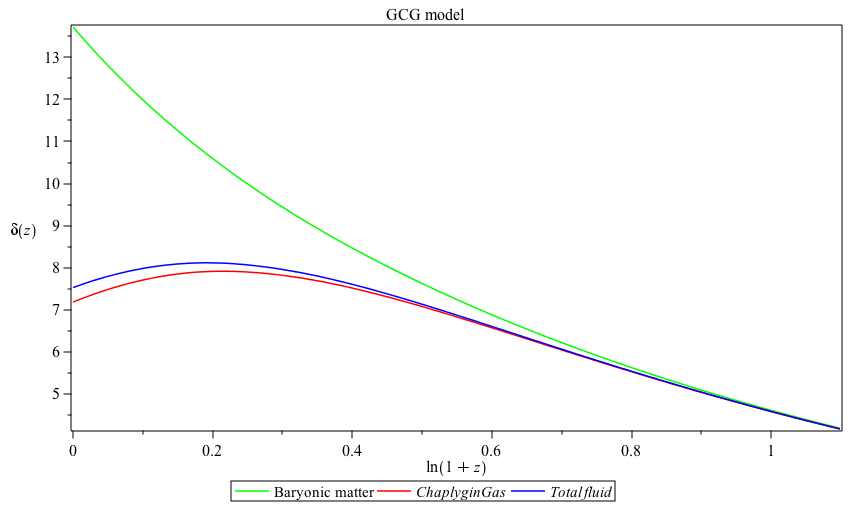}
		\caption{Left: The density contrasts of the baryonic-Chaplygin gas with $z$ for the short-wavelength mode at $\kappa =10$ for the GCG model. Right: The density contrasts of baryonic matter - Chaplygin gas with $z$ for long-wavelength mode at $\kappa =0$ for the GCG model. We use $\alpha = 0.6424$ from Table \ref{tab: best-fitting parameter values} for illustrative purposes.}
		\label{fig:baryonic2}
	\end{figure}
	
	\begin{figure}[ht!]
		\includegraphics[width=6cm,height= 5cm ]{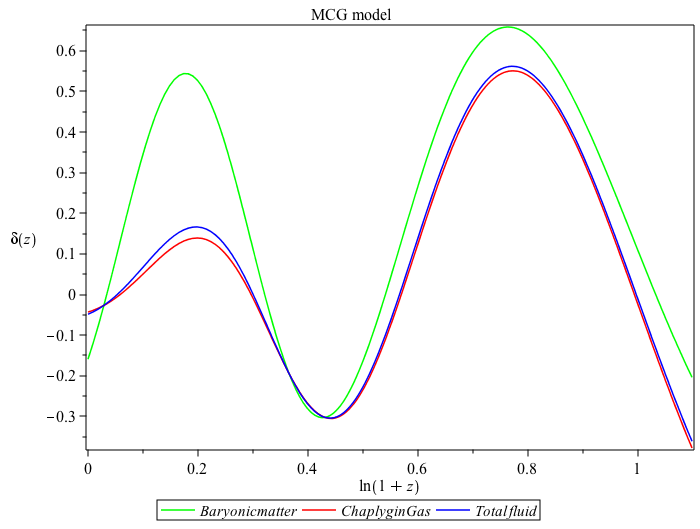}
		\includegraphics[width=6cm,height= 5cm ]{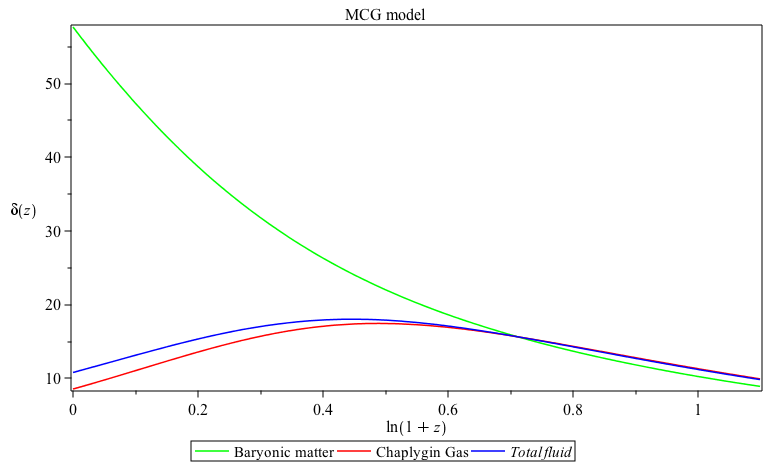}
		\caption{The growth of density contrasts in the baryonic matter-Chaplygin gas mixture for the short-wavelength mode at $\kappa =100$ for MCG model (left) and long-wavelength mode at $\kappa =0$ (right). We use $B= 0.1448$ and $\alpha = 0.5694$ from Table \ref{tab: best-fitting parameter values} for illustrative purposes.}
		\label{fig:baryonic1}
	\end{figure}
	\subsection{Various CG models with growth rate data}\label{sigma}
	
	{In this section, the linear growth of the structure has been studied using the most recent $f_{\sigma 8}$ data using the frame of $1+3$ covariant formalism. We admit the definition of the growth rate $f$ as presented in: \cite{barros2019coupled, sagredo2018internal, song2009reconstructing}
		\begin{equation}
			f(z) = \frac{\Delta'_m(z)}{\Delta_m(z = 0)}\;,
		\end{equation}
		and the root mean square (RMS) normalization of the matter power-spectrum $\sigma_8$ as:
		\begin{equation}
			\sigma 8 = \sigma_{8,0}\frac{\Delta'_m(z)}{\Delta_m(z=0)}\;,
		\end{equation}
	}
	
	{For illustrative purpose we have used $\sigma_{8,0} = 0.829$ \cite{ade2016planck}. As presented in \cite{sagredo2018internal}, the combination of $f(z)$ and $\sigma_{8,0}$ yield as:
		\begin{equation}
			f_{\sigma 8}(z) = \sigma_{8,0}\frac{\Delta'_m(z)}{(1+z)\Delta_m(z=0)}\;, \label{fsigma}
		\end{equation}
		We considered the first-order evolution equations of $\Lambda$CDM, OCG, GCG, and MCG models to present the contributions of CG for the growth structure with the $f_{\sigma 8}(z)$ data as presented in Fig. \ref{fig:sigma} using Eq. \ref{fsigma}.  We also present the numerical plots of the three models with data to see the significant contributions to the cosmic growth structure. For further analysis, the statistical values of $\chi^2$, $AIC$, $\Delta AIC$, $BIC$, and $\Delta BIC$ are calculated as presented in Tab. \ref{Tab:sigma8}.  From our statistical results, the values of $\Delta IC$ for the OCG and GCG models are selected but MCG mode 
		is out of range of the Jeffrey scale \cite{nesseris2013jeffreys} and it is going to be ruled out in the perturbative level. 
		\begin{figure}[ht!]
			\centering
			\includegraphics[width=12cm,height=8cm]{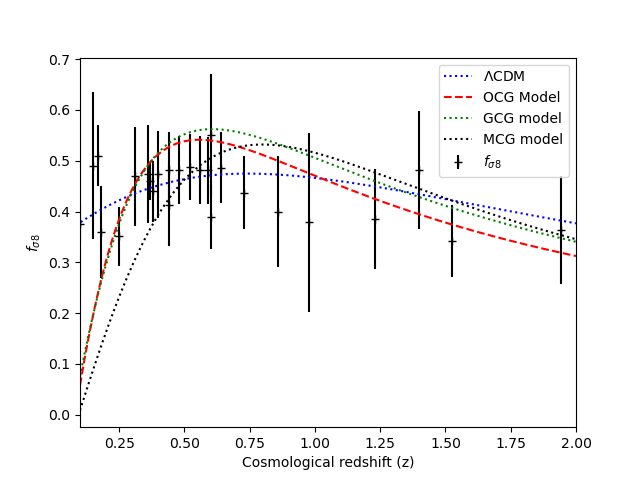}
			\caption{Evolution of $f_{\sigma 8}(z)$ with cosmological redshift ($z$) in $\Lambda$CDM, OCG, GCG and MCG models using $f_{\sigma 8}(z)$ data.}
			\label{fig:sigma}
		\end{figure}
	}
	\begin{table}
		\centering
		\scalebox{0.7}{
			\begin{tabular}{llllllll}
				\hline\noalign{\smallskip}
				\textbf{Model} & \textbf{$\mathcal{L}(\hat{\theta}|data)$} & \textbf{$\chi ^{2}$} & \textbf{Red. $\chi ^{2}$} & \textbf{$AIC$} &\textbf{$|\Delta AIC|$} & \textbf{$BIC$} & \textbf{$|\Delta BIC|$}\\
				\noalign{\smallskip}\hline\noalign{\smallskip}
				$\Lambda$CDM & -116.6867 & 233.3734 & 8.3347 &237.3734 & 0 & 240.1758 & 0\\
				\noalign{\smallskip}
				Original CG &-116.6867& 233.3734 & 8.9759 & 241.3734 & 4.0014 & 240.1758& 0\\
				\noalign{\smallskip}
				General CG & -114.3674 & 228.7348 & 8.7974 & 236.7348 & 0.6382 & 235.5372& 4.6386\\
				\noalign{\smallskip}
				MCG & -260.4506 & 520.9013 & 21.7042 & 532.90133 & 295.5279 & 527.7037 &287.5279\\
				\noalign{\smallskip}\hline
		\end{tabular}}
		\caption{The best fit for $\Lambda$CDM model, OCG, GCG, MCG models using $f_{\sigma 8}$.}
		\label{Tab:sigma8}
	\end{table}
	
	\section{Conclusions}
	{In this manuscript, the background cosmological parameters have been studied and compared against supernova cosmological data for different Chaplygin gas models, namely: OCG, GCG, MCG, MGCG, ECG, and GCCG. A comparison of the deceleration parameters of the various CG models with $\Lambda$ CDM has been made to identify which model can best explain the late-time accelerating universe. From Fig. \ref{fig: Best-fitting deceleration parameters} both the specific solution for the MGCG, ECG, and GCCG models can be ruled out, while our AIC and BIC criteria analyses supported this conclusion for the GCCG model. We also considered the comparison of the residuals of the various CG models and the $\Lambda$CDM as well, and from the results, we observe that the ECG model is unrealistic compared to the other CG models considered in the paper, due to the hard cut-off at $z=1$, which also caused the problems from Fig. \ref{fig: Best-fitting deceleration parameters}. However, our statistical analysis test indicated that the OCG model, and to an extent the GCG can be alternative models relative to the ``accepted/true model" by using the AIC and BIC selection criteria. Based on the analysis, the MCG model has substantially/less observational support using OHD and SNIa data, respectively, and is considered for the perturbation level. This means that although this specific ECG model is ruled out, statistical analysis indicated that the ECG model as a whole deserves further testing, just not for this particular solution.}
	
	Linear cosmological perturbations have also been investigated in a matter-Chaplygin gas multifluid setting around the FLRW background using the $1+3$ covariant formalism. On the basis of our statistical analysis for the background cosmology, the four Chaplygin gas models that performed the best, namely OCG, GCG, and MCG, were considered for the study of large-scale structure formation. From Figs. \ref{fig:baryonic} - \ref{fig:baryonic1}, we observe the contribution of baryonic gas, Chaplygin gas, and total fluid to the formation of large-scale structures. The results show that as the density contrast of baryonic matter increases with cosmic time (decreases with redshift), the density contrast of the Chaplygin gas decays with cosmic time. It can be seen that the Chaplygin gas supports the formation of large-scale structures (as it behaves like dark matter) in the early universe but discourages it at late times as it behaves like dark energy. We also consider these models to study the growth of cosmic structure with $f_{\sigma 8}$ data, as presented in Fig. \ref{fig:sigma}.
	
	In general, both the OCG and the GCG provide cosmological scenarios that are more or less consistent with the $\Lambda$CDM model, both at the background and perturbation levels, but the former has better observational support based on the statistical criteria described here. From our statistical analysis, the MCG model has $\Delta IC > 7$, which means, although not ruled out in the background but rejected at the perturbation level, that it does not have significant observational support. We went a step further and investigated the perturbations of this model and found that even at this level, the growth rates are not in line with the expected results $\Lambda$CDM. The MGCG and GCCG models that were rejected by the background statistical criteria were not included in our analysis of the perturbations.
	
	
	\section{Acknowledgments}
	RH also acknowledges the NRF for financial support for this work with grant number 120850. AA acknowledges that this work is based on research supported in part by the NRF with grant number 112131.

	\section{References*}
	\bibliographystyle{unsrt}
	
\end{document}